\documentclass[pdflatex,iicol,sn-mathphys-num]{sn-jnl}% Math and Physical Sciences Numbered Reference Style
\usepackage{graphicx}%
\usepackage{multirow}%
\usepackage{amsmath,amssymb,amsfonts}%
\usepackage{amsthm}%
\usepackage{mathrsfs}%
\usepackage[title]{appendix}%
\usepackage{xcolor}%
\usepackage{textcomp}%
\usepackage{manyfoot}%
\usepackage{booktabs}%
\usepackage{algorithm}%
\usepackage{algorithmicx}%
\usepackage{algpseudocode}%
\usepackage{listings}%
\usepackage{xspace}
\usepackage{enumitem}
\usepackage{placeins}
\usepackage{mdframed}
\usepackage{tabularx}
\usepackage{fancyvrb}

\newcommand{\ruletitle}[2]{%
\vspace{0.6em}%
\noindent\textbf{Rule #1} -- #2.\par\vspace{0.3em}}

\newcommand{\uschema}{U-Schema\xspace}

\newcommand{\setS}{\texttt{S}\xspace}
\newcommand{\setM}{\texttt{M}\xspace}
\newcommand{\setL}{\texttt{L}\xspace}

\lstdefinelanguage{athena}{
  morekeywords={
    FSet, Schema, Root, Entity, Relationship, Common, Variation, Identifier,
    Date, String, List, Number, Ref, Aggr, Timestamp, Integer,
    Decimal, Boolean, SQL, CREATE, TABLE, VARCHAR, NOT, NULL,
    FOREIGN, PRIMARY, KEY, REFERENCES, as, U, in, Import, Option
  },
  sensitive=true,
  morecomment=[l]{//},
  morecomment=[s]{/*}{*/},
  morestring=[b]"
}

\lstdefinestyle{athenaStyle}{
  language=athena,
  keywordstyle=\bfseries\color{blue},
  commentstyle=\itshape\color{gray},
  stringstyle=\color{black}
}

\lstdefinelanguage{orion}{
  morekeywords={
    USING, ADD, DELETE, RENAME, SPLIT, EXTRACT, MERGE,
    COPY, MOVE, NEST, UNNEST,
    ATTR, CAST, PROMOTE, DEMOTE,
    REF, MULT, MORPH,
    AGGR, ENTITY, RELATIONSHIP,
    ADAPT, DELVAR, UNION,
    TO, INTO, WHERE, AS,
    rmId, rmEntity
  },
  sensitive=true,
  morecomment=[l]{//},
  morecomment=[s]{/*}{*/},
  morestring=[b]"
}

\lstdefinestyle{orionStyle}{
  language=orion,
  keywordstyle=\bfseries\color{blue},
  commentstyle=\itshape\color{gray}
}

\lstdefinelanguage{json}{
  morestring=[b]",
  sensitive=true
}

\lstdefinelanguage{sql}{
  morekeywords={
    CREATE, TABLE, PRIMARY, KEY, FOREIGN, REFERENCES,
    NOT, NULL, DEFAULT, BOOLEAN, CHAR, VARCHAR, INT, DATE
  },
  sensitive=false,
  morecomment=[l]{--},
  morestring=[b]'
}

\lstdefinestyle{sqlStyle}{
  language=sql,
  keywordstyle=\bfseries\color{blue}
}

\definecolor{jsonBrace}{RGB}{190,45,45}       % { }
\definecolor{jsonBracket}{RGB}{110,80,165}    % [ ]
\definecolor{jsonColon}{RGB}{45,95,180}       % :
\definecolor{jsonString}{RGB}{40,145,75}       % valores string
\definecolor{jsonQuote}{RGB}{20,105,55}        % comillas
\definecolor{jsonBoolean}{RGB}{150,70,150}     % true/false/null
\definecolor{jsonNumber}{RGB}{190,105,20}      % números
\definecolor{jsonComment}{RGB}{120,120,120}    % comentarios

\lstdefinestyle{jsonStyle}{
    language=json,
   commentstyle=\color{jsonComment}\itshape,
    morecomment=[l]{//},
    morestring=[b]",
    stringstyle=\color{jsonString},
    morekeywords={true,false,null},
    keywordstyle=\color{jsonBoolean},
    literate=
        *{\{}{{{\color{jsonBrace}\{}}}{1}
         {\}}{{{\color{jsonBrace}\}}}}{1}
         {[}{{{\color{jsonBracket}[}}}{1}
         {]}{{{\color{jsonBracket}]}}}{1}
         {:}{{{\color{jsonColon}:}}}{1},
    breaklines=true,
    showstringspaces=false,
    columns=fullflexible
}

\theoremstyle{thmstyleone}%
\theoremstyle{thmstyletwo}%

\theoremstyle{thmstylethree}%

\begin{document}

\title[A Model-Driven Approach to Database Migration]
{A Model-Driven Approach to Database Migration with a Unified Data Model}

% Add * to the corresponding author
\author*[1]{\fnm{Mar\'ia-Jos\'e} \sur{Ort\'in}}
\email{mjortin@um.es}

\author[1]{\fnm{Jos\'e R.} \sur{Hoyos}}
\email{jose.hoyos@um.es}

\author[1]{\fnm{Jes\'us J.} \sur{Garc\'ia-Molina}}
\email{jmolina@um.es}

\affil*[1]{%
  \orgdiv{Facultad de Informática},
  \orgname{Universidad de Murcia},
  \orgaddress{
    \street{Campus de Espinardo},
    \city{Murcia},
    \postcode{30100},
    \country{Spain}
  }
}

%% Abstract para Sosym

\abstract{
Database migration is a key task in software modernization that
increasingly requires transformations across heterogeneous data models.
Existing approaches are commonly tailored to specific source--target
combinations, making extension costly as the number of supported
models grows.

This paper presents a model-driven database migration approach based on
\uschema{} as a pivot metamodel. Explicit mappings are defined between
each supported data model and \uschema{} and composed to derive
source-to-target schema transformations, reducing the number of required
mappings from quadratic to linear in the number of supported models.
These mappings formalize the structural patterns required to transform
schemas across data models. During schema transformation, fine-grained
trace links are generated between model elements and later used to drive
data migration through model-independent source-data access. The
intermediate \uschema{} model also provides an adaptation point, where
selected schema changes are applied together with consistent updates to
the trace.

The current implementation supports relational and document data models
through four mappings. The evaluation focuses on relational-to-document
migration using a synthetic case study, the Northwind sample dataset,
and the TPC-C benchmark. It assesses generated schema models against
expected models, trace-aware schema adaptation, schema-information
preservation through round-trip source-schema reconstruction,
data-semantic preservation through query-result equivalence, and
operational feasibility and scaling behavior. The generated schemas
matched the expected models for the evaluated cases, the round-trip
experiments showed high structural preservation, and corresponding
source and target queries produced equivalent results. The largest
experiment processed nearly 50 million source rows without throughput
degradation over the evaluated range.
% while flat-copy baselines provide a reference for the additional
% execution cost relative to direct relational-to-document copying.
}

%% Keywords

\maketitle

%%%%%%%%%%%%%%%%%%%%%%%%%%%%%%%%%%%%%%%%%%%%%%%%%%%%%%%%%%%%
\section{Introduction}
\label{sec:introduction} 

Database migration is a common task in software modernization, where data
must be moved from a source system to a target system that better satisfies
current requirements. For decades, most migrations were performed within
the relational model. This situation changed with the emergence of NoSQL
(\emph{Not-only SQL}) systems, which introduced alternative data models
designed to overcome some limitations of relational databases.

NoSQL systems were developed to support data-intensive applications,
providing flexibility in schema evolution, horizontal scalability, and
high availability. Many of them are schemaless and follow a
schema-on-read approach, supporting data abstractions such as column
families, documents, key–value pairs, or graphs~\cite{fowler-nosql2012}.
In contrast to the relational model, where foreign keys are the main
mechanism to relate data, NoSQL models provide constructs such as nested
objects, aggregation hierarchies, or explicit relationship types.
However, these models lack a standard specification, and each paradigm—
and often each system—defines its own data model.

As a consequence of this evolution, the database landscape has become heterogeneous,
as illustrated by the DB-Engines ranking~\cite{dbengines},
although relational systems remain the most widely used by a considerable margin. 
This ranking reflects not only the
growing relevance of NoSQL systems, but also the evolution of major relational
DBMSs towards multi-model platforms with native support for document data
(e.g., PostgreSQL JSONB, Oracle Multimodel, MySQL JSON).
In parallel, research on \emph{polystore} systems 
has gained increasing attention~\cite{panse2022-polyglot},
demonstrating unified query and data management across heterogeneous models.
Moreover, the adoption of \emph{polyglot persistence}~\cite{fowler-nosql2012}
has grown, with different data models coexisting within the same system
to address diverse application requirements~\cite{roy-hubaraSS2022}.

In this context, database migration increasingly involves transformations
between different data models. Existing approaches are usually designed
for specific source–target combinations, most often relational to
document databases~\cite{Jia2016,Kim2020,kuszera2019,Rocha2015,scavuzzo2014,Schreiner2020}.
However, migrations may involve different combinations of data models and defining
dedicated transformations for each combination does not scale, as the
number of mappings grows quadratically with the number of supported
paradigms. Traditionally, this issue has been tackled by using a generic or
unified data model as a common intermediate representation, reducing
the number of required mappings from quadratic to linear in the
number of supported data models~\cite{hainaut1994,atzeni2009,
carlos-uschema-2022}.
For example, the DB-Main approach~\cite{hainaut1994,Hainaut2008}
applied this idea in the mid-1990s, when alternatives to relational
systems such as XML and object-oriented databases began to emerge.
More recently, unified data models have been proposed 
to provide uniform access to NoSQL systems~\cite{atzeni2009} 
and \uschema{}~\cite{carlos-uschema-2022} to
have a unified view of relational and NoSQL data models 
(document, column-family, graph, and key–value).

This paper presents a model-driven database migration approach based on
\uschema{}. The work introducing this unified data model~\cite{carlos-uschema-2022}
defined basic bidirectional correspondences showing how the core
constructs of relational and NoSQL data models can be represented in
\uschema{}, and identified database migration as one of its potential
application scenarios. Building on this foundation, this work develops
substantially richer, migration-oriented mappings that address the
structural configurations required for automated schema and data
migration across heterogeneous data models. 

The approach focuses on logical migration, independently of deployment concerns 
such as cloud migration or platform-specific configuration.

The mappings are composed into a model transformation chain 
in which a target schema is systematically derived from the source schema 
through a \uschema{} intermediate schema.
The transformations also generate fine-grained trace links between
source, \uschema{}, and target model elements. These traces are used
during data migration to access the source database in a
model-independent way and to guide the construction of target data.

The resulting target schema corresponds to a canonical representation,
obtained by applying predefined conventions when several alternative
target structures are possible, independently of application-specific
optimization or workload considerations. In practice, such canonical
schemas may require adaptation to meet application-specific requirements.
In our proposal, this adaptation is performed at the level of the
\uschema{} model, allowing designers to apply changes before target-schema
generation while consistently updating the corresponding trace links.
Such adaptations can be expressed using the generic schema evolution
language Orion~\cite{alberto-orion-2024}, and the current prototype
implements a selected set of these operations.

The approach currently supports relational and document data models,
with mappings defined in both directions through the \uschema{} pivot
model. The empirical evaluation focuses on the relational-to-document
migration scenario and uses three datasets: a synthetic Music Streaming
case study, the Northwind sample dataset, and the TPC-C benchmark.
The experimentation includes schema-level validation by comparing the
generated schema models with expected models, evaluation of trace-aware
schema adaptation, data-level validation through query-result equivalence, 
and an assessment of operational feasibility and scaling behavior. 
Schema-information preservation is further assessed through round-trip
schema reconstruction using the mappings involved in the
document-to-relational migration, whereas data migration in this
direction is left for future empirical evaluation.

\paragraph{Contributions}
The contributions of this work can be summarized as follows:

\begin{itemize}[leftmargin=1.2em]

    \item \textit{Composition of model transformations for heterogeneous
    database migration.}
    The proposal uses the \uschema{} model as a pivot representation,
    with explicit migration-oriented mapping rules defined between each
    supported data model and \uschema{}. 
    These rules address the structural patterns relevant to automated
    database migration. Source-to-target schema transformations are obtained by composing these mappings rather than by defining a dedicated transformation
    for each source--target combination, reducing the number of required
    mappings from quadratic to linear in the number of supported data
    models.

    % \item \textit{A compositional model-driven architecture for heterogeneous database migration.}
    % The proposal uses the \uschema{} model as a pivot representation
    % between heterogeneous data models, enabling schema and data migrations
    % to be obtained by composing reusable mappings defined independently of
    % specific source--target combinations. 
    % This improves extensibility and reuse in multi-model migration
    % scenarios.

    % By avoiding dedicated transformations for each pair of models, the approach improves
    % extensibility and reuse in multi-model migration scenarios.

% To the best of our knowledge, this work is the first to provide a unified, model-driven approach 
% to database migration across heterogeneous data models, supporting both schema and data transformation 
% within a single framework. This generality is achieved through the use of the \uschema{} model as a pivot 
% representation, which enables the definition of transformations independently of specific source–target combinations.

\item \textit{A trace-driven mechanism for decoupled data migration.}
  The approach generates fine-grained trace links during schema
  transformation, which are later used to guide data migration
  independently of the native source model. 
  This enables model-independent access to source data 
  through a generic adapter mechanism.

  \item \textit{A separation between canonical schema generation and
trace-aware schema adaptation.}
  The proposal distinguishes between the systematic derivation of a
  canonical target schema through mapping rules and the application of
  designer-controlled schema adaptations. These adaptations are performed
  on the intermediate \uschema{} model while consistently updating the
  corresponding trace links, allowing controlled refinement without
  changing the mapping rules.

  % \item \textit{A validation covering structural, semantic, and performance aspects.}
  %   The proposal is evaluated through unit and integration testing of transformation
  %   rules, schema-level validation via round-trip reconstruction, data-level validation
  %   using representative queries, and performance and scalability analysis over datasets of
  %   increasing size.

\end{itemize}

\paragraph{Paper organization}
The remainder of this paper is organized as follows.
Section~\ref{sec:background} introduces the necessary background and key
concepts, including database migration, the \uschema{} model,
model-driven schema transformations and traceability, and the Orion language.
Section~\ref{sec:relatedwork} contrasts our proposal with the most
relevant work on database migration using several comparison criteria.
Section~\ref{sec:framework} presents the proposed migration approach,
including schema migration, trace-aware schema adaptation, trace
management, and data migration.
Section~\ref{sec:mappings} defines the mappings involved in the
relational-to-document migration scenario.
Section~\ref{sec:validation} presents the empirical evaluation of the
approach, covering schema and data validation, trace-aware adaptation,
round-trip reconstruction, and performance assessment.
The mappings used for the round-trip experiment are provided in the appendices.
Finally, Section~\ref{sec:conclusions} concludes the paper and outlines
directions for future work.

\section{Background\label{sec:background}}

This section briefly introduces the key concepts required to understand
the proposal developed in the following sections.
It first outlines the main characteristics of database migration and
then describes \uschema{}, the unified metamodel used as the pivot
representation in our approach.
Schema mappings, model transformations, and traceability are subsequently discussed as the modeling foundations of the proposed migration process.
Finally, Orion~\cite{alberto-orion-2024} is presented as the
schema-evolution language used to express adaptations of the intermediate
\uschema{} model.

\subsection{Elements of a database migration process}
\label{sec:elems-migration}

A database migration is the process by which a target database~$T$ 
is obtained from a source database~$S$. Migrations are commonly considered
\emph{homogeneous} if $S$ and $T$ share the same database technology, 
and \emph{heterogeneous} otherwise~\cite{google-migration}. 
In this work, however, this distinction is defined in terms of the 
data model.

In heterogeneous migrations, a \emph{schema conversion} step is always required,
whereas in homogeneous migrations this step is only necessary when the target design
introduces structural changes beyond a direct copy of the schema.
Whenever schema conversion takes place, data extracted from $S$ must be transformed
before being written to~$T$, and application code accessing the data
must be adapted accordingly.

A database migration process therefore typically includes
three main activities: (i) schema conversion, (ii) data migration,
and (iii) adaptation of the application code. Validation of the
migrated database constitutes an additional activity aimed at
assessing the correctness and consistency of the migration result.
These activities can be partially or fully automated by specialized tools 
or services, which we refer to as \emph{migrators}.
In this paper, we focus on schema and data conversion, as is common 
in the literature on database migration.

The \emph{cardinality} of a migration refers to the number 
of source and target systems involved. While the most common case is 1:1 
(one source migrated to one target), other scenarios are possible~\cite{google-migration}. 
Cardinality~1:n occurs when a single source is migrated to several heterogeneous targets,
as in deployments involving polyglot persistence. 
Conversely, cardinality~n:1 appears when several databases 
are consolidated into a single target system. The general n:m case, 
although conceptually feasible, is less frequent and normally appears in large-scale data integration or modernization initiatives.

Migration processes often create an opportunity to improve the database design. 
For example, the target schema may be customized to better reflect 
the usage patterns of the application, for instance by merging tables 
in relational systems to optimize read-intensive workloads, 
or by transforming references 
into aggregations in document-oriented databases to reduce the need for joins.
We refer to this design refinement step as \emph{schema customization} or \emph{schema adaptation}.

\subsection{ The unified metamodel \uschema{}}
\label{uschema}

\begin{figure*}[!t]
  \centering
  \includegraphics[width=\linewidth]{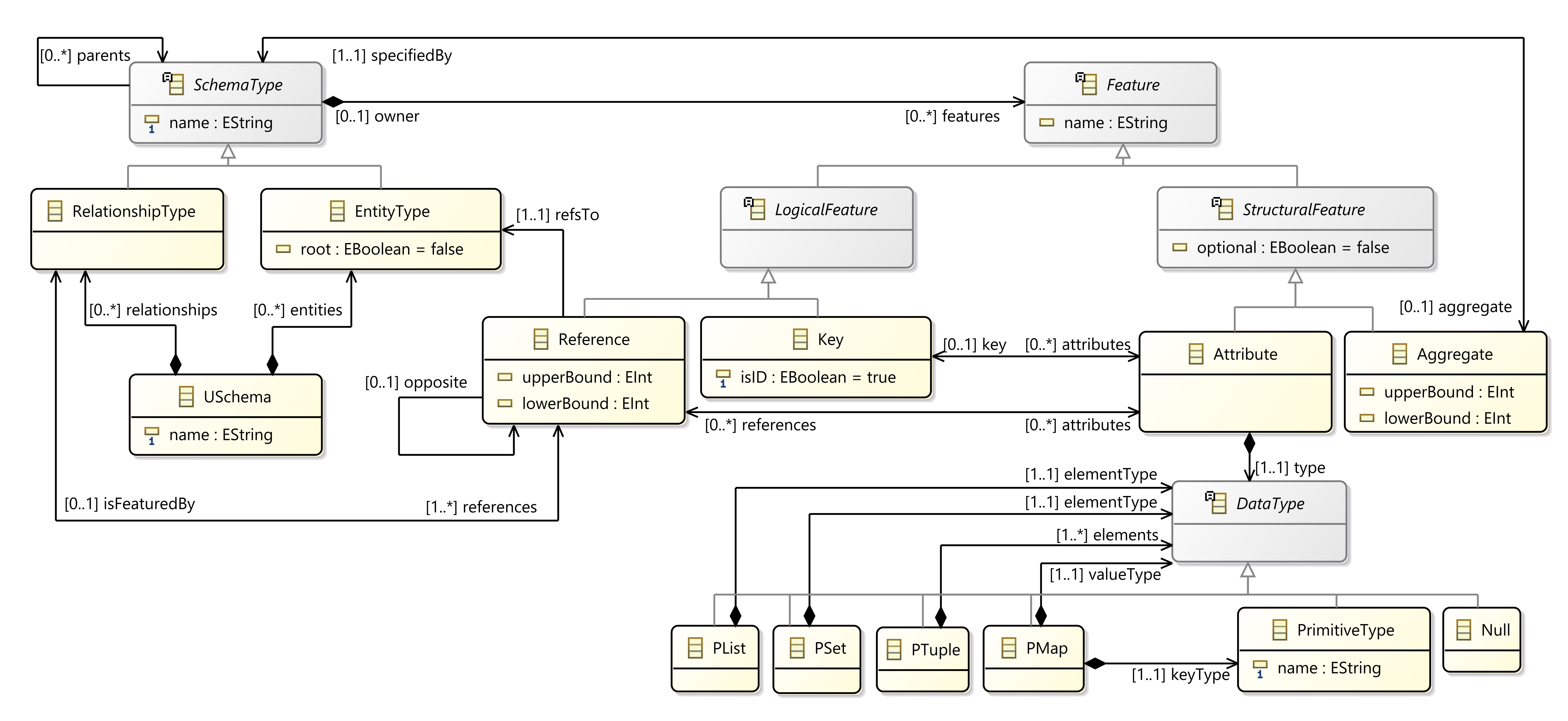}
  \caption{Simplified U-Schema metamodel (union-schema flavor).}
  \label{fig:uschema-metamodel}
\end{figure*}

A data model defines the constructs and structuring rules used to
represent data, whereas a database schema applies those constructs
to a particular domain. In software modeling terms, a data model
can be represented as a metamodel and a database schema as a model
conforming to that metamodel~\cite{carlos-uschema-2022}. 
This model-based representation has long been exploited 
in database engineering to address problems involving 
schema correspondences and transformations~\cite{bernstein2000,hainaut2006}.

% A data model defines the data structuring rules that can be used to
% represent real world data in a database, as well as the operations that
% can be applied on the data. A schema results of applying a data model to
% a particular domain or problem. The link between a schema and its
% data model, can formally be established through the \emph{conforms-to} or
% \emph{instance-of} relationship that is characteristic of the software
% modeling: a schema is represented as a model that is an instance of the
% metamodel representing the data model~\cite{brambilla2012}. Thus,
% metamodeling has traditionally been applied to define data models, 
% and transformational approaches have been proposed to tackle problems
% involving schema mappings~\cite{bernstein2000, hainaut2006}.

Over the years, the co-existence of the relational model with other data
models motivated the definition of generic or unified data models,
e.g., EER~\cite{elmasri-2015} and DB-Main~\cite{hick2003}. 
With the emergence of NoSQL systems, new unified proposals were introduced
to integrate NoSQL and relational paradigms, such as
\uschema{}~\cite{carlos-uschema-2022}, PartiQL~\cite{PartiQL-spec}, and
SOS~\cite{atzeni2012}, each differing in the supported NoSQL paradigms and
their underlying features. 
The migration approach presented in this paper is
based on \uschema{}, whose main characteristics are 
the following, as described in
detail in~\cite{carlos-uschema-2022}: 
(i) it integrates the four most popular NoSQL paradigms
together with the relational model; 
(ii) it distinguishes between entity types and relationship types;
(iii) it provides aggregation and reference constructs to model
relationships between entity types; and
(iv) it supports structural variations of
types, since the possible absence of a declared schema 
allows an entity or relationship to
be stored with different data structures. 

\uschema{} represents the logical structure of a database through a unified
set of constructs that apply across relational, document, columnar, graph,
and key–value paradigms~\cite{carlos-uschema-2022}. 
Table 1 shows how the main constructs of the relational and document data models 
map to U-Schema abstractions, evidencing its role as a unified metamodel.
\emph{N/A} entries denote the absence of a native
counterpart in the corresponding model, not the impossibility of representing
the construct: an aggregation, for instance, can be expressed in the relational
model through a weak-to-strong table relationship.

% \FloatBarrier
\begin{table*}[t!] %[ht!]
\centering
{\fontsize{8.5}{10}\selectfont
\scalebox{.95}[.95]{
\begin{tabular}{p{.25\textwidth}p{.35\textwidth}p{.35\textwidth}}
\toprule
\raggedright\textbf{\uschema{}} & \textbf{Relational} & \textbf{Document}\\
\midrule

\textit{Schema} 
& Schema 
& Optional (schema-on-read) \\[0.5em]

\textit{Entity Type} 
& Table 
& Collection and nested objects\\[0.5em]

\textit{Relationship Type} 
& Associative table %Relationship table 
& N/A \\[0.5em]

\textit{Structural Variation} 
& No (fixed structure of tables) 
& Yes \\[0.5em] %(root and nested documents) \\[0.5em]

\textit{Key} 
& Primary key 
& Document identifier \\[0.5em]

\textit{Reference} 
& Foreign key
& Reference -limited support (e.g., MongoDB) \\[0.5em]

\textit{Aggregation} 
& N/A 
& Nested object \\[0.5em]

\textit{Attribute} 
& Column 
& Document property field\\[0.5em]

\textit{Primitive Types} 
& Scalar types 
& Scalar types \\[0.5em]

\textit{Structured Types} 
& N/A 
& Arrays \\[0.5em]

\bottomrule
\end{tabular}
}
}%end fontsize
\caption{Correspondence between \uschema{} constructs and their native counterparts 
in relational and document models.}
\label{table:uschema-mappings}
\end{table*}

Next, we describe \uschema{}, as illustrated by the excerpt shown in 
Figure~\ref{fig:uschema-metamodel}. 
A \texttt{schema} is defined as a set of \texttt{SchemaType}s, either 
\texttt{EntityType}s or  \texttt{RelationshipType}s, which describe the domain 
entities and their relationships.

\texttt{EntityType}s represent domain concepts whose instances are the data
objects stored in the database. They contain \texttt{Feature}s, which may be
structural or logical. Structural features describe the stored values and
include simple or multivalued \texttt{Attribute}s and \texttt{Aggregation}s,
the latter used to embed objects of other entity types. Logical features
capture links among data and consist of \texttt{Key}s (unique identifiers)
and \texttt{Reference}s, which represent associations through key values.
\texttt{EntityType}s may be declared \emph{root} (top-level objects) or
\emph{non-root} (embedded through aggregations).

A \texttt{RelationshipType} denotes an explicit association that connects two 
(or more) \texttt{EntityType}s and may contain its own attributes. 
Such elements usually originate from associative tables representing M:N relationships 
in relational schemas and %to [MJ] asumo que "to" es errata y debe ser "from"
from edge types in graph data models. 
Similarly to association classes in UML, a \texttt{RelationshipType} models both
the connectivity between entities and the properties attached to the association itself.

% \mjose{Llevar este párrafo antes de describir la Fig. 1? }
\uschema{} supports two flavors~\cite{carlos-uschema-2022}. 
The \emph{full-variability} flavor records all structural variations of each schema type, 
while the \emph{union-schema} flavor merges them into a single structure by marking
optional features. In this work we adopt the union-schema flavor. This
choice simplifies the presentation of our migration approach without
affecting its generality. 

To illustrate the expressive capabilities of \uschema{}, Figure~\ref{fig:athena} 
presents a fragment of a simple music–streaming domain specified using 
\textit{Athena}, a generic schema definition language that provides a textual 
notation for \uschema{}~\cite{alberto-comonos2021}.
% The schema defines root entity types such as \texttt{User}, \texttt{Song}, and 
% \texttt{MusicalStyle}, together with non-root entity types \texttt{PlayList} and 
% \texttt{Listening}, which are aggregated within \texttt{User}.
The schema defines several root entity types, together with the non-root
entity type \texttt{PlayList}, which is aggregated within \texttt{User}, and the 
relationship type \texttt{Listening}, which connects \texttt{User} and \texttt{Song}
and has its own attributes.
The example is intentionally small and serves solely as an introductory illustration 
of \uschema{}. A more detailed version of this domain will be reused later 
(Section~\ref{sec:running-example}) as the running example for the mapping rules. 
Understanding the full syntax of Athena is not required here, as the modeling 
constructs used are self-explanatory.

\begin{figure}[!t]
  \centering
  \begin{minipage}{0.9\textwidth}
  \begin{lstlisting}[style=athenaStyle, xleftmargin=1cm, basicstyle=\ttfamily\footnotesize]
Schema MusicStreaming:1

Root entity User {
  +id:           String,
  name:          String,
  isPremium:     Boolean,
  registerDate:  Date,
  playLists:     Aggr<PlayList>,
  mostRecentlyListened: Ref<Song>*
}
Entity PlayList {
  +id:           String,
  name:          String,
  creationDate:  Date,
  songs:         Ref<Song>*
}
Root entity Song {
  +id:           String,
  title:         String,
  duration:      Decimal(4,2),
  artist:        String,
  playsCount:    Integer,
  styles:        Ref<MusicalStyle>*
}
Root entity MusicalStyle {
  +id:           String,
  name:          String
}
Relationship Listening {
  user:          Ref<User>,
  song:          Ref<Song>,
  playsCount:    Integer,
  status:        String
}
\end{lstlisting}
  \end{minipage}
  \caption{\uschema{} schema example declared with \textit{Athena}.}
  \label{fig:athena}
\end{figure}

\subsection{Schema Mappings, Model Transformations, and Traceability}
\label{mde}

Schema mappings establish correspondences between constructs of
source and target schemas and support schema transformation.
In heterogeneous database migration, such mappings must bridge
data models with different structural constructs, such as tables
and foreign keys in relational models or embedded objects and
references in document models.

MDE provides a natural setting for schema transformations by building
on the model-based view introduced in Section~\ref{uschema}: data models
are represented as metamodels and database schemas as conforming
models~\cite{brambilla2012}. Within this setting, schema mappings
can be realized as model-to-model (M2M) transformations between
schema models, while model-to-text (M2T) transformations can
serialize the generated models into executable schema definitions.
Mappings can also be composed through an intermediate metamodel,
avoiding a dedicated transformation for each source--target
combination.

In our approach, schema migration is realized by composing M2M
transformations through the \uschema{} pivot metamodel. The source
schema model may be obtained from an explicit schema definition,
such as SQL DDL, or inferred from data or application
code~\cite{carlos-uschema-2022}. 

% The transformations also generate
% trace information recording the correspondences between source,
% \uschema{}, and target schema elements. These correspondences are
% subsequently exploited to guide data migration consistently with
% the generated target schema.

% In our setting, each database schema is represented as a model
% conforming to the metamodel of its data model, and schema migration
% is realized by composing m2m transformations through the
% \uschema{} pivot metamodel. The source schema model may be obtained
% from an explicit schema definition, such as SQL DDL, or inferred
% from data or application code~\cite{carlos-uschema-2022}. The
% resulting transformation chain additionally produces trace
% information that records the correspondences established between
% source, \uschema{}, and target schema elements. These
% correspondences are subsequently exploited during data migration.

The conceptual correspondences introduced in
Table~\ref{table:uschema-mappings} provide the foundation for these
transformations, but should not be interpreted as the
migration-oriented mappings defined in this work. The table
identifies basic counterparts between constructs of the relational,
document, and \uschema{} models, whereas each transformation is
direction-specific and operates at a substantially finer level of
granularity. In particular, the mappings defined in
Section~\ref{sec:mappings} address migration-specific concerns such
as relationship patterns, weak and associative structures,
identifier representation, alternative target representations, and
the generation and maintenance of trace links.

All metamodels used in this work ---relational, document, and
\uschema{}--- are implemented in Ecore within the Eclipse Modeling
Framework (EMF)~\cite{steinberg-emf2009}. The M2M and M2T
transformations are implemented in Xtend, a Java-based
general-purpose language integrated with the Xtext
workbench~\cite{bettini2016}.

\paragraph{Traceability in model transformations}
Traceability in model transformations captures the correspondences
established between source and target model elements during
transformation execution~\cite{brambilla2012,vara2014traceability}.
In our approach, such traceability plays a central role because the
schema-level correspondences linking source, \uschema{}, and target
schema elements are subsequently used to guide the data-migration stage.

During the execution of the M2M transformations, these correspondences
are explicitly recorded as trace links using the representation detailed
in Section~\ref{sec:trace}. Since schema migration is realized as a
composition of transformations, trace information is maintained across
the transformation chain and updated when adaptations are applied to
the intermediate schema.

\subsection{The Orion Schema Evolution Language}
\label{sec:orion}

Orion is a generic schema-evolution language that implements a taxonomy 
of schema-change operations defined over \uschema{}~\cite{alberto-orion-2024}.
The language provides a declarative mechanism for expressing schema
changes independently of the underlying data model or database
technology. These changes include renaming, type casting, 
attribute deletion, and conversions between references and aggregations.

Given the \textit{Music Streaming} schema introduced earlier, 
the script shown below 
illustrates several Orion schema evolution operations.

\begin{lstlisting}[style=orionStyle, xleftmargin=1cm, basicstyle=\ttfamily\footnotesize]
RENAME ENTITY User TO User_app
RENAME Song::duration TO length
CAST ATTR Song::length TO Integer
MORPH REF Song::styles TO styles
DELETE Listening::status
\end{lstlisting}

Within the proposed migration approach, Orion provides the language
for expressing schema-level adaptations on the intermediate
\uschema{} model. These adaptations allow designers to guide target
schema generation according to specific design requirements. For
instance, references and aggregations can be converted into one another
when a different representation is preferred in the target design.

In its original setting, given a schema and an Orion script, the Orion
execution engine applies the specified schema-change operations and
produces an evolved schema together with data-evolution scripts expressed
in the language of the target database. 
Our migration approach is intended to reuse Orion's notation and its
schema-evolution component for modifying the \uschema{} model, but does not
require the data-evolution scripts generated by Orion, since data migration
is subsequently performed by our data migrator. 
We have implemented a prototype to validate the feasibility of
trace-aware schema adaptation before integrating it with Orion.
The current implementation supports several schema-change operations,
updating both the \uschema{} model and the corresponding trace links.

% In its original setting, given a schema and an Orion script, the Orion
% execution engine applies the specified schema-change operations and
% produces an evolved schema together with data-evolution scripts expressed
% in the language of the target database. Our migration approach reuses
% Orion's notation and its schema-evolution capabilities, but does not
% require these data-evolution scripts, since data migration is subsequently
% performed by the data migrator. Instead, integrating Orion into the
% migration pipeline requires the trace generated during schema
% transformation to be updated consistently with the changes applied to
% the \uschema{} model. Therefore, the Orion engine must be
% adapted so that schema-change operations also maintain the corresponding
% trace links.

\section{Related Work\label{sec:relatedwork}}

In this section, we compare representative approaches proposed to
automate database migrations. Our review focuses on proposals that
involve at least one NoSQL data model and address schema and/or data
migration across heterogeneous data models. 
We nevertheless include DB-Main because of its conceptual
closeness to our proposal, although it predates the NoSQL era.
Schema-evolution approaches are outside the scope of the comparison
because they address changes to an existing database schema rather
than migration between heterogeneous data models. This body of work
was reviewed in detail in our previous work on
Orion~\cite{alberto-orion-2024}, which predates the recent MDE-based
approach EvolveDB~\cite{eckwert2025evolvedb}.

% The database-migration literature is dominated by relational-to-document
% conversion. 
% %, most commonly targeting MongoDB 
% \mjose{$\leftarrow$ quitar? se dice lo mismo en el 1er criterio}. 
% From these works, we selected those most technically significant and conceptually sound for our review. 

To structure the comparison, we defined a set of criteria based on
the characterization of database migration presented in
Section~\ref{sec:elems-migration}, complemented with modeling,
transformation, and data-migration aspects relevant to the comparison.
For schema migration, these criteria cover the supported source and
target data models, data-model independence, schema representations,
the definition and implementation of schema mappings, migration
cardinality, the use of additional information to guide mapping
decisions, and the degree of schema-conversion customization.
For data migration, they cover whether data migration is supported,
the inputs used to drive the migration, the ETL strategy, and the
extent to which application-code adaptation is addressed.
The criteria are discussed below and summarized in
Tables~\ref{tab:comparison-schema} and~\ref{tab:comparison-data}.

\paragraph{Data models}
This criterion refers to which data models are supported as source and target 
during migration. Most existing proposals focus on a single direction, 
namely relational-to-document conversion, typically targeting MongoDB. 
Representative examples include the automated or semi-automated approaches 
by Jia et al.~\cite{Jia2016}, Rocha et al.~\cite{Rocha2015}, and 
Zhao et al.~\cite{zhao2014schema}, all of which assume a relational source 
and a document-store target. Metamorfose~\cite{kuszera2019} extends 
this common scenario by also supporting column-family stores, 
making it one of the few proposals addressing more than a single NoSQL target.

Support for data models different from document stores is less common. 
Scavuzzo et al.~\cite{scavuzzo2014} consider the migration from relational 
databases to HBase (a columnar store) by defining a columnar metamodel.
A more general approach is represented by the Dynamite system~\cite{Wang2020dynamite},
which addresses heterogeneous data transformation 
by synthesizing Datalog programs capable of handling relational, document, and graph data. 
However, it focuses on data transformation rather than schema migration.

In our case, the use of \uschema{} provides a unified representation that can
accommodate both relational and NoSQL data models, as illustrated in
Section~\ref{sec:mappings} for the relational-to-document migration and
in Appendices~A and~B for the document-to-relational direction.

% In our case, the use of \uschema{} provides a unified representation that can 
% accommodate both relational and NoSQL data models, as illustrated in 
% Section~\ref{sec:mappings} for the relational-to-document example.

\paragraph{Data model independence}
Data model independence refers to whether a migration approach is intrinsically
coupled to specific source and target data models or whether it defines
abstractions that enable its application across different database paradigms.
Most existing proposals are model-specific: their transformation logic is hard-coded for a particular migration direction (e.g., relational-to-MongoDB) and cannot be reused for other NoSQL models without redesigning the mapping rules or the extraction and loading procedures. Approaches of this kind include 
Jia et al.~\cite{Jia2016}, Rocha et al.~\cite{Rocha2015}, and 
Zhao et al.~\cite{zhao2014schema}, whose algorithms are tailored to the 
document model.

Only a few works move toward a more general perspective. 
Metamorfose~\cite{kuszera2019} introduces a transformation pipeline based on an 
intermediate structure, although its rules remain specific to the document and 
column-family models it supports. Dynamite~\cite{Wang2020dynamite}  achieves a higher degree of generality by synthesizing Datalog programs that can operate over 
different data representations, but it does not define 
a unifying schema model and focuses solely on data transformation.

Earlier attempts at broader platform independence can be found in tools such as 
DB-Main, which supported schema evolution and migration 
across relational, XML, and object databases~\cite{hainaut1994}. 
However, DB-Main was never extended to incorporate NoSQL 
models, and to the best of our knowledge, no existing approach provides a unified 
and model-independent framework for schema and data migration across relational 
and NoSQL systems. In contrast, our proposal relies on \uschema{} as a common 
intermediate representation, enabling schema mappings and transformations to be 
defined independently of the specific source or target data model.

\paragraph{Representations in schema conversion}
Migration approaches differ substantially in the abstractions they use to 
represent source and target schemas, as well as in whether an intermediate 
representation is employed. Intermediate models are typically introduced for 
two reasons: (i) to reduce the semantic gap between heterogeneous data models, 
thus simplifying the conversion process, and (ii) to provide a universal format 
that reduces the number of transformations from $N \times M$ to $N + M$ for 
$N$ source and $M$ target models. Schema conversion can also operate at the 
physical or logical level~\cite{Hainaut2008}; in the former case, reverse 
engineering is applied to recover a logical schema that carries richer 
semantic information.

A logical intermediate representation is used in Jia et al.~\cite{Jia2016}, 
where an ER model is extracted from the relational source and enriched with 
usage-based annotations (e.g., join frequency). This ER model is then converted 
into a directed graph whose nodes correspond to entity types (tables) and whose 
edges represent foreign keys, forming the basis for generating the MongoDB 
schema. Zhao et al.~\cite{zhao2014schema} similarly employ a graph-based representation
proposing a schema conversion algorithm over a relational-to-NoSQL graph model.
Bansal et al.~\cite{bansal2024} adopt a workload-aware hypergraph-based representation, 
combined with incidence matrices, to capture query-driven relationships 
among tables and derive a denormalized document schema.

Several works define explicit intermediate metamodels. Scavuzzo et 
al.~\cite{scavuzzo2014} introduce a columnar metamodel that provides a 
canonical format for HBase schemas, while Metamorfose~\cite{kuszera2019} 
uses a DAG-based representation as the core structure for expressing 
relational-to-document and relational-to-column-family transformations. 
Schreiner et al.~\cite{Schreiner2020} propose a hierarchical canonical model 
that captures relational schemas in a semi-structured format to facilitate 
their mapping to aggregate-based NoSQL schemas. 

In contrast, Dynamite~\cite{Wang2020dynamite} does not rely on a unifying schema 
representation: instead, it synthesizes Datalog programs from examples, 
focusing solely on data transformation and leaving schema abstraction 
implicit.

Our approach is explicitly model-based: both source and target schemas are
represented using metamodels specific to each data model, and \uschema{}
acts as a conceptual pivot model that provides the common abstraction on
which executable migration mappings are defined.

\paragraph{Schema mapping}

When automating schema migration, the target schema may be derived in two
ways: either generated automatically from explicit mapping rules, or
hard-coded within the transformation procedures themselves. Most existing
approaches fall into the latter category. Although some of them distinguish
basic relationship patterns, such as one-to-one, one-to-many, and
many-to-many associations, the corresponding mapping decisions are encoded
procedurally and remain implicit in the implementation. As a result, the
mapping logic cannot be easily reused, analyzed, or adapted to other
source and target data models. This applies to approaches such as
Jia et al.~\cite{Jia2016}, Zhao et al.~\cite{zhao2014schema},
%Schreiner et al.~\cite{Schreiner2020}, 
and Bansal et al.~\cite{bansal2024}.

A more structured transformation architecture is adopted by
Metamorfose~\cite{kuszera2019}, which defines a pipeline over a
DAG-based representation. However, the mapping rules remain implicit
in the graph traversal and schema construction procedures.
Similar implicit strategies are followed in the
guidelines proposed by Kim et al.~\cite{Kim2020} 
and in the approach of Rocha et al.~\cite{Rocha2015}.

Dynamite~\cite{Wang2020dynamite} departs entirely from rule-based mappings:
the system synthesizes Datalog transformation programs from examples,
making the schema mapping implicit and preventing a declarative
specification of the transformation space.

In our approach, schema mappings are explicitly specified and
realized as M2M transformations implemented in Xtend.
% In our approach, schema mappings are formally specified over \uschema{}
% and implemented using model transformation languages (e.g., Xtend). 
This allows alternative mappings to be expressed explicitly and ensures that
the mapping logic is defined independently of the specific source and
target data models. The approach supports executable mappings in both
directions between each supported data model and the \uschema{} pivot
model.

\paragraph{Cardinality}
Most approaches assume a 1:1 migration scenario, where a single source 
database is mapped to a single target system. Only Dziedzic et 
al.~\cite{Dziedzic2016} address more complex situations, supporting 1:n 
migrations in the context of a polystore architecture. Although this area 
remains largely unexplored, the use of a common intermediate representation 
in our approach facilitates supporting 1:n migration scenarios, as mappings 
can be defined independently for each target model.

\paragraph{Parameterized mappings}
Beyond the structural information contained in the schema, some approaches 
parameterize their mappings with additional information provided by users 
or derived from database usage statistics. Jia et al.~\cite{Jia2016} exploit 
several indicators—such as frequent joins, very large tables, and the 
frequency of insertions and updates—to decide whether a given foreign key 
should be mapped to an embedded document or to a reference in MongoDB. 
Kim et al.~\cite{Kim2020} and, more recently, QBDNH~\cite{bansal2024}
use workload analysis to guide denormalization decisions when
generating target schemas for HBase and MongoDB, respectively. Other proposals, including Metamorfose~\cite{kuszera2019} and the canonical-model approach by Schreiner et al.~\cite{Schreiner2020}, rely mainly on schema 
structure and do not incorporate such statistical parameters. 
Currently, our approach does not use workload or statistical data to 
parameterize mappings, although it could be extended to support this
capability.

\paragraph{Schema-conversion customization}
Another relevant aspect is the degree to which the schema conversion 
performed by a tool can be customized or adapted to user requirements. 
Most existing approaches apply fixed, hard-coded conversion rules that 
leave little room for variation (e.g., Zhao et al.~\cite{zhao2014schema}, 
Schreiner et al.~\cite{Schreiner2020}, Kuszera et al.~\cite{kuszera2019}). 
Although Jia et al.~\cite{Jia2016} use database statistics to influence 
some decisions (e.g., embedding vs.~referencing), these heuristics do not 
expose alternative mappings to the user and therefore do not constitute 
genuine customization. 

In our proposal, customization is addressed by separating
schema generation from schema adaptation. A canonical target schema
is first derived through the \uschema{} intermediate model, while
adaptations can be applied at the \uschema{} level
prior to target generation, with the corresponding trace links
updated consistently. Orion is intended to provide the notation and
schema-evolution component for specifying and applying such adaptations.

% In our proposal, customization is addressed by separating
% schema generation from schema adaptation. A canonical target schema
% is first derived through the \uschema{} intermediate model, while
% design-oriented adaptations can be applied at the \uschema{} level
% prior to target generation, with the corresponding trace links
% updated consistently. Orion is intended to provide the notation and
% schema-evolution component for specifying and applying such adaptations.

\paragraph{Data migration}
A key distinction among the reviewed approaches is whether they support
data migration in addition to schema conversion. Several proposals handle
schema migration only (e.g., Zhao et al.~\cite{zhao2014schema}, 
Rocha et al.~\cite{Rocha2015}), while others implement both schema and 
data transformation (e.g., Jia et al.~\cite{Jia2016}, Scavuzzo et 
al.~\cite{scavuzzo2014}, Schreiner et al.~\cite{Schreiner2020}, 
Kuszera et al.~\cite{kuszera2019}). Dynamite~\cite{Wang2020dynamite} focuses 
exclusively on data transformation. Our proposal supports both schema 
and data migration. 
For those approaches that support data migration, two additional aspects 
are relevant for comparison and are analyzed below: the kinds of inputs 
used to drive the migration process, and the ETL strategies applied to 
materialize the target data.

\paragraph{Data-migration inputs}
Approaches differ in the kinds of artefacts and information they use to drive 
data extraction, transformation, and loading. In DB-Main~\cite{hick2003}, ETL 
flows are derived from explicitly specified schema mappings. 
Jia et al.~\cite{Jia2016} rely on an annotated ER-derived graph, enriched with 
usage statistics (e.g., join frequency, table size, and update rates), 
which guides both schema decisions and the materialization of MongoDB documents.
Schreiner et al.~\cite{Schreiner2020} use their canonical model as the primary 
input for generating target data instances, whereas Scavuzzo et 
al.~\cite{scavuzzo2014} and Kuszera et al.~\cite{kuszera2019} base their ETL 
logic mainly on the structural information captured in their intermediate 
models (columnar metamodel and DAG, respectively). 
Dynamite~\cite{Wang2020dynamite} takes a different route by synthesizing Datalog 
programs from input–output examples, effectively specifying data 
transformations by example rather than through explicit mappings. 

In our approach, the inputs driving data migration are the generated
target schema and the trace links produced during schema transformation.
Together, they identify the source data corresponding to the target
schema elements and the structural correspondences required for their
transformation.

% In our approach, data migration is driven by the trace links 
% generated during the execution of the schema transformation pipeline. 
% This trace, together with the generated target schema, 
% is used to identify and transform the source data corresponding 
% to the elements of the target schema, as explained below. 
% This eliminates the need for independently specified ETL logic.

\paragraph{ETL strategies}
The strategies adopted to execute data migration also vary significantly. 
Some approaches implement custom, procedural ETL code tightly coupled to a 
specific source–target pair, as in Jia et al.~\cite{Jia2016} or 
Schreiner et al.~\cite{Schreiner2020}. 
Scavuzzo et al.~\cite{scavuzzo2014} use a queue-based producer–consumer 
architecture to populate HBase, while Metamorfose~\cite{kuszera2019} 
generates Apache Spark commands to transform relational data into document or 
column-family structures, thus leveraging data-parallel execution. 
Dynamite~\cite{Wang2020dynamite} executes synthesized Datalog programs on top of a 
logic engine, providing a declarative execution model for data 
transformation. Other approaches, such as Rocha et al.~\cite{Rocha2015} or 
Kim et al.~\cite{Kim2020}, focus mainly on query rewriting and denormalization 
guidelines rather than on a dedicated ETL pipeline.

In contrast to approaches based on source--target-specific ETL
implementations, our proposal uses a generic data-migration mechanism
that interprets the target schema and transformation traces rather than
requiring independently specified transformation scripts.
Source data are accessed through a \uschema{}-based adapter that
decouples data access from technology-specific APIs and provides a
uniform execution mechanism across heterogeneous systems.

% In contrast to these approaches, our proposal does not rely
% on custom, procedural ETL logic tied to specific source--target
% pairs. \mjose{Esta frase se ha dicho muy parecida en el apartado anterior $\rightarrow$ Quizá se podrían fusionar los párrafos en los que describimos nuestro trabajo en las dos dimensiones (ETL y Data-migration inputs...}Instead, data migration is driven by the trace links
% generated during schema transformation and by the resulting
% target schema, without requiring independently specified
% transformation scripts. The execution is supported by a
% \uschema{}-based adapter that provides model-independent access
% to the source data, decoupling data access from technology-specific
% APIs and enabling uniform execution across heterogeneous systems.

\paragraph{Code adaptation}
Some proposals also address the problem of adapting existing application 
code to the target NoSQL system. Several techniques have been explored in 
this context: (i) translating SQL queries into HBase API calls 
\cite{Kim2020,scavuzzo2014} or into MapReduce jobs executed on HBase 
\cite{chung2014jackhare}; (ii) generating REST API calls from DML statements 
(\texttt{SELECT}, \texttt{INSERT}, \texttt{UPDATE}, \texttt{DELETE}) 
as in Schreiner et al.~\cite{Schreiner2020}; and (iii) wrapping original 
SQL queries so that they are intercepted and redirected as calls to the 
target NoSQL store \cite{Rocha2015}. These techniques vary in scope and 
automation level but share a common focus on query and API translation. 
More recently, commercial migration tools have begun to incorporate
generative-AI support for code adaptation; for example, AWS integrates
Amazon Q Developer with DMS Schema Conversion to automate the conversion
of embedded SQL in Java application code for supported migration scenarios.
In our case, code evolution is outside the scope of the present work. 

%%%%%%%%Cuando la tabla era una sola y estaba girada:
%Tables~\ref{tab:comparison} summarizes the comparison of all reviewed 
%approaches with respect to the proposed criteria. Due to its size and 
%horizontal layout, the table is placed at the end of the paper to 
%preserve readability and avoid interrupting the flow of the discussion. 
%The cardinality and parameterized mappings criteria are excluded, as 
%they are not supported by most of the analyzed approaches.

%%%%%%%%Ahora la comparación está repartida en dos tablas:

Tables~\ref{tab:comparison-schema} and~\ref{tab:comparison-data} 
summarize the comparison with respect to schema- and data-migration criteria, 
respectively. For readability and to keep the tables within the page margins, 
migration cardinality and parameterized mappings are omitted.
Migration cardinality is 1:1 in all the approaches
considered except Dziedzic et al.~\cite{Dziedzic2016}. Regarding
parameterized mappings, some approaches use workload or usage
information to influence mapping decisions, whereas our current
approach does not yet support such parameterization.

\begin{table*}[!t]
\centering
\footnotesize
\caption{Comparison of migration approaches: schema migration criteria.}
\label{tab:comparison-schema}
\resizebox{\linewidth}{!}{%
{\renewcommand{\arraystretch}{1.2}%
\begin{tabular}{p{0.08\textwidth}p{0.16\textwidth}p{0.13\textwidth}p{0.13\textwidth}p{0.15\textwidth}p{0.13\textwidth}p{0.13\textwidth}p{0.13\textwidth}}
\toprule
\textbf{Approach} &
\textbf{Data models (source~→~target)} &
\textbf{Data model independence} &
\textbf{Schema reps.: source/target} &
\textbf{Schema reps.: intermediate} &
\textbf{Mapping explicitness} &
\textbf{Mapping implementation} &
\textbf{Customization} \\
\midrule

\cite{Dziedzic2016} &
Rel → Polystore &
Not model-independent &
DDL (parsed) &
Binary/CSV formats &
Implicit &
ETL-layer logic &
No \\[0.4em]

\cite{hick2003} &
Rel/ORel/XML →Same models &
Multi-model (no NoSQL) &
DDL + logical schemas &
Extended ER (DB-Main) &
Explicit &
Model transformations &
Limited (manual) \\[0.4em]

\cite{Jia2016} &
Rel → Doc &
Model-specific &
ER / MongoDB schema &
Annotated ER + directed graph &
Procedural rules &
Algorithmic traversal &
Limited (heuristics only) \\[0.4em]

\cite{Kim2020} &
Rel → Col &
Model-specific &
DDL + workload graphs &
Transaction/query graphs &
Implicit (guidelines) &
Heuristics, no mapping language &
No \\[0.4em]

\cite{kuszera2019} &
Rel → Doc, Col &
Partially model-independent (Doc+Col) &
DDL + target logical schema &
DAG per entity &
Implicit in DAG logic &
DAG traversal procedures &
No (fixed pipeline) \\[0.4em]

\cite{bansal2024} &
Rel → Doc &
Model-specific &
DDL + workload queries &
Hypergraph &
Implicit in hypergraph traversal &
Hypergraph algorithm &
No (workload-driven only) \\[0.4em]

\cite{Rocha2015} &
Rel → KV/Doc/Col &
Model-specific &
DDL / MongoDB &
None &
Partially described &
Procedural rules + wrappers &
No (fixed strategy) \\[0.4em]

\cite{scavuzzo2014} &
Rel → Col &
Model-specific &
DDL / HBase &
Columnar metamodel &
Implicit in columnar mapping &
Transformations to column model &
No \\[0.4em]

\cite{Schreiner2020} &
Rel → Doc/Col/KV &
Partially model-independent (Doc/Col/KV) &
Relational DDL / NoSQL schema &
Canonical hierarchical model &
Procedural mapping over canonical model &
Procedural canonical transformations &
No (canonical fixed) \\[0.4em]

\cite{zhao2014schema} &
Rel → Doc/Col/KV &
Model-specific &
Relational DDL / NoSQL model &
Rel→NoSQL conversion graph &
Algorithmic (implicit variability) &
Graph algorithm &
No \\[0.4em]

\cite{Wang2020dynamite} &
Rel/Doc/Graph →Rel/Doc/Graph &
Multi-model (data only) &
No explicit schema abstraction &
None (example-based) &
Implicit in synthesized Datalog &
Datalog program synthesis &
No (implicit mappings) \\[0.4em]

\midrule
\textbf{Our approach} &
Rel/NoSQL →Rel/NoSQL &
Model-independent via \uschema{} &
Metamodels per model &
\uschema{} universal model &
Explicit mappings over \uschema{} &
Model transformations (Xtend) &
Yes: alternative mappings + trace-aware U-Schema adaptation
%Orion-based adaptation
\\

\bottomrule
\end{tabular}%
}%
}
% \end{table*}

% \begin{table*}[t]
% \centering
% \footnotesize

\vspace{1em}
\caption{Comparison of migration approaches: data migration criteria.}
\label{tab:comparison-data}
\resizebox{\linewidth}{!}{%
{\renewcommand{\arraystretch}{1.2}%
\begin{tabular}{p{0.08\textwidth}p{0.25\textwidth}p{0.20\textwidth}p{0.20\textwidth}p{0.18\textwidth}}
\toprule
\textbf{Approach} &
\textbf{Data migration supported} &
\textbf{Data-migration inputs} &
\textbf{ETL strategies} &
\textbf{Code adaptation required} \\
\midrule

\cite{Dziedzic2016} &
Yes &
Schema + binary formats &
Binary-format ETL &
No \\[0.4em]

\cite{hick2003} &
Yes &
Explicit schema mappings &
Mapping-driven ETL &
Some code generation \\[0.4em]

\cite{Jia2016} &
Yes &
Annotated ER-derived graph &
Traversal-based ETL &
No \\[0.4em]

\cite{Kim2020} &
Partial (denormalization only) &
Schema + workload stats &
No dedicated ETL &
SQL→HBase API adaptation \\[0.4em]

\cite{kuszera2019} &
Yes &
DAG + columnar metamodel &
Spark-based ETL from DAG &
No \\[0.4em]

\cite{bansal2024} &
No &
N/A &
N/A &
No \\[0.4em]

\cite{Rocha2015} &
No &
Relational schema + SQL queries &
No ETL (query rewriting) &
SQL wrapping to NoSQL \\[0.4em]

\cite{scavuzzo2014} &
Yes &
Relational schema + columnar model &
Producer–consumer queue ETL &
HBase API adaptation \\[0.4em]

\cite{Schreiner2020} &
Yes &
Canonical model &
Traversal of canonical model &
REST API generation \\[0.4em]

\cite{zhao2014schema} &
No &
Relational schema + NoSQL target model &
Not described &
SQL→KV middleware \\[0.4em]

\cite{Wang2020dynamite} &
Yes (data only) &
I/O examples &
Execute Datalog programs &
No \\[0.4em]

\midrule
\textbf{Our approach} &
Yes &
% \uschema{} mappings + \uschema{}-based adapter &
Target Schema + transformation traces &
Trace-driven ETL via \uschema{}-based adapter &
Not addressed \\

\bottomrule
\end{tabular}%
}%
}
\end{table*}

The tables highlight the heterogeneity of current approaches: 
most focus on a single source--target direction, rely on fixed mapping 
rules, and implement migration logic tightly coupled to specific NoSQL 
technologies. Only a few works employ intermediate models or provide 
partial forms of model independence, and support for customizable schema 
conversion or systematic data migration remains limited.

In contrast, our approach builds on a unified intermediate
representation (\uschema{}) and explicit executable mappings,
enabling schema conversion across heterogeneous data models
together with a structured data migration process. The combination
of a metamodel-based foundation and trace-driven migration provides
a unified and extensible architecture for schema and data
migration.

That said, some approaches offer advantages in specific
dimensions not explicitly addressed by our proposal.
Kim et al.~\cite{Kim2020} focus on workload-aware denormalization
decisions for HBase schemas, while Dynamite~\cite{Wang2020dynamite}
supports the automatic synthesis of data transformations from
examples without requiring explicit mapping specifications.
These complementary ideas could be incorporated into our
architecture in future work.

\section{Migration Approach}
\label{sec:framework}

This section presents our migration approach.
We describe how each of the two phases of the proposed migration
process---schema migration and data migration---is performed, and
how the transformation traces are managed.

\begin{figure}[!t]
\centering
    \includegraphics[width=\linewidth]{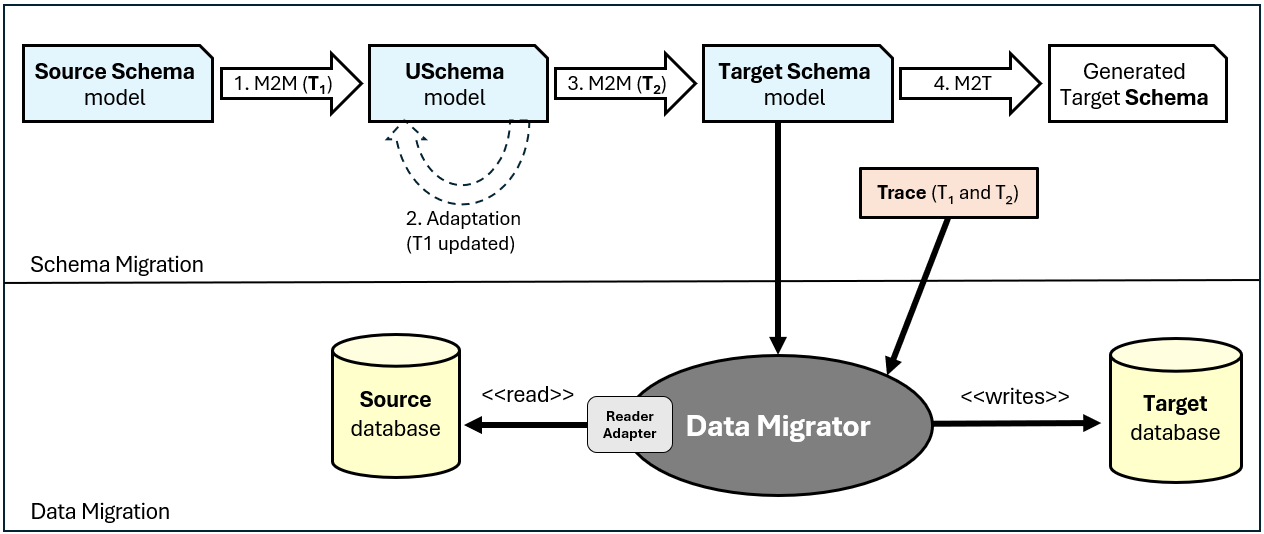}
    \caption{\uschema{}-based migration process.}
    \label{fig:workflow}
\end{figure}

Figure~\ref{fig:workflow} summarizes this process. 
The \emph{Schema Migrator} composes the source-to-\uschema{} and
\uschema{}-to-target transformations, 
with an optional trace-aware adaptation stage
at the intermediate model level. 
The \emph{Data Migrator} then uses the
generated target schema and the traces established during schema
migration to transform source instances into target instances. Source
data are accessed through the \emph{Reader Adapter}. The following
subsections describe these components and the traceability mechanism
in detail.

\subsection{Schema Migrator Component}
\label{sec:schema-migrator}

Before the migration starts, a model 
representing the source schema must be available. 
When an explicit schema definition is available, for instance in
SQL DDL, JSON Schema, or the Cassandra Query Language (CQL), the
source model is obtained through a model injection process.
Otherwise, it can be inferred from the data
\cite{carlos-uschema-2022} or from application code
\cite{Carlos-JSS-2026}.

The schema migration process consists of four steps:

\textit{Step 1. Translation to \uschema{}.}
An M2M transformation derives the \uschema{} model from the source schema 
model according to a set of predefined mapping rules.
These rules map elements of the source data model metamodel to elements 
of the \uschema{} metamodel.
When multiple representations are possible, default conventions are applied.
For example, a 1:N relationship in a relational schema is represented as a
\texttt{Reference} by default. This step generates the trace $T_1$ between
source and \uschema{} elements.

\textit{Step 2. Trace-aware schema adaptation in \uschema{}.}
The generated \uschema{} model may be adapted to reflect
application-specific design decisions. The designer can inspect its
textual representation (i.e., an Athena script) and specify the desired
changes through an Orion script (Section~\ref{sec:orion}).
Such adaptations require trace $T_1$ to be updated consistently with
the changes applied to the \uschema{} model.

% \mjose{Eliminar. Ha sido sustituido por el de abajo, entiendo.$\rightarrow$ }In $T_1$, the source elements previously associated with the aggregate are then associated with the newly created reference. 
% The subsequent \uschema{}-to-Document transformation 
% maps the adapted model without requiring changes to the
% mapping rules.

For example, an \texttt{Aggregate} in the generated \uschema{} model
may be replaced by a \texttt{Reference} when a referenced representation
is preferred in the target document design. 
In $T_1$, the source elements previously associated
with the aggregate are then associated with the newly created reference. 
The subsequent \uschema{}-to-Document transformation 
maps the adapted model without requiring changes to the
mapping rules.

% For example, in a relational-to-document migration, if the
% relational-to-\uschema{} mapping represents a 1:N relationship as a
% \texttt{Reference}, a designer may specify an Orion \texttt{MORPH}
% operation to transform this reference into an \texttt{Aggregate} when
% an embedded representation is preferred in the target document design.
% In the proposed integration, applying this schema change requires the
% corresponding entries in $T_1$ to be updated so that they refer to the
% newly created aggregate. The subsequent \uschema{}-to-Document
% transformation then maps the aggregate to an embedded document
% structure. The updated trace allows the Data Migrator to follow this
% adapted design without modifying the mapping rules.

The current trace-updater prototype implements trace-aware adaptation
for selected schema-change operations, including the renaming of any
named schema element, attribute deletion, and conversion of an
\texttt{Aggregate} into a \texttt{Reference}.

% The trace-aware adaptation of the Orion engine required to automate this
% update, described in Section \ref{sec:orion}, is not yet fully implemented
% in the current prototype; therefore, Orion-based schema adaptation is not
% included in the empirical evaluation.

\textit{Step 3. Generation of the target schema.}
An M2M transformation derives the target schema model from the \uschema{} model, applying the mapping rules and conventions defined
for the target data model. This step generates the trace $T_2$ between
\uschema{} and target schema elements.

% \textit{Step 3. Generation of the target schema.}
% The adapted (or original) \uschema{} model is transformed \mjose{(m2m)} into the
% target schema, generating the trace $T_2$ between \uschema{} and target schema
% elements.
% \mjose{[MJ: ¿Indicar que en este paso tb se aplican reglas de mapping? si no, ¿puede parecer que sólo se aplican en el Step 1?}

%Incluir (quizá reducirlo un poco) este párrafo:
%A m2m transformation derives the target schema from the
%adapted (or original) \uschema{} model, according to a set of predefined
%rules which map elements of the \uschema{} metamodel to elements of the
%target data model metamodel. 
%This step generates the trace $T_2$ between \uschema{} and target
%elements.

\textit{Step 4. Target schema serialization.}
When the target database requires or supports an explicit schema
definition, the resulting target schema model can be translated through
an M2T transformation into a target-specific representation
(e.g., SQL DDL, JSON Schema, or CQL) and materialized in the target
database. For schemaless systems, this step may be omitted, while the
generated target schema model is still used to guide the subsequent
data migration.

% \textit{Step 4. Target schema serialization.} 
% \mjose{ Este paso es opcional, no? en MongoDB no lo hacemos, puesto que el esquema está en los datos. O al menos lo que es opcional es la materialización del esquema *previa a ser poblado con datos*. }
% The resulting target schema model is translated into an executable schema
% definition (e.g., SQL DDL, JSON Schema, or CQL) through a M2T transformation.
% This step materializes the schema in the target database, which is 
% subsequently populated during the data migration phase.

Although presented as a linear workflow, the process may involve multiple
iterations. Designers may inspect the generated target schema and, if 
necessary, adapt the intermediate \uschema{} model and repeat the 
transformation until the desired target schema is obtained.

\subsection{Traceability Mechanism}
\label{sec:trace}

In our approach, traceability connects schema and data migration.
The traces $T_1$ and $T_2$ record the correspondences established
between source, \uschema{}, and target schema elements during schema 
transformation. These correspondences are subsequently used by the
Data Migrator to access and transform the source data according to
the generated target schema.

\paragraph{Trace representation} 
Conceptually, a trace link relates one or more elements of a source
model to one or more elements of a target model, preserving
their metamodel types. In this work, the trace mechanism is independent of the particular source and target metamodels and supports bidirectional navigation between corresponding elements. It also provides direct access to model elements during schema transformation and data-migration execution.

Each source and target element is identified by a qualified
name derived from its path in the aggregation hierarchy of the
corresponding schema model. For example,
\texttt{user.is\_premium} identifies column \texttt{is\_premium}
of table \texttt{user}, and
\texttt{user.is\_premium.Boolean} the
\texttt{PrimitiveType} associated with the corresponding \uschema{}
attribute.

In the implementation, each correspondence is represented in two
complementary forms. The \emph{symbolic trace} associates source
element identifiers with the identifiers of their corresponding
target elements, providing lightweight identifier-based lookup.
The \emph{object trace} associates these identifiers with the
corresponding target model elements (\texttt{EObjects} in EMF), thereby
supporting direct navigation during transformation and data-migration
execution. These associations are maintained bidirectionally, allowing
navigation from source to target elements and vice versa. Since a source
element may give rise to several target elements, an identifier may be
associated with a collection of target identifiers or target objects.

\begin{figure}[htbp]
  \centering
  \includegraphics[width=\linewidth]{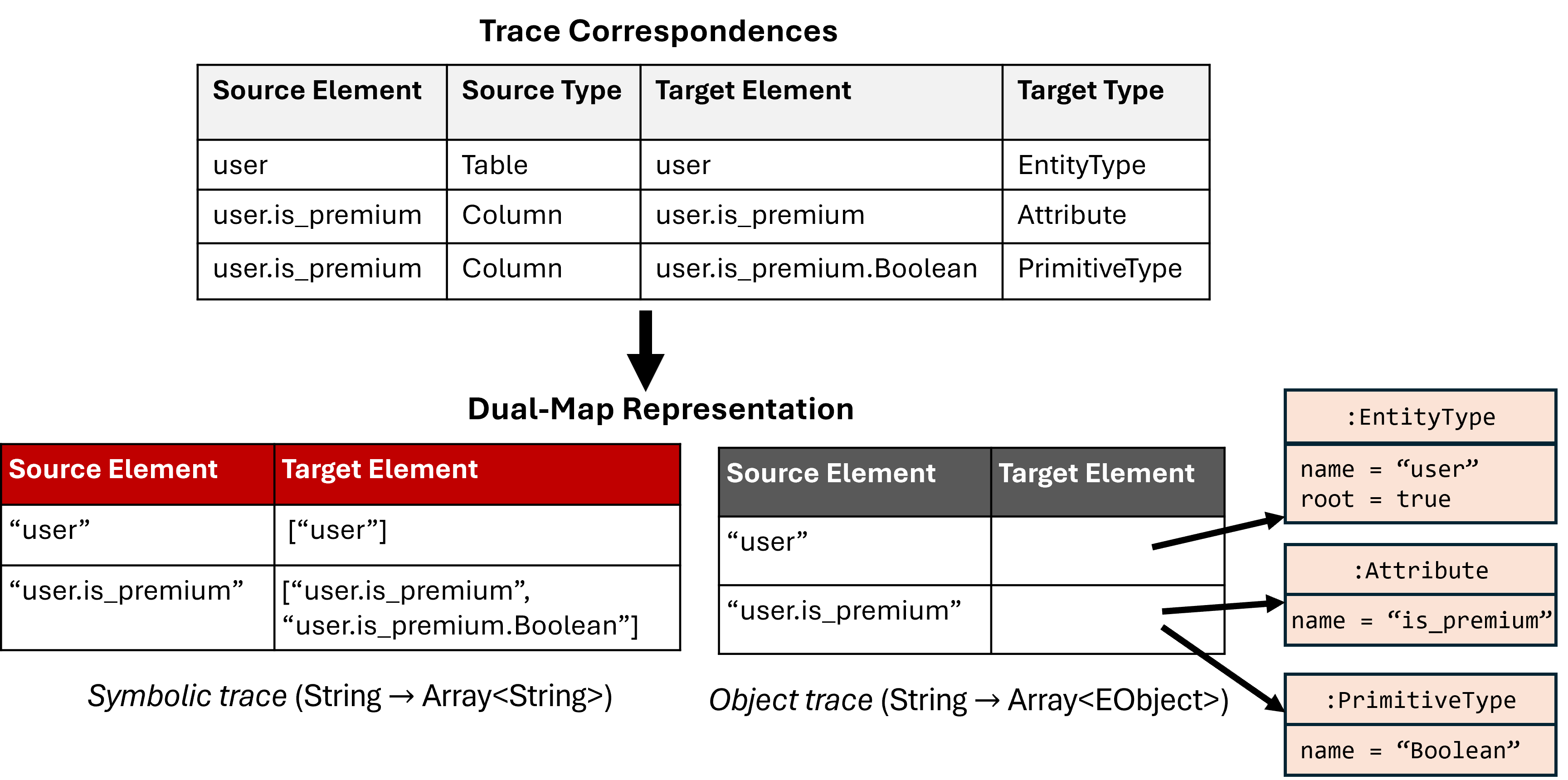}
 \caption{Trace correspondences and their symbolic and object
representations (Relational to~\uschema{}).}
  \label{fig:trace-example}
\end{figure}

Figure~\ref{fig:trace-example} illustrates these
correspondences for the Relational-to-\uschema{} transformation.
The relational table \texttt{user} corresponds to the \uschema{}
\texttt{EntityType} \texttt{user}. The relational column
\texttt{user.is\_premium}, in turn, gives rise to both the U-Schema
\texttt{Attribute} \texttt{user.is\_premium} and its associated
\texttt{PrimitiveType} \texttt{user.is\_premium.Boolean}. 
Both correspondences therefore share the same source element. 
% This occurs because, in the relational
% metamodel, the datatype is represented by the \texttt{datatype}
% property of the \texttt{Column}, rather than by a separate model
% element.
The figure also shows how these conceptual correspondences are
materialized in the two trace representations. In the symbolic trace,
the source identifier \texttt{``user.is\_premium"} is associated with
the identifiers of both generated target elements. In the object
trace, the same source identifier is associated with the corresponding
\texttt{Attribute} and \texttt{PrimitiveType} instances. Thus, the
trace supports one-to-many correspondences.

\paragraph{Trace persistence}
The trace can be persisted and subsequently reloaded together with
the corresponding models. Since references to \texttt{EObjects} are
valid only for the model instances loaded in memory, object
associations are persisted using their containment paths. When
the trace is reloaded, these paths are resolved against the loaded
models to reconstruct the object-level associations.

\paragraph{Trace generation and update}
During schema migration, the source-to-\uschema{} transformation
generates trace $T_1$. If the intermediate \uschema{} model is adapted,
$T_1$ is updated accordingly so that its correspondences remain
consistent with the modified model. The \uschema{}-to-target
transformation then generates trace $T_2$ from the resulting
\uschema{} model. Together, $T_1$ and $T_2$ establish the schema-level correspondences required by the Data Migrator, as explained in the following subsection.

% \hilight{Ver si se funde con lo dicho en la 4.4} Together, $T_1$ and $T_2$ establish the schema-level
% correspondences required by the Data Migrator. At runtime, $T_1$
% allows the Reader Adapter to relate U-Schema elements to their
% corresponding source-schema elements and thereby access the source
% data through a model-independent interface. Trace $T_2$ rela
tes
% U-Schema elements to the generated target-schema elements and guides
% the construction of target instances, as described in
% Section~4.4.

\subsection{Data Migrator Component}
\label{sec:data-migrator}

\begin{figure}[htbp]
  \centering
  \includegraphics[width=0.8\linewidth]{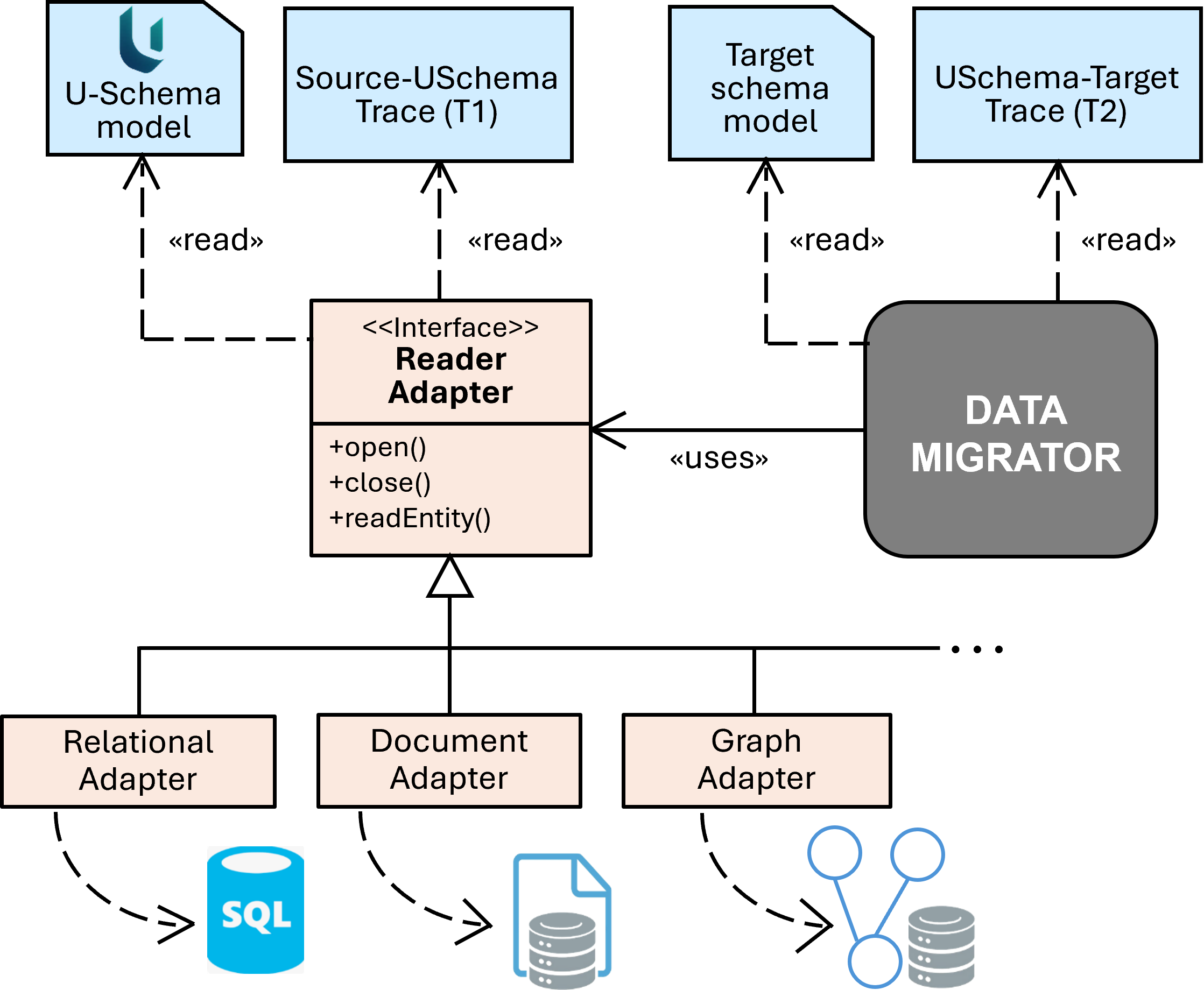}
  \caption{\uschema-based adapter for reading source data.}
  \label{fig:adapter-uschema}
\end{figure}

The Data Migrator executes the instance-level transformation between
the source and target databases according to traces $T_1$ and $T_2$.
Source instances are accessed through a \uschema{}-based adapter,
whose interface is defined in terms of \uschema{} concepts.
For a given migration, the adapter uses the resulting \uschema{} model
and its corresponding trace $T_1$ to expose source data through this
interface. Trace $T_2$ then determines how these data are materialized according
to the target schema. Figure~\ref{fig:adapter-uschema} summarizes this design.

\subsubsection{The Unified Reader Adapter}
\label{sec:adapter}

The adapter follows a \textit{driver–cursor pattern},
abstracting query execution and iterative access to source instances.
Its interface defines a minimal, technology-agnostic contract for 
opening connections, retrieving all instances of a given 
\uschema{} \texttt{EntityType}, and closing sessions:

{\footnotesize %\small
\begin{verbatim}
public interface ReaderAdapter {
    //open and terminate session
    String openConnectionService(
            String server, String service, 
            String username, String password);
    void closeConnection(); 
    // Iterate over all instances
    USchemaCursor readEntity(EntityType st);  
}
\end{verbatim}
} %fin de \footnotesize %\small

Through the \texttt{readEntity} operation, the Data Migrator obtains
a \texttt{USchemaCursor}, which provides uniform access to the
identifiers, attributes, references, and aggregates associated with
each source instance.

{\footnotesize %\small
\begin{verbatim}
public interface USchemaCursor {
    String getEntityName();
    Object getAttributeValue(Attribute att);
    String getAttributeType(Attribute att);
    USchemaID getID();
    USchemaCursor getReference(Reference ref);
    USchemaCursor getAggregate(Aggregate agg);
    boolean hasData();
    void next();
    void close();
}
\end{verbatim}
} %fin de \footnotesize %\small

The \texttt{getReference} and \texttt{getAggregate} operations 
provide a cursor for navigating through the instances of the referenced or aggregated entity.
Each adapter implementation translates these abstract operations
into the native access language of the source database
(e.g., SQL queries, MongoDB cursors, or Cypher traversals),
returning data in a form consistent with the \uschema{} model.
This abstraction enables the Data Migrator to process source instances
uniformly across heterogeneous database technologies while preserving
cursor-based streaming access and avoiding full dataset materialization.

% The \texttt{USchemaAdapter} API therefore provides:

% \begin{itemize}
%     \item A logical interface expressed in terms of \uschema{} elements
%           (\texttt{EntityType}, \texttt{Attribute}, \texttt{Reference}, \texttt{Aggregate});
%     \item Iterative, cursor-based access that abstracts system-specific result formats;
%     \item Extensibility through the implementation of new adapters
%           for additional database technologies;
%     \item Seamless integration with the trace mechanism,
%           since \uschema{} identifiers align data instances
%           with their schema-level correspondences.
% \end{itemize}

\subsubsection{Execution Pipeline}

The Data Migrator executes a trace-driven pipeline that transforms 
source data into target instances according to the schema correspondences 
established during schema migration. 
It iterates over the entity types defined in the target schema and performs the
following stages: 

\textit{1. Initialization and connection.}
The concrete adapter corresponding to the source database is initialized.
A connection is established through the \texttt{openConnectionService}
method, enabling the migration process to operate either in batch or
streaming mode depending on data volume.

\textit{2. Preparing source data access through a cursor.}
The entities in the target schema are traversed and, for each entity $te$
in the target schema, the trace $T_2$ is consulted to determine the
corresponding \texttt{EntityType} $ue$ in the \uschema{} model.
The adapter operation \texttt{readEntity} is then invoked to obtain a
\texttt{USchemaCursor} for $ue$, which provides access to the
corresponding source instances. Let $src_i$ denote a source instance
returned by this cursor.

\textit{3. Instance transformation guided by the trace.}
Each source instance $src_i$ retrieved through the cursor is transformed
into a target instance $tgt_i$ of $te$. For each feature of $te$, the
trace $T_2$ is consulted to identify the corresponding \uschema{} elements
(attribute, key, reference, or aggregate) and apply the required
transformation.

Attribute and key values are converted to the corresponding target
representations. References and aggregates are resolved recursively
through the adapter, which provides cursors over the corresponding
referenced or embedded source instances. In this way, the migrator
constructs target instances according to the structural 
correspondences recorded in the trace.

% In this way, the migrator can
% materialize referenced entities, relationship attributes, and nested
% aggregates according to the structural correspondences recorded in the
% trace.

\textit{4. Writing data to the target system.} 
Each transformed instance $tgt_i$ is serialized according to the target schema
and written using the native access mechanisms of the target database.
To improve performance, instances may be streamed directly, buffered
for batch insertion, or exported to intermediate files (e.g., JSON or
CSV) for bulk loading.

\textit{5. Finalization and integrity checking.}
After all target entities have been processed,
connections to the source and target databases are properly closed.
% \hilight{LLEVAR A TRABAJO FUTURO}
% The Data Migrator may optionally perform consistency checks,
% including record counts, referential completeness,
% and verification that all entities defined in the trace
% have been materialized in the target system.

\medskip
\noindent The pipeline can be abstracted as:
{\footnotesize 
\begin{center} 
\begin{BVerbatim}[fontsize=\footnotesize] 
 // Step 1: Initialization
sourceAdapter = initializeAdapter(T1)
sourceAdapter.openConnectionService()
targetConnection = open()

// Step 2: Preparing source data access
for each target entity te in targetSchema:
    ue = lookupUSchemaEntity(T2, te)
    // retrieve source instances corresponding to ue
    cursor = sourceAdapter.readEntity(ue)

// Step 3: Instance transformation
    while cursor.hasData():
        src = cursor.current()
        tgt = transformInstance(T2, src)

// Step 4: Writing to target
        writeNative(targetConnection, te, tgt)     
        cursor.next()
        
// Step 5: Finalization
targetConnection.close()
sourceAdapter.closeConnection()
\end{BVerbatim}
\end{center}
}%end de footnotesize

The \texttt{transformInstance(T2, src)} operation may involve recursively
processing aggregates and references.
This iterative, cursor-based execution strategy enables scalable data processing 
while preserving independence from the source data model through the adapter abstraction.

\section{Schema Mapping Rules: the Relational to Document case }
\label{sec:mappings}

As previously indicated, the proposed approach relies on \uschema{} as a
platform-independent pivot representation. Accordingly, two complementary
mappings are defined for each supported data model: (i) a data-model-to-\uschema{} mapping, which transforms a schema model
conforming to that data model into \uschema{}, and
(ii) a \uschema{}-to-data-model mapping, which transforms a
\uschema{} model into a schema model conforming to that data model.
A migration can therefore be implemented as the composition of the
source-model-to-\uschema{} mapping and the
\uschema{}-to-target-model mapping.

This section presents the mappings defined for the relational
and document data models, which enable both homogeneous migrations
within each data model and heterogeneous migrations in both directions
between the two data models.
We first introduce dedicated metamodels for both models,
together with the running example used throughout the section.

% This section presents the mappings defined for the relational 
% and document data models, which enable homogeneous and heterogeneous migrations 
% in both directions between systems conforming to these two models. 

Although the experimental evaluation focuses primarily on the
relational-to-document migration scenario, all four mappings
have been fully implemented.
The rules for the inverse direction are included in
Appendices A and B and are used in the round-trip validation
experiments described in Section~\ref{sec:roundtrip}

\subsection{Relational and Document metamodels}
\label{sec:metamodels}

\begin{figure}[htbp]
  \centering
  \includegraphics[width=\linewidth]{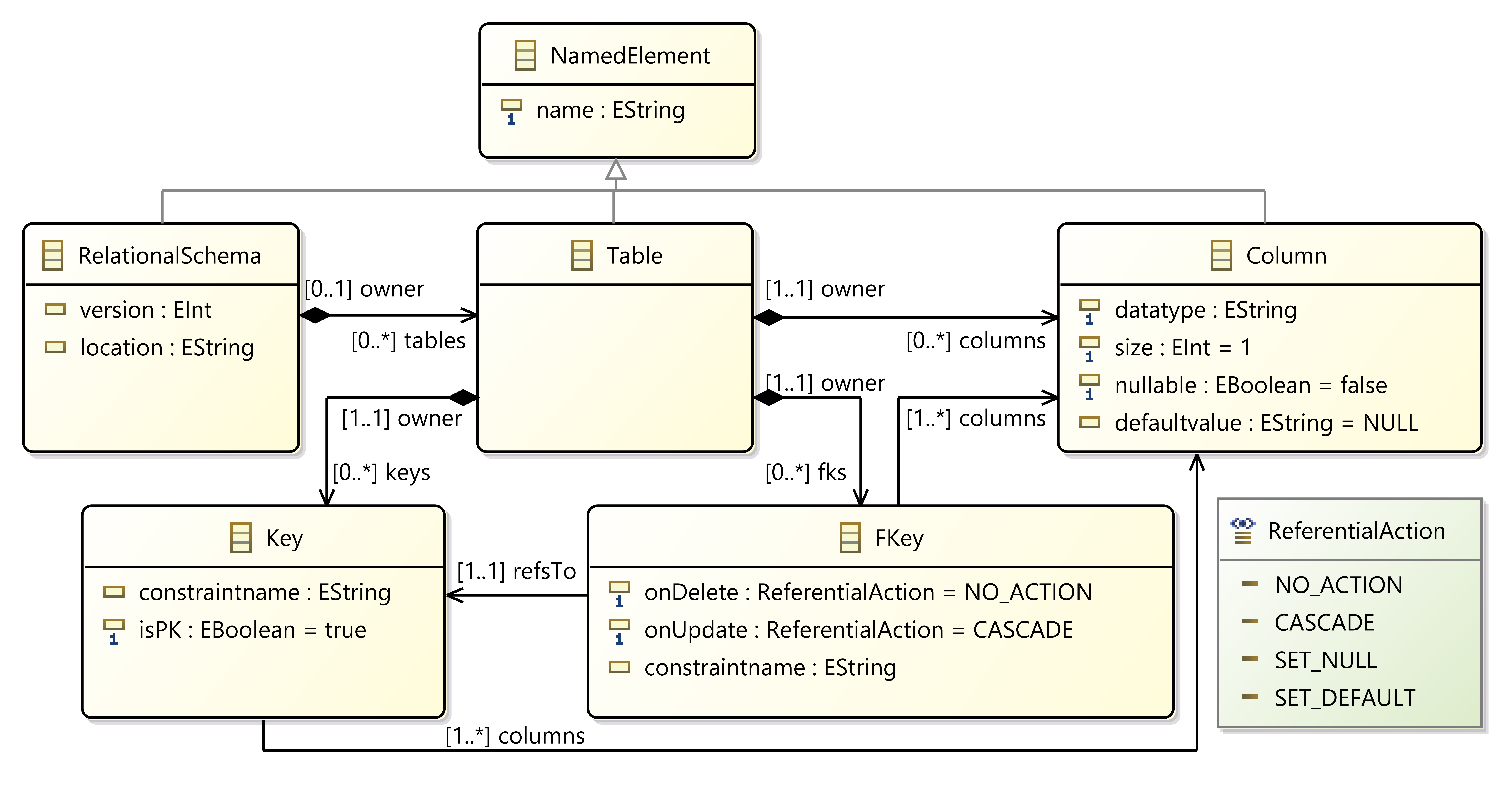}
  \caption{Relational metamodel.}
  \label{fig:relational-MM}
\end{figure}

The relational metamodel (Figure~\ref{fig:relational-MM}) captures the
structural components of a classical relational schema.
A \texttt{RelationalSchema} contains one or more \texttt{Table}s.
Each table may contain a set of \texttt{Column}s, \texttt{Key} elements and \texttt{FKey} elements.
Primary keys and unique key constraints are represented by the class
\texttt{Key}, while foreign key dependencies are explicitly modeled by the
class \texttt{FKey}. A foreign key references a target key and specifies
the referential actions applied on deletion or update through the
enumeration \texttt{ReferentialAction} (e.g., \textit{CASCADE},
\textit{SET\_NULL}).

\begin{figure}[htbp]
  \centering
  \includegraphics[width=\linewidth]{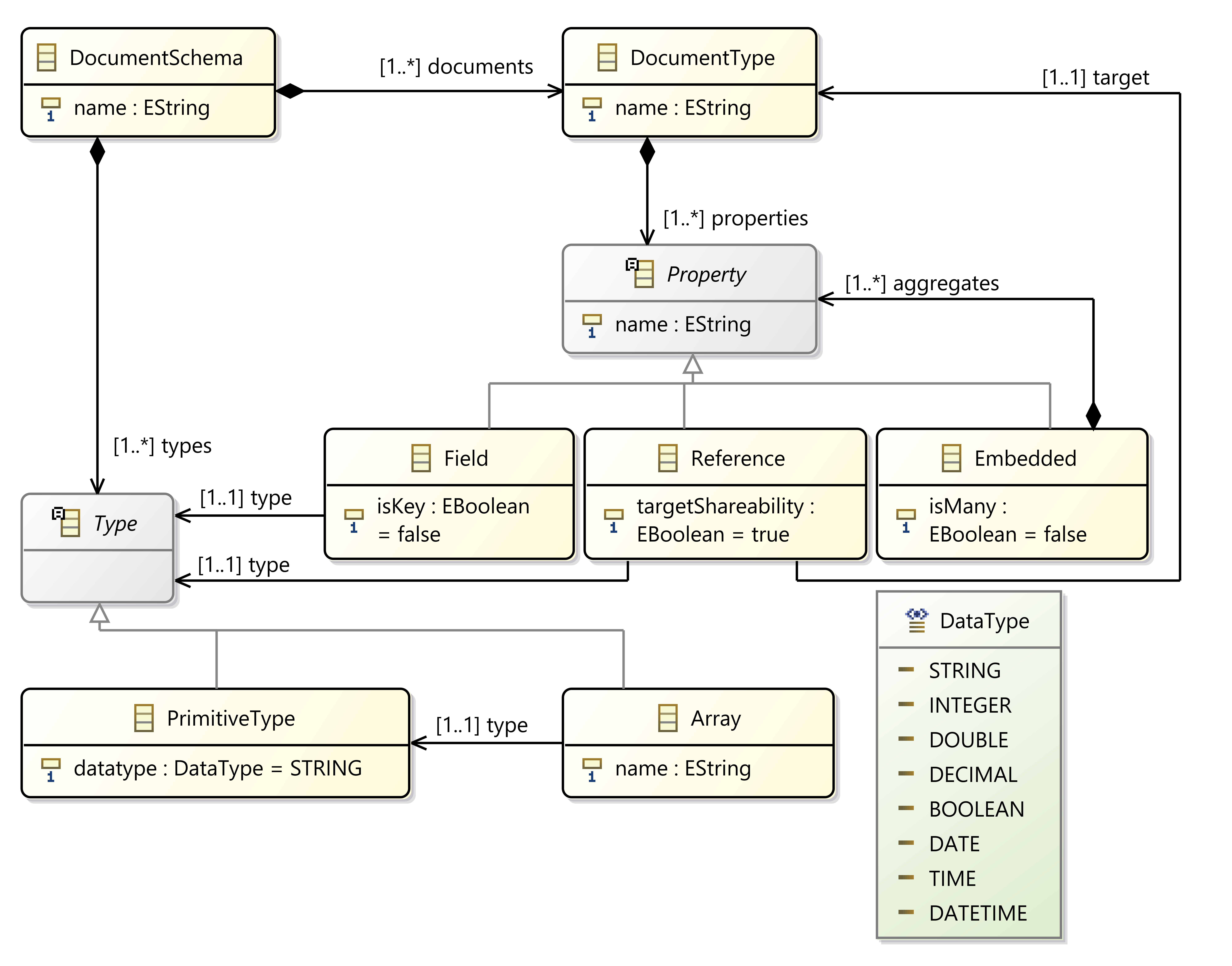} 
  \caption{Document metamodel.}
  \label{fig:document-MM}
\end{figure}

The document metamodel (Figure~\ref{fig:document-MM}) represents the
structural components of document-oriented databases, which store
documents as JSON-like objects.
A \texttt{DocumentSchema} contains one or more \texttt{DocumentType}
elements, each of which defines a set of \texttt{Property} instances.
A property can be either a \texttt{Field}, a \texttt{Reference}, or an
\texttt{Embedded} element.
\texttt{Field}s represent properties whose type is defined through the
\texttt{Type} hierarchy, either as a \texttt{PrimitiveType} or as an
\texttt{Array}. This differs from the relational metamodel, where column
datatypes are represented as \texttt{String} values.
% \texttt{Field}s  represent properties whose value is one of the scalar types
% defined by the metamodel or an array of scalar values.
% % \texttt{Field}s represent properties whose value is a scalar
% % (\textit{BOOLEAN}, \textit{INTEGER}, \textit{DOUBLE}, or \textit{STRING})
% % or an array of scalar values. 
A field may optionally be marked as a key
through the attribute \texttt{isKey}.
\texttt{Embedded} properties represent nested document structures,
and the flag \texttt{isMany} indicates whether the embedded object
appears as a single element or as an array.

\texttt{Reference}s represent links between document types and specify
the \texttt{target} document type. The cardinality of a reference is
determined through the \texttt{Type} hierarchy: a \texttt{PrimitiveType}
represents a single-valued reference, whereas an \texttt{Array} represents
a multi-valued reference. Following common practices in document-oriented
databases, we assume that references are represented by single-field
identifiers rather than composite ones.

This direct cardinality does not indicate whether a target object may be
referenced by several source objects. In relational schemas, this inverse
multiplicity is encoded by the foreign key and uniqueness constraints,
whereas in \uschema{} it can be represented through an opposite
\texttt{Reference}. Since document schemas do not usually express this
information, the Boolean attribute \texttt{targetShareability} indicates
whether the same target object may be referenced by more than one distinct
source object. It is set to \texttt{true} by default, and the database
designer must identify the references whose targets are not shareable.
In \uschema{}, the non-shareability of the target of a reference is
represented by an opposite \texttt{Reference} with upper bound one.
% \mjose{Lo que sigue quizá es muy específico y podría quitarse ¿?. Aquí está bien explicar que esa opposite sirve para representar la shareability, y más adelante en las reglas se explicará cómo y cuándo se crea y usa $\rightarrow$ }
% If no such opposite is already linked to the reference through the
% \texttt{opposite} association, the mapping that creates the reference
% must synthesize an attribute-less one (e.g., Rule~RU7.1). 
% Its sole purpose is to record this information so that subsequent mappings can
% recover it when needed.

\medskip

\subsection{Running Example: Music Streaming Service}
\label{sec:running-example}

To illustrate the mappings defined in the following subsections,
we use a simplified Music Streaming service deliberately designed
to include the main schema structures and mapping cases addressed
by the proposed rules.
% \mjose{Llevar lo que sigue  al final del párrafo siguiente (donde se habla de las figuras y las secciones 5.5 y 5.6), o traer aquello aquí.} 
% The example covers the mapping rules considered in this work, and its relational, \uschema{},
% and document representations are shown together in
% Figure~\ref{fig:running-example-mapping}.
% } \hoyos{CIERTO. La figura está 5 páginas más adelante, demasiado separada}

% To illustrate the mappings defined in the following subsections,
% we use a simplified Music Streaming service that captures a representative
% variety of structural elements.
% The example covers the main constructs of \uschema{}, including entities, 
% attributes, references, and aggregates.
% It therefore serves as a suitable domain to demonstrate both relational
% and document transformations.

\begin{figure}[htbp]
  \centering
  \includegraphics[width=0.9\linewidth]{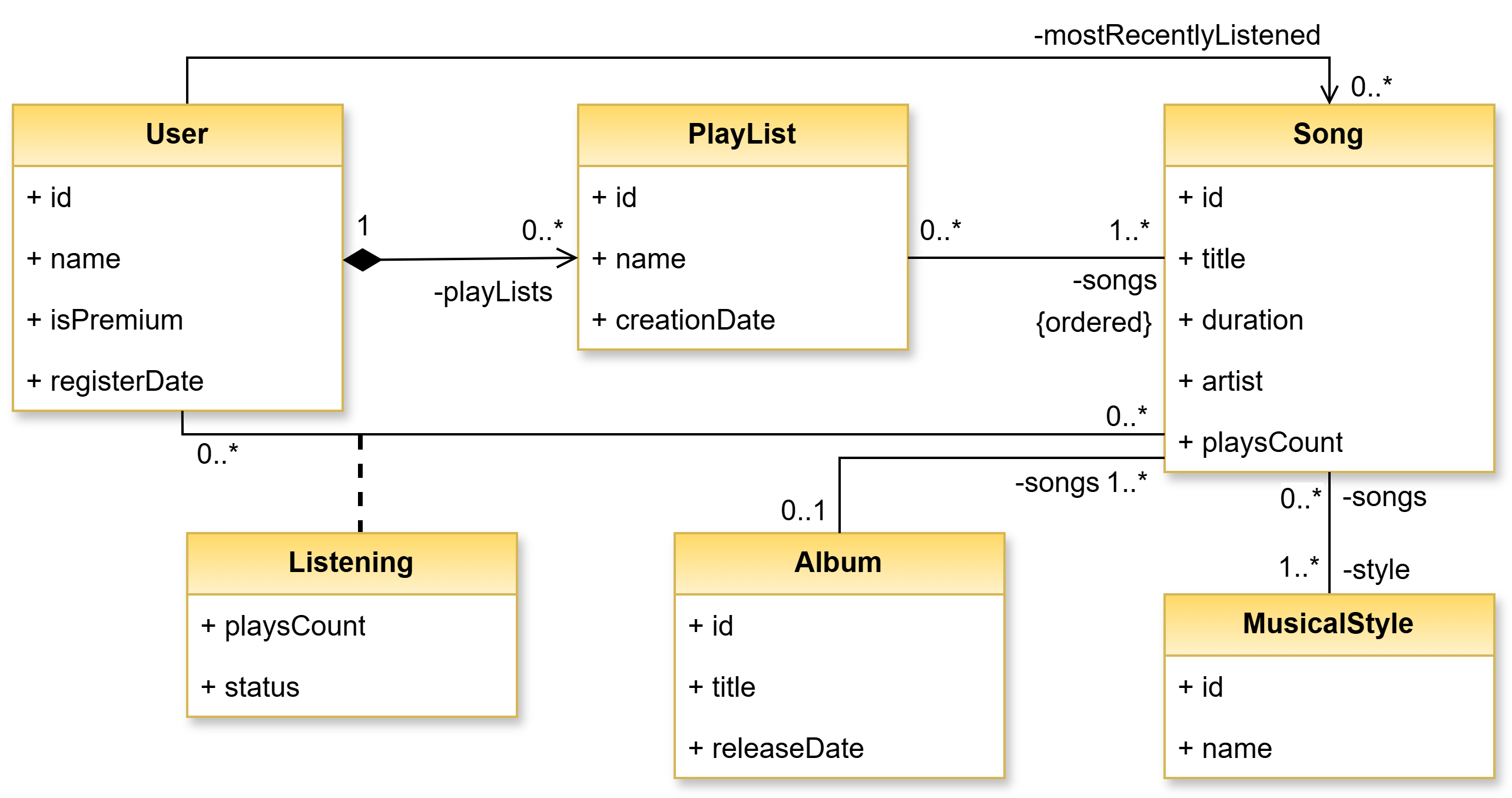}
  \caption{Conceptual schema of the Music Streaming service used as running example.}
  \label{fig:music-uml}
\end{figure}

The domain involves users, playlists, songs, albums, and musical styles,
as well as listening activity. Each user has a unique identifier, can be
premium or standard, and can create multiple playlists containing an
ordered list of songs. Each song has a title, duration, and one or more
musical styles (e.g., \textit{Rock}, \textit{Jazz}, or \textit{Pop}),
and may optionally belong to an album. For each user, the system also
maintains a list of the most recently played songs. Listening activity is
recorded for each user--song pair, including the number of times the user
has played the song and the current listening status.
Figure~\ref{fig:music-uml} shows the conceptual schema of this domain as
a UML class diagram. This example is used in
Sections~\ref{sec:rel2uschema-rules} and
\ref{subsec:uschema-to-document} to illustrate the corresponding
mapping rules, while Figure~\ref{fig:running-example-mapping} presents
the relational, \uschema{}, and document models together with 
the mapping rules applied to their transformation.

% This example is used in Sections~\ref{sec:rel2uschema-rules} and \ref{subsec:uschema-to-document} 
% to illustrate the mapping rules between relational, \uschema{}, and document models. 

% \hoyos{En UML, no habría una relación desde User a Song. Habría una relación de User a Listening, con cardinalidades "0..1"(caja user) y "0..*"(caja listening) y otra de Listening a Song con cardinalidades "0..*" y "1". Tal y como están las asociaciones, si fueran unidireccionales (ej. Listening no conoce a user) entonces sólo se pondría la segunda cardinalidad en los dos casos. NOTA: Las cardinalidades bidireccionales se representan en UML como líneas sin flechas. Cambiar las de PlayList-Song y Album-Song}

% Table~\ref{tab:rel-doc-example} contrasts the relational and document representations 
% of the Music Streaming service. 
% The relational schema (left) defines normalized tables for users, playlists, and songs, 
% with some tables (e.g., \texttt{song}, \texttt{album}, and \texttt{musical\_style}) omitted for clarity. 
% The document model (right) reorganizes the same information around the user, 
% embedding playlists as nested arrays while maintaining references to songs stored in a separate collection. 
% User listening activity is also stored in a separate collection due to its potentially large size.

Table~\ref{tab:rel-doc-example} contrasts the relational and document
representations of the Music Streaming service.
% The relational schema (left)
% includes \texttt{user} as a regular entity table;
% \texttt{playlist} and \texttt{playlist\_song} as weak tables whose
% identifiers combine the key of their owner with a local discriminator \mjose{(partial key)};
% and \texttt{listening} as an associative table with attributes, whose
% composite primary key consists of the foreign keys referencing
% \texttt{user} and \texttt{song}. 
% Moreover, \texttt{playlist\_song} contains a non-identifying foreign key 
% to \texttt{song}. Some regular entity tables, including \texttt{song},
% \texttt{album}, and \texttt{musical\_style}, are omitted from the DDL
% excerpt for clarity.}
The relational schema (left) includes 
the entity \texttt{user} as a regular table;
\texttt{playlist} and \texttt{playlist\_song} as weak tables whose
identifiers combine the key of their owner with a local discriminator (partial key);
and \texttt{listening} as an associative table with attributes, whose
composite primary key consists of the foreign keys referencing
\texttt{user} and \texttt{song}. 
Moreover, \texttt{playlist\_song} contains a non-identifying foreign key 
to \texttt{song}. Some regular tables, including \texttt{song},
\texttt{album}, and \texttt{musical\_style}, are omitted from the DDL
excerpt for clarity.

Following the mapping conventions described in
Section~\ref{sec:conventions}, the weak tables
\texttt{playlist} and \texttt{playlist\_song} are embedded, whereas the
associative table \texttt{listening} is represented as a separate document
type. Consequently, each user document contains an array of playlists,
and each playlist in turn contains an array of ordered
\texttt{playlist\_song} entries, forming a two-level nesting hierarchy.
Each \texttt{playlist\_song} entry stores its position and a reference
to a song, while songs remain as separate documents.

\begin{table*}[t]
\centering
\caption{Relational and document simplified representations of the Music Streaming service.}
\label{tab:rel-doc-example}
\begin{tabular}{p{0.47\textwidth} p{0.47\textwidth}}
\toprule
\textbf{(a) Relational schema (SQL DDL)} &
\textbf{(b) Document model (JSON excerpts)} \\
\midrule

\vspace{-1.8em}
\begin{minipage}[t]{\linewidth}
\begin{lstlisting}[style=sqlStyle, basicstyle=\ttfamily\footnotesize]
CREATE TABLE user (
  user_id       CHAR(36)    PRIMARY KEY,
  name          VARCHAR(80) NOT NULL,
  is_premium    BOOLEAN     DEFAULT false,
  register_date DATE        NOT NULL
);

CREATE TABLE playlist ( -- Weak table
  playlist_id   CHAR(36)    NOT NULL,
  user_id       CHAR(36)    NOT NULL,
  name          VARCHAR(30),
  creation_date DATE        NOT NULL,
  PRIMARY KEY (user_id, playlist_id),
  FOREIGN KEY (user_id) REFERENCES user(user_id)
);

CREATE TABLE playlist_song ( -- Weak table
  playlist_id  CHAR(36) NOT NULL,
  user_id      CHAR(36) NOT NULL,
  position_idx INT      NOT NULL,
  song_id      CHAR(36) NOT NULL,
  PRIMARY KEY (user_id, playlist_id, position_idx),
  FOREIGN KEY (user_id, playlist_id) 
    REFERENCES playlist(user_id, playlist_id),
  FOREIGN KEY (song_id) REFERENCES song(song_id)
);

CREATE TABLE listening ( -- Associative table (MN)
  user_id     CHAR(36) NOT NULL,
  song_id     CHAR(36) NOT NULL,
  plays_count INT      NOT NULL,
  status      VARCHAR(10),
  PRIMARY KEY (user_id, song_id),
  FOREIGN KEY (user_id) REFERENCES user(user_id),
  FOREIGN KEY (song_id) REFERENCES song(song_id)
);
\end{lstlisting}
\end{minipage}
&
\vspace{-1.8em}
\begin{minipage}[t]{\linewidth}
\begin{lstlisting}[style=jsonStyle, basicstyle=\ttfamily\footnotesize]
// Example of "user" document
{ user_id: "u001",
  name: "Alice",
  is_premium: true,
  register_date: "2025-02-25 14:09:30",
  playlists: [
    { playlist_id: "p001",
      name: "indi90s",
      creation_date: "2026-08-26 18:20:00",
      playlist_songs: [
        { position_idx: 1,
          song_id: "s002"
        },
        { position_idx: 2,
          song_id: "s007"
        }
      ]
    },
    { playlist_id: "p002",
      name: "moviesOST",
      creation_date: "2025-05-20 19:50:00",
      playlist_songs: [
        { position_idx: 1,
          song_id: "s054"
        }
      ]
    }
  ]
}
// Example of "listening" document
{ listening_id: "72af18c52a0f8162fc1ae15b",
  user_id: "u001",
  song_id: "s007",
  plays_count: 7,
  status: "completed"
}
\end{lstlisting}
\end{minipage}

\\[-0.8em]%[-1.0em]
\bottomrule
\end{tabular}
\end{table*}

The structural correspondences illustrated in
Table~\ref{tab:rel-doc-example} are formally specified by the mapping
rules presented in the following subsections. These rules define how the
relational structures are first represented in \uschema{} and subsequently
transformed into the canonical document schema.

% The structural correspondences between schemas based on different data models,
% such as those illustrated in Table~\ref{tab:rel-doc-example},
% can be formally captured through mapping rules.
% The following subsections present the mapping rules defined in our approach
% for relational-to-document migration.

% \subsection{Mapping Conventions}
% \hilight{As we mentioned earlier, when multiple alternatives are possible for a schema transformation, we apply predefined conventions, which are described below.
% +Foreign key
% - If the foreign key is part of the primary key in a strong/weak relationship, then it is transformed into an aggregate, unless it is referenced from another table.
% - Otherwise, it is mapped into a reference stored in the same entity if is 1:1 and stored in the referenced entity if is 1:N.
% +Associative Table
% - no aggregate - Its foreign keys are mapped into references (Mapped into a relationshiptype)
% +Composite key: Concatenate value of fields components of the composite key 
% } 

\subsection{Mapping Rule Notation, Auxiliary Functions and Predicates}
\label{sec:mapping-rule-notation}

To formally express the schema mappings in a concise and
uniform way, we adopt a declarative notation based on
\emph{mapping expressions}. Each mapping rule is described using a
structured format consisting of three parts: a short \emph{preamble},
the \emph{mapping expression} itself, and an optional set of final
remarks. In addition, a small set of auxiliary functions and predicates
is used across all mappings. This subsection introduces the notation
together with two auxiliary functions and four predicates.

\paragraph{1) Preamble}
Each rule begins with a brief textual preamble that introduces the
intent of the mapping and identifies the kinds of source and target
elements involved. Its
purpose is purely descriptive, helping the reader understand the scope
and applicability of the rule. 

\paragraph{2) Mapping expression}
The core of each rule is written using the correspondence operator
$\rightarrow$:
\[
    s \;\rightarrow\; t \;\|\; \{\, C_1, C_2, \ldots, C_n \,\}
\]
where $s$ denotes an instance of a source metaclass, $t$ denotes one
or more instances of target metaclasses, and each $C_i$ is a clause
expressing a declarative constraint that must be satisfied by the
elements involved in the mapping.

% where $s$ denotes an element of the source metamodel, $t$ denotes
% one or more elements of the target metamodel, ...

Mapping expressions are purely declarative: they specify properties that
must hold between source and target elements, without prescribing any
execution order, update semantics, or operational behavior.

Clauses may take one of the following forms:

\vspace{-0.3cm}
\begin{itemize}
    \item \textit{Equality constraint}: $p_1 = p_2$,
    stating that two properties have the same value
    (e.g., $s.name = t.name$).

    \item \textit{Value constraint}: $p = v$,
    stating that property $p$ has the fixed value $v$
    (e.g., $t.root$ = \texttt{true}).

    \item \textit{Conditional clause}: $Q \Rightarrow C$,
    meaning that clause $C$ must hold whenever condition $Q$ is satisfied.

    \item \textit{Universal clause}:
    $\{C_1, C_2, \ldots, C_n\} : \forall x,\; Q(x)$,
    meaning that clauses $C_1, C_2, \ldots, C_n$ are instantiated
    and applied for every element $x$ satisfying condition $Q(x)$.

    \item \textit{Existential condition}:
    $\exists x,\; Q(x)$,
    meaning that there exists at least one element $x$ satisfying
    condition $Q(x)$.

    % \item \textit{Universal clause}:
    % %$\forall x \;\text{such that}\; Q(x) : \{C_1, C_2, \ldots, C_n\}$,
    % $\{C_1, C_2, \ldots, C_n\} : \forall x,\; Q(x)$
    % meaning that the clauses $C_i$ must hold for all elements $x$
    % satisfying condition $Q$.

    % \item \textit{\mjose{**NEW**Existential Clause}:}
    % $\exists\,x,\; Q(x):\{C_1, C_2, \ldots, C_n\}$
    % \mjose{meaning that at least there exits an element $x$ such that condition $Q$ and clauses $C_i$ are satisfied.}
    
\end{itemize}

Clauses may include set operators (e.g., $\cup$, $\in$, $\subseteq$)
and may be composed of subclauses connected by boolean operators
such as $\land$, $\vee$, and $\neg$.

\paragraph{3) Final remarks (optional)}
Following the mapping expression, certain rules include short remarks to
clarify notational details or to indicate the specific collections of
the target model where the resulting elements are placed. These remarks
are complementary and do not introduce additional formal constructs.

\paragraph{Auxiliary functions}
We define the following two common auxiliary functions:

%\begin{itemize}

%\item 
\textit{Mapping lookup}. When a rule needs to refer to the target element
corresponding to a source element $s$, we use the function 
\texttt{map}(\textit{s}), which denotes the target element produced from 
$s$ by the transformation. When more than one target element is generated
from the same source element, the typed form \texttt{map}(\textit{s}, 
\textit{T}) denotes the target element of metaclass $T$ generated from $s$.

%\item
\textit{Primitive type mapping}. The \texttt{primitiveTypeMap}(\textit{p})
function returns, for the current transformation, the target primitive datatype
value corresponding to a source primitive datatype value $p$.
This function is used for datatype values, such as relational type names, 
\uschema{} \texttt{PrimitiveType} names, or document primitive-type literals.

% \item \textit{Primitive type mapping.}
% The function \texttt{typeMap}(\textit{dt}) returns 
% ``the target data type associated with a data type $dt$'',
% according to the appropriate mapping.
% It provides a uniform treatment of data domains for all supported paradigms.
%\end{itemize}

\paragraph{Auxiliary predicates}
We introduce four predicates to characterize relational
structures: tables with a primary key, 
foreign keys that are part of the primary key of
their table, weak tables, and associative tables.

%\begin{itemize}

%\item
\textbf{HasPK(t).}
Although relations in the classical relational model necessarily
have a candidate key, SQL schemas may contain tables without an
explicitly declared primary key, a situation that also occurs in
real-world databases. The predicate $\mathrm{HasPK}(t)$ holds when
table $t$ has an explicitly declared primary key.

\vspace{-0.5cm}
{\fontsize{9}{10}\selectfont
\[
\mathrm{HasPK}(t) \equiv
\exists k \in t.keys :
k.isPK == \text{true}
\]
}

%\item
\textbf{FkInPk(t, fk).}
The predicate $\mathrm{FkInPk}(t, fk)$ holds true when a foreign key $fk$ in a table $t$ is part of the primary key $pk$ of $t$. 

\vspace{-0.5cm}
{\fontsize{9}{10}\selectfont\[
\begin{aligned}
\mathrm{FkInPk}(t,fk) \equiv {}&
\mathrm{HasPK}(t)\\
&{}\land fk.columns \subseteq  \texttt{pk}(t).columns
\end{aligned}
\]
}
% \[
% \begin{aligned}
% \text{FkInPk($t, fk$)}\;\equiv\;
% %&\exists \;
% %&fk \in t.fks \land \; 
% &fk.columns \subseteq \texttt{pk}(t).columns \\
% %&\text{where} \; pk \in t.keys \land pk.isPK
% %\;)
% \end{aligned}
% \]
The function \texttt{pk}($t$) returns the primary key of table \textit{t}.

\vspace{-0.5cm}
{\fontsize{9}{10}\selectfont\[
\begin{aligned}
\texttt{pk}(t) := k \;| \; k \in t.keys,\, k.isPK == \text{true}
\end{aligned}
\]
}
%\item 
\textbf{Weak(t).}
The predicate $\text{Weak}(t)$ holds true when a table $t$ represents a \emph{weak table}. 
This situation occurs when the primary key $pk$ of $t$ includes exactly one foreign key $fk$ that references another table $s$ (the strong one).
In this case, the foreign key contributes to the identification of each row in $t$, 
meaning that the existence of the records in $t$ depends on the corresponding records in $s$.

\vspace{-0,5cm}
{\fontsize{9}{10}\selectfont
\[
\begin{aligned}
\text{Weak} &(t) \equiv\;\exists \, fk \in t.fks : \{\,\text{FkInPk}(t,fk)\\
&\land \not\exists fk'\in t.fks : \{\,fk'\neq fk \land \text{FkInPk}(t,fk')\,\}\,\}\\
%\;\land &(\;\forall\,t'\in t.owner\;(t'\neq t\;\land \\ 
%                &\hspace{1cm}(\not\exists\;r \in t'.fks\; (r.refsTo.owner==t)\;)\;)\;)\\
%&\hilight{\text{where} \; pk \in t.keys \land pk.isPK}
\end{aligned}
\]
}

%\item
\textbf{MN(t).}
The predicate $\text{MN}(t)$ holds true when a table $t$ represents an associative
table. This occurs when there exist two foreign keys $fk_1$ and $fk_2$ in $t$ such 
that $fk_1$ references a table $t_1$ and $fk_2$ references a table $t_2$, and 
the primary key of $t$ is defined by the two foreign keys,
thereby materializing a many-to-many relationship between $t_1$ and $t_2$. 
Moreover, $pk$ could include additional columns.

\vspace{-0.5cm}
{\fontsize{9}{10}\selectfont
\[
\begin{aligned}
&\text{MN}(t) \equiv\;
\exists fk_1, fk_2 \in t.fks\,:\\
&\qquad \{\,fk_1 \neq fk_2 \land \text{FkInPk}(t, fk_1) \land \text{FkInPk}(t, fk_2) \,\}
\end{aligned}
\]
}
%\end{itemize}

\subsection{Mapping Conventions}
\label{sec:conventions}
% ESTAS CONVENCIONES SON ESTRUCTURALES, NO SEMÁNTICAS.

When multiple alternatives exist for representing a given schema structure, 
our approach applies predefined conventions to derive a canonical target schema. 
These conventions ensure deterministic transformations and avoid the need 
for additional configuration during schema migration.
The main conventions used in the relational to document migration are as follows:

\paragraph{Foreign keys}

If $\mathrm{Weak}(t)$ holds, $t$ is mapped to an entity embedded
in the entity corresponding to the table referenced by its
identifying foreign key $fk$, unless $t$ is referenced by a
non-weak table. To express this exception precisely, we introduce
the mapping-specific predicate $\mathrm{Aggregable}(t)$.
For a table $t$ in a source relational schema $rS$, 
the predicate $\mathrm{ReferencedByNonWeak}(t)$ holds 
when a distinct non-weak table $t'$ owns a foreign key that references $t$:

\vspace{-0.5cm}
{\fontsize{9}{10}\selectfont
\[
\begin{aligned}
&\mathrm{ReferencedByNonWeak}(t) \equiv {}\\
&\quad \exists \, t' \in rS.tables,\; 
        fk' \in t'.fks \colon \\[-1mm]
&\qquad\{\,t' \not= t \,\land\,\neg\mathrm{Weak}(t') \land fk'.refsTo.owner == t \,\}
\end{aligned}
\]
}

A weak table is aggregable when it is not referenced by any distinct non-weak table:

\vspace{-0.5cm}
{\fontsize{9}{10}\selectfont
\[
\mathrm{Aggregable}(t)
\equiv
\mathrm{Weak}(t) \land \neg\mathrm{ReferencedByNonWeak}(t)
\]
}

Unlike the structural predicates introduced in
Section~\ref{sec:mapping-rule-notation},
$\mathrm{Aggregable}(t)$ is a mapping-specific condition that
determines whether the identifying relationship of a weak table
can be represented through embedding in the target document
model.

% Foreign keys are mapped to references in the remainder cases.
% For a 1:1 relationship, the reference is added to the entity mapped from
% the referencing table, preserving the foreign key direction; for a 1:N
% relationship, it is added to the entity mapped from the referenced table,
% reversing that direction.

Foreign keys not used to represent an aggregation or an
associative relationship are mapped to references. A unique
foreign key is represented by a single-valued reference in the
entity mapped from the referencing table, preserving the
foreign key direction. A non-unique foreign key also preserves this direction when its
owner table is aggregable or has no primary key; otherwise, it is
represented by a multi-valued reference in the entity mapped from
the referenced table, reversing the foreign key direction.

\paragraph{Associative tables}
Associative tables are mapped to separate collections rather than embedded documents. Their foreign keys are mapped to references to the
corresponding document types.

\paragraph{Composite keys}
Each composite key is represented in the document schema by a derived identifier field whose value is obtained by concatenating the values of
its component fields.

% Each composite key is mapped to a derived key attribute whose
% value is obtained by concatenating the values of its component
% attributes.

\vspace{0.4em}
Table~\ref{tab:relational-patterns-rules} provides a compact
roadmap of the source-schema patterns addressed by the 
relational-to-document transformation chain. 
For each pattern, it summarizes its recognition condition,
the rules responsible for its transformation, and the resulting
U-Schema and document representations. The table is intended as
an overview rather than as an execution specification: the
mappings are declarative, and several rules may contribute to
the transformation of the same source table.
%\jesus{source-schema structures and mapping-specific cases en vez de patterns??}\hoyos{Sí. Me parece mucho más claro. Además en la Table 3, en el encabezado de la primera columna se podría poner en vez de "Relational pattern" , un "Relational structure" y luego dentro de la tabla eliminar "patterns" al final de los apartados en negrita "Baic table, column, and keys", "Weak-table and associative-table", "foreign key relationship", etc...}

\begin{table*}[!htb]
\centering
\footnotesize
\setlength{\tabcolsep}{4pt}
\renewcommand{\arraystretch}{1.12}

\caption{Relational schema patterns and their treatment in the
relational-to-document mapping chain.}
\label{tab:relational-patterns-rules}

\begin{tabularx}{\textwidth}
{p{2.6cm} p{4.0cm} p{1cm} p{3.0cm} p{1cm} p{3.0cm}}
\toprule
\textbf{Relational pattern} &
\textbf{Recognition condition} &
\textbf{RU rule} &
\textbf{U-Schema representation} &
\textbf{UD rule} &
\textbf{Document representation} \\
\midrule

\addlinespace[3pt]
\multicolumn{6}{l}{\textbf{Basic table, column, and key}}\\
%\multicolumn{6}{@{}l}{\textbf{Basic table, column, and key patterns}}\\
\addlinespace[2pt]

Regular table &
$\neg\mathrm{MN}(t)
 \land
 \neg\mathrm{Aggregable}(t)$ &
RU2 &
Root \texttt{EntityType} &
UD2 &
\texttt{DocumentType} \\

\addlinespace[3pt]

Column not participating in a foreign key &
$\nexists fk \in t.fks \colon c \in fk.columns$ &
RU3 &
\texttt{Attribute} and associated \texttt{PrimitiveType} &
UD3 &
\texttt{Field} and associated \texttt{PrimitiveType} \\

\addlinespace[3pt]

Simple primary key &
$|\texttt{pk}(t).columns|==1$ &
RU4 &
Identifying \texttt{Key} with one attribute &
UD4 &
Existing field marked as identifier \\

Composite primary key &
$|\texttt{pk}(t).columns|>1$ &
RU4 &
Identifying \texttt{Key} with multiple attributes &
UD4 &
Derived single-field identifier; component fields are retained \\

\addlinespace[3pt]
\multicolumn{6}{l}{\textbf{Weak-table and associative-table}}\\
\addlinespace[2pt]

Aggregable weak table with a partial key &
$\mathrm{Aggregable}(t) \land
\mathrm{FkInPk}(t,fk) \land
fk.columns \subsetneq \texttt{pk}(t).columns$ &
RU2,
RU5 &
Non-root \texttt{EntityType} and multi-valued
\texttt{Aggregate} &
UD6 &
Array of embedded documents \\

\addlinespace[3pt]

Aggregable weak table without a partial key &
$\mathrm{Aggregable}(t) \land
\mathrm{FkInPk}(t,fk) \land 
fk.columns == \texttt{pk}(t).columns$ &
RU2,
RU5 &
Non-root \texttt{EntityType} and single-valued
\texttt{Aggregate} &
UD6 &
Single embedded document \\

\addlinespace[3pt]

Non-aggregable weak table &
$\mathrm{Weak}(t)
 \land
 \neg\mathrm{Aggregable}(t)$ &
RU2,
RU7.1 / RU7.3 &
Root \texttt{EntityType}; its identifying foreign key is
represented through references according to its uniqueness &
UD2,
UD5 &
\texttt{DocumentType}; the identifying
relationship is represented by a single reference or an array
of references, depending on the foreign key uniqueness \\

Associative table &
$\mathrm{MN}(t)$ &
RU2,
RU6 &
\texttt{RelationshipType} and references connecting the
participating entity types &
UD7 &
\texttt{DocumentType} with references and
relationship attributes \\

\addlinespace[3pt]
\multicolumn{6}{l}{\textbf{Foreign key }}\\
\addlinespace[2pt]

Unique foreign key &
$\exists uk \in t.keys \colon
uk.columns == fk.columns$, and $fk$ is not handled by
RU5 or RU6 &
RU7.1 &
Single-valued \texttt{Reference} preserving the foreign key
direction, together with a synthesized single-valued opposite;
the foreign key attributes are added to the corresponding
\texttt{Key} &
UD5 &
Single-valued document reference; the synthesized opposite
is used to characterize the inverse multiplicity \\

Non-unique foreign key in an aggregable table &
$\mathrm{Aggregable}(t) \land
\neg\mathrm{FkInPk}(t,fk) \land
\nexists uk \in t.keys \colon uk.columns == fk.columns$ &
RU7.2 &
Single-valued N:1 \texttt{Reference} preserving the
foreign key direction; no opposite reference is synthesized &
UD5 &
Single-valued document reference \\

Non-unique foreign key in a non-aggregable table &
$\mathrm{HasPK}(t) \land
\neg\mathrm{Aggregable}(t) \land
\nexists uk \in t.keys \colon$

$uk.columns == fk.columns$,
and $fk$ is not handled by RU6 &
RU7.3 &
Reverse multi-valued \texttt{Reference} and synthesized
single-valued opposite; identifying foreign key attributes are
retained in the corresponding \texttt{Key} when applicable &
UD5 &
Array of document references in the document type
corresponding to the referenced table \\

Non-unique foreign key in a table without a primary key
&
$\neg\mathrm{HasPK}(t) \land
\nexists uk \in t.keys \colon$

$uk.columns == fk.columns$
&
RU7.4
&
Single-valued \texttt{Reference} preserving the foreign key direction
&
UD5
&
Single-valued document reference
\\

\bottomrule
\end{tabularx}
\end{table*}

\subsection{Relational to \uschema Mapping}
\label{sec:rel2uschema-rules}

\begin{figure*}[p]
  \centering
  \includegraphics[width=\textwidth, height=0.9\textheight, keepaspectratio]%
    {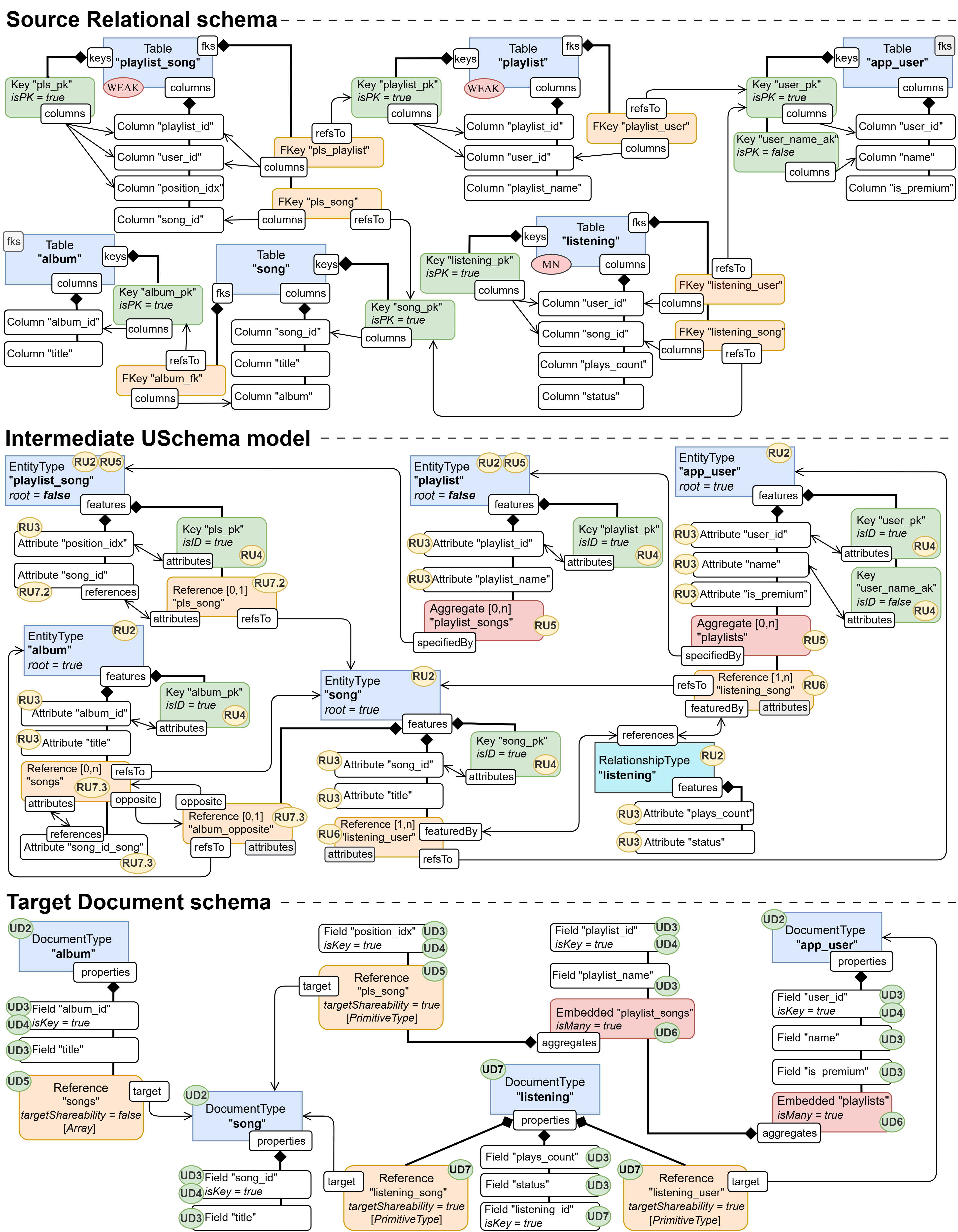}
  \caption{Transformation of the running example across the relational, 
  \uschema{}, and document models.}
  \label{fig:running-example-mapping}
\end{figure*}

Given a relational schema represented as a model conforming
to the Relational metamodel (Figure~\ref{fig:relational-MM}), 
the following rules are defined to map the relational constructs
to \uschema (Figure~\ref{fig:uschema-metamodel}). 
Both metamodels have a root element denoting the schema, \texttt{RelationalSchema}
for the relational model and \texttt{USchema} for the \uschema model.
In what follows, such root elements denote the source and target schema.
Figure~\ref{fig:running-example-mapping} is used throughout this section 
to illustrate the application of the mapping rules. This figure omits some elements,
such as \texttt{MusicalStyle} entity, \texttt{mostRecentlyListened} association, and
some attributes, for the sake of clarity.

\ruletitle{RU1}{Schema Mapping}

A root \texttt{RelationalSchema} element $rS$ in a relational model maps to
a root \texttt{USchema} element $uS$ in a \uschema{} model. 
The mapping preserves the schema name and serves as the entry point for subsequent 
element-level correspondences.

{\fontsize{9}{10}\selectfont
\[
rS \rightarrow uS \;\|\; \{\, uS.name = rS.name \,\}
\]
}

\ruletitle{RU2}{Table Mapping}

Each \texttt{Table} $t$ in $rS$ is mapped to a \texttt{SchemaType} $st$ in $uS$, 
which is a root \texttt{EntityType} except when $\text{MN}(t)$ predicate is satisfied, 
%(i.e., $t$ represents a many-to-many join table), 
in which case $t$ corresponds to a \texttt{RelationshipType} as specified by Rule RU6.
This distinction captures the semantic difference between \textit{entity} tables and 
%\textit{many-to-many} join tables.
associative tables.

\vspace{-0.5cm}
{\fontsize{9}{10}\selectfont
\[
\begin{aligned}
&t \rightarrow st \;\|\;
\{\, 
st.name= t.name,\\ 
&\hspace{0.5cm}\neg \text{MN}(t)\Rightarrow \{\,\texttt{class}(st) = \texttt{EntityType},\\
&\hspace{2.3cm} st.root=\text{true}\;\},\\
&\hspace{0.5cm}\;\;\text{MN}(t)\Rightarrow \texttt{class}(st) = \texttt{RelationshipType}\,\} 
\end{aligned}
\]
}

The \texttt{class($x$)} function returns the metaclass of $x$.
The instance $st$ is added to $uS.entities$ if $\neg \text{MN}(t)$, 
and to $uS.relationships$ otherwise.

In the running example (Fig.~\ref{fig:running-example-mapping}),
tables \texttt{app\_user}, \texttt{album} and \texttt{song} are mapped to
entity types with $root$=true, and the associative table 
\texttt{listening} is mapped to a \texttt{RelationshipType}.

\ruletitle{RU3}{Column Mapping}

Let $c$ be a \texttt{Column} in a \texttt{Table} $t$ that is
\emph{not} part of a foreign key. Then, $c$ is mapped to an
\texttt{Attribute} $at$ and a \texttt{PrimitiveType} $pt$.
The attribute has the same name as the column, and its optionality
(optional or mandatory) is determined by the \textit{nullable}
property of the column. The primitive type corresponding to the
column datatype is assigned as the type of the attribute.

\vspace{-0.5cm}
{\fontsize{9}{10}\selectfont
\[
\begin{aligned}
c \rightarrow &(at,pt) \;\|\; \{\,\\
&at.name=c.name,\\
&at.optional=c.nullable, \\
&pt.name=\texttt{primitiveTypeMap}(c.datatype),\\
&at.type=pt\,\
\end{aligned}
\]
}

The \texttt{Attribute} instance $at$ is added to the collection
$st.features$, where $st$ is the schema type corresponding to
table $t$ (i.e., $st=\texttt{map}(t)$).

In our example, the rule applies to all columns except
\texttt{user\_id} in table \texttt{playlist},
\texttt{user\_id} and \texttt{playlist\_id} in table
\texttt{playlist\_song}, and \texttt{album} in table \texttt{song},
since these columns participate in foreign keys and are handled
through the corresponding relationship-mapping rules (RU5, RU6, RU7). 
\texttt{PrimitiveType} elements associated with the generated
\texttt{Attribute}s are not shown in Figure~\ref{fig:running-example-mapping}
to avoid overcomplicating the diagram.

\FloatBarrier

% \begin{figure*}[!t]
%   \centering
%   \includegraphics[width=\textwidth]{figures/schema-conversion-rel-uschema-doc.jpeg}
%   \caption{Transformation of the running example across the relational, 
%   \uschema{}, and document models.}
%   \label{fig:running-example-mapping}
% \end{figure*}

% =========================
% Relational → U-Schema (R1–R7)
% =========================

\ruletitle{RU4}{Key Mapping}

Let $ck$ be a \texttt{Key} (primary or alternative) of $t$, 
composed of a set of columns. Assume that $t$ is a table 
that does \emph{not} satisfy the MN($t$) predicate, 
and $t$ is associated with entity type $et$ by Rule~RU2, i.e., 
$et$ = \texttt{map}($t$).

Then, $ck$ maps to a \texttt{Key} $k$, which is added to
$et.features$. If $ck$ is the primary key of $t$, then $k$ is
marked as identifier. The attributes corresponding 
to the columns of $ck$ that are 
mapped by Rule~RU3 are included in $k.attributes$. 

\vspace{-0.5cm}
{\fontsize{9}{10}\selectfont
\[
\begin{aligned}
ck \rightarrow k \;\|\; \{&\,
k.name = ck.constraintName,\\
&k.isID = ck.isPK,\\
&\{\,\texttt{map}(c,\texttt{Attribute}) \mid
c \in ck.columns \, \land \\
&\qquad\qquad \nexists fk \in t.fks : c \in fk.columns \\
&\}
\subseteq k.attributes
\,\}
% &k.attributes =
% \{\jesus{\texttt{map}(c,\texttt{Attribute}) \mid
% c \in ck.columns,}\\
% \}
% \,\}
\end{aligned}
\]
}
%\mjose{Notación no conforme con el resto de reglas. }
%The \texttt{Key} instance $k$ is added to $et.features$.

In Figure~\ref{fig:running-example-mapping}, for table \texttt{playlist\_song}, 
for instance, the composite key \texttt{pls\_pk} is mapped to a \texttt{Key}. Only the attribute \texttt{position\_idx}, 
corresponding to the key column handled by Rule RU3, is included in its
\texttt{attributes} collection. The remaining primary-key columns 
participate in foreign keys and are handled by the corresponding 
relationship-mapping rules.

% In Fig~\ref{fig:running-example-mapping}, for table \texttt{playlist\_song}, 
% for instance, the composite key \texttt{pls\_pk} is mapped to a \texttt{Key} 
% in the \uschema{} model.
% Only the attribute \texttt{position\_idx}, 
% \jesus{corresponding to the key column handled
% by Rule RU3, belongs to its \texttt{attributes} collection.}
%Only the attribute \texttt{position\_idx} is added to its attributes collection.

% \mjose{LO CAMBIARIA POR ESTO: The primary keys for tables \texttt{song} and \texttt{app\_user}, and the alternative key in \texttt{app\_user} are mapped to keys in the USchema model. For tables \texttt{playlist} and \texttt{playlist\_song}, only one attribute is included in the corresponding key.}

% \mjose{The primary keys of tables \texttt{song} and \texttt{app\_user}, 
% as well as the candidate key of \texttt{app\_user}, 
% are mapped to \texttt{Key}s in the \uschema{} model. For tables \texttt{playlist} and \texttt{playlist\_song}, only one attribute is mapped to the corresponding key.}

\ruletitle{RU5}{Aggregable Weak Table Mapping}

Let $w$ be a \texttt{Table} satisfying $\mathrm{Aggregable}(w)$, 
let $fk$ be its identifying foreign key satisfying $\mathrm{FkInPk}(w,fk)$, 
and let $s$ be the table referenced by $fk$. 
%\jesus{Let \texttt{pk}(w) be the primary key of $w$}
%$p_k$ be the primary key of $w$, i.e., $p_k \in w.keys$ and $p_k.isPK = true$.
Assume that $w$ and $s$ are mapped by Rule~RU2 to entity types $ew$ and $es$,
respectively, i.e., $ew=\texttt{map}(w)$ and $es=\texttt{map}(s)$.

% This rule applies when $w$ \mjose{holds the condition Aggregable($w$).}
% \sout{is not referenced by any non-weak table according to the convention for 
% foreign key introduced in Section~\ref{sec:conventions}.
% If $w$ is referenced by a non-weak table, }
% \mjose{Otherwise, $w$ remains mapped to a root entity type and 
% its identifying foreign key $fk$ is mapped by the applicable variant of Rule~RU7.}

Then, \textit{i)} $ew$ is marked as non-root, and \textit{ii)}
the strong--weak relationship is represented by an
\texttt{Aggregate} $ag$ belonging to $es$ and specified by $ew$.
The upper bound of $ag$ depends on whether the primary key of
$w$ contains additional columns beyond those of $fk$. If both
sets of columns coincide, each instance of $es$ may aggregate
at most one instance of $ew$; otherwise, the additional
primary key columns act as a partial key, allowing multiple
instances of $ew$.
% \mjose{To express how many instances of $ew$ may be aggregated in each $es$ is necessary to check if the primary key of $w$ contains a partial key, that is, whether the columns of $fk$ are the same as the columns of the primary key of $w$.}

\vspace{-0.5cm}
{\fontsize{9}{10}\selectfont
\[
\begin{aligned}
%w \rightarrow &(ew, ag) \;\|\;\{\\
w \rightarrow \,&ag \;\|\;\{\\
& ew.root = \text{false},\\
& ag.name = \texttt{plural}(ew.name),\\
& ag.lowerBound = 0,\\
& fk.columns == \texttt{pk}(w).columns \\
    &\qquad\qquad\qquad \Rightarrow ag.upperBound = 1,\\
& fk.columns \subsetneq \texttt{pk}(w).columns\\
    &\qquad\qquad\qquad\Rightarrow ag.upperBound = n,\\
& ag.specifiedBy = ew
\;\}
\end{aligned}
\]
}

The function \texttt{plural} returns the plural form of its string argument.
\texttt{Aggregate} instance $ag$ is added to the collection $es$.features.

As shown in Fig.~\ref{fig:running-example-mapping}, weak tables such as
\texttt{playlist} and \texttt{playlist\_song} are mapped to entity types
with \textit{root}=false, and to aggregates. Each aggregate belongs to the
features of the corresponding root entity type, 
capturing the containment of weak entities, 
and is specified by the non-root entity type.

\ruletitle{RU6}{Associative Table Mapping} %{Many-to-Many Table Mapping}

As established in Rule~RU2, a \texttt{Table} $m$ that satisfies the $\mathrm{MN}(m)$
predicate is mapped to a \texttt{RelationshipType} $rm$, and by Rule~RU3, $rm$ contains the attributes corresponding to all columns of $m$, except for the foreign key components.
Assume that $m$ contains two foreign keys $fk_1$ and $fk_2$ referencing tables 
$t_1$ and $t_2$, respectively. And let $et_1$ and $et_2$ be their corresponding 
entity types. i.e. $et_1$ = \texttt{map}($t_1$) and $et_2$ = \texttt{map}($t_2$). 
Those foreign keys are mapped to two \texttt{Reference} $r_1$ and $r_2$,
which denote mutual references between $et_1$ and $et_2$. 
This representation explicitly captures the semantics of many-to-many 
relationships in \uschema{}.

\vspace{-0.5cm}
{\fontsize{9}{10}\selectfont
\[
\begin{aligned}
m \rightarrow \,&(rm, r_1, r_2) \; \| \;\{\\
& r_i.name = fk_i.constraintName, \\
& r_i.lowerBound = 1, \; r_i.upperBound = n, \\
& r_i.refsTo = \texttt{map}(fk_i.refsTo.owner),\\
& r_i.isFeaturedBy = rm
%& r_i.attributes = \{ at \mid at = map(c), \forall c \in fk_i.columns\}\\
\;\}:\forall i,\; i \in \{1,2\}\;
%& r_i \in rm.\mathrm{references},\\
\}
\end{aligned}
\]
}
\texttt{Reference} $r_1$ is added to the features of the entity type $et_2$ 
and $r_2$ is added to the features of the entity type $et_1$. 
Both references $r_1$ and $r_2$ are added to $rm.references$ collection.
The component columns of $fk_1$ and $fk_2$ are not mapped to
attributes according to Rule~RU3, as they are not needed in the
target entity types $et_1$ and $et_2$, since references in
\uschema{} are sufficient to connect them.

In the running example, the associative table \texttt{listening} is mapped
to a \texttt{RelationshipType} (by Rule~RU2). Rule~RU6 introduces two references,
\texttt{listening\_user} and \texttt{listening\_song},
within the features of entity types \texttt{song} and \texttt{app\_user} respectively. 
These references are associated with the relationship type through 
\texttt{isFeaturedBy} and point to the corresponding entity types via \texttt{refsTo}.

\ruletitle{RU7}{Foreign Key Mapping}

This rule defines how foreign keys not handled by
Rules~RU5 or RU6 are mapped to \texttt{Reference} instances
between the corresponding entity types.

For an aggregable table, Rule~RU7 handles only foreign keys
that do not participate in its primary key, since its identifying
foreign key is handled by Rule~RU5. For a non-aggregable table
with a primary key, the rule also handles foreign keys that
participate in its primary key. Tables without a primary key
require a separate variant, since the reverse representation used
for non-aggregable tables cannot be constructed. Foreign keys
defining an associative table are handled by Rule~RU6.
Foreign keys targeting associative tables are not supported by
the current mapping.

Four variants are distinguished:
\textit{(i)} a unique foreign key;
\textit{(ii)} a non-unique foreign key in an aggregable table;
\textit{(iii)} a non-unique foreign key in a non-aggregable table with a primary key; and
\textit{(iv)} a non-unique foreign key in a table without a
primary key.

% The generated reference preserves the direction of $fk$ in
% the first two variants and reverses it in the third. In the first
% (1:1) and third (1:N) variants, an attribute-less opposite
% \texttt{Reference} with upper bound one is synthesized to
% represent the inverse multiplicity. The second (N:1) variant
% requires no opposite, since its target may be shared.

The generated reference preserves the direction of $fk$ in
the first, second, and fourth variants and reverses it in the third.
In the first (1:1) and third (1:N) variants, an attribute-less
opposite \texttt{Reference} with upper bound one is synthesized
to represent the inverse multiplicity. The second and fourth
(N:1) variants require no opposite, since their target may be
shared.

Let $s$ be the table referenced by $fk$, and let $et$ and
$es$ be the entity types mapped from $t$ and $s$,
respectively, i.e., $et$ = \texttt{map}($t$) and
$es$ = \texttt{map}($s$).

For Rules RU7.1, RU7.2, RU7.4, and also for Rule RU7.3 when 
FkInPk($t,fk$) holds, let $A_{fk}$ denote the set of \texttt{Attribute}
instances corresponding to the columns in $fk.columns$. Their names, 
optionality, and associated \texttt{PrimitiveType} instances satisfy 
the same mapping constraints as those specified in Rule~RU3 for column mapping.
%\[
%\begin{aligned}
%A_{fk} = \{\, at \mid
%\{\, &c \rightarrow \jesus{(at,pt)}
%     &at.name = c.name,\\
%     &at.optional = c.nullable,\\
%     &pt.name = \texttt{primitiveTypeMap}(c.datatype),\\
%     &at.type = pt
%\,\}: \forall c \in fk.columns
%\}
%\end{aligned}
%\]

% \[
% \begin{aligned}
% A_{fk}=\{\, at \mid
% \{\, &c \rightarrow at,\\
%     &at.name = c.name,\\
%     &at.type = \texttt{typeMap}(c.datatype),\\
%     &at.optional = c.nullable
% \,\}:\forall c \in fk.columns
% \,\}.
% \end{aligned}
% \]

%\jesus{where $c \rightarrow (at,pt)$ is used as a correspondence clause,
%indicating that an \texttt{Attribute} $at$ and its associated
%\texttt{PrimitiveType} $pt$ are generated from each \texttt{Column}
%$c$; the remaining clauses specify their properties.} \mjose{¿Hace falta esta explicación?}

% where $c \rightarrow at$ is used as a
% correspondence clause, indicating that a new \texttt{Attribute} $at$
% is created from each \texttt{Column} $c$; the remaining clauses specify
% the properties of $at$.
The four variants of the rule are defined below.

\vspace{1ex}
\textbf{RU7.1}: \emph{Foreign Key Is Unique.}

This variant applies when the columns of $fk$ are subject to a
uniqueness constraint. This also includes the identifying foreign
key of a non-aggregable weak table when its columns constitute the
entire primary key of the table.

The foreign key $fk$ is mapped to a single-valued
\texttt{Reference} $r_s$ in $et$ whose target is $es$.
Since the columns of $fk$ are subject to a uniqueness constraint,
the relationship has a maximum 1:1 cardinality. Therefore, an
attribute-less opposite \texttt{Reference} $r_i$ with upper bound
one is synthesized in $es$ to record that an instance of $es$
cannot be referenced by more than one instance of $et$ through
this relationship. 
%to record that the target of $r_s$ is not shareable.

Let $uk$ be the unique \texttt{Key} in $t$, possibly its primary
key, such that $uk.columns == fk.columns$.

\vspace{-0.5cm}
{\fontsize{9}{10}\selectfont
\[
\begin{aligned}
&fk \rightarrow (r_s,r_i)\;\|\;\{\\[-0.2ex]
&\quad r_s.name = fk.constraintName,\\
&\quad r_s.refsTo = es,\\
%&\quad r_s.lowerBound = 0,\; r_s.upperBound = 1,\\
&\quad\neg\mathrm{FkInPk}(t,fk)
   \Rightarrow r_s.lowerBound = 0,\\
&\quad\mathrm{FkInPk}(t,fk)
   \Rightarrow r_s.lowerBound = 1,\\
&\quad r_s.upperBound = 1,\\
&\quad r_s.attributes = A_{fk},\\
&\quad \texttt{map}(uk).attributes =\\
&\qquad \texttt{map}(uk).attributes
          \cup r_s.attributes,\\
&\quad r_i.name =
  \texttt{concat}(fk.constraintName,
                  \text{`\_opposite'}),\\
&\quad r_i.refsTo = et,\\
&\quad r_i.lowerBound = 0,\; r_i.upperBound = 1,\\
&\quad r_i.attributes = \emptyset,\\
&\quad r_i.opposite = r_s,\; r_s.opposite = r_i
\;\}
\end{aligned}
\]
}

The attributes in $A_{fk}$ are created by this rule because
Rule~RU3 excludes columns that participate in foreign keys.
The \texttt{Reference} $r_s$ and the \texttt{Attribute} instances
in $r_s.attributes$ are added to $et.features$. These attributes
are also added to \texttt{map}$(uk).attributes$, thereby preserving
the uniqueness constraint. The synthesized opposite reference $r_i$
is added to $es.features$.

% \texttt{FKey} $fk$ is mapped to a 1:1 \texttt{Reference} $r_s$ in $et$, which references to $es$. 
% The reference $r_s$ is associated with new attributes corresponding to the columns of the 
% primary key of $s$.
%%% \hilight{MJ. explicación: para que se pueda activar la RU3, los atributos deben generarse a partir de columnas que no sean componentes de foreign keys}

% \[
% \begin{aligned}
% fk \rightarrow r_s \;\|\;
% \{\, 
% %&\text{FkNotPk}(t,s,fk) \wedge \\
% %&(\exists k \in t.\text{keys} \mid fk.\text{columns} = k.\text{columns}), \\ %\text{unique}(fk),\\
% &r_s.name = s.\text{name}, \\
% &r_s.refsTo=es,\\
% &r_s.lowerBound=0,\; r_s.upperBound=1, \\
% &r_s.attributes = \{at \mid c \rightarrow at, \forall c \in \texttt{pk} (s).columns
% %&\hspace{2.2cm} \text{where} \; pk_s \in s.keys \wedge pk_s.isPK 
% \} \; \}
% \end{aligned}
% \]

% Since $fk$ is unique, there exists a \texttt{Key} $uk$ in $t$ whose columns are the same as 
% the columns of $fk$, and therefore each $at$ is also added to the collection \texttt{map}($uk$).$attributes$.
% Here, the correspondence operator $\rightarrow$ also appears within a clause.

%The weak table \texttt{playlist\_song} provides an example of this rule,
%as the foreign key \texttt{pls\_song} does not participate in the primary key
%and is therefore mapped to a reference in the corresponding entity type.

\vspace{1ex}
\ruletitle{RU7.2}{\emph{Non-Unique Foreign Key in an Aggregable Table}}

This variant applies when $fk$ is a non-unique foreign key
owned by a table $t$ satisfying $\mathrm{Aggregable}(t)$, and
$\neg\mathrm{FkInPk}(t,fk)$ holds. The identifying foreign key
of an aggregable table is handled by Rule~RU5.

The foreign key $fk$ is mapped to a single-valued
\texttt{Reference} $r_s$ in $et$ whose target is $es$,
preserving the direction of the foreign key. The reference
represents an N:1 relationship, since multiple instances of
$et$ may reference the same instance of $es$. Therefore, no
opposite reference is synthesized, since an instance of $es$
may be referenced by multiple instances of $et$.

As in Rule~RU7.1, the columns of $fk$ are mapped to the
attributes associated with $r_s$, since Rule~RU3 excludes
columns that participate in foreign keys.

\vspace{-0.5cm}
{\fontsize{9}{10}\selectfont
\[
\begin{aligned}
fk \rightarrow r_s\;\|\;\{\,%\\[-0.2ex]
&r_s.name = fk.constraintName,\\
&r_s.refsTo = es,\\
&r_s.lowerBound = 0,\\
&r_s.upperBound = 1,\\
&r_s.attributes = A_{fk}
\,\}
\end{aligned}
\]
}

The \texttt{Reference} $r_s$ and the \texttt{Attribute}
instances in $r_s.attributes$ are added to $et.features$.

In the running example, the non-unique foreign key
\texttt{pls\_song} in the aggregable weak table
\texttt{playlist\_song} is mapped to a single-valued reference
in the entity type \texttt{playlist\_song} whose target is
\texttt{song}. Since the same song may occur in multiple
playlists, an instance of \texttt{song} may be referenced by
multiple instances of \texttt{playlist\_song}; therefore, no
opposite reference is synthesized.

\ruletitle{RU7.3}{Non-Unique Foreign Key in a Non-Aggregable Table
with a Primary Key}

This variant applies when $fk$ is a non-unique foreign key
owned by a table $t$ satisfying $\neg\mathrm{Aggregable}(t)$
and $\mathrm{HasPK}(t)$, and $fk$ is not handled by Rule~RU6.
This includes both non-weak tables with a primary key and weak
tables that are non-aggregable because they are referenced by a
non-weak table.
% This variant applies when $fk$ is a non-unique foreign key
% owned by a table $t$ satisfying
% $\neg\mathrm{Aggregable}(t)$, and $fk$ is not handled by Rule~RU6. 
% This includes both non-weak tables and weak tables
% that are non-aggregable because they are referenced by a
% non-weak table.

The foreign key $fk$ is mapped to a multi-valued
\texttt{Reference} $r_t$ in $es$ whose target is $et$, thus
reversing the direction of the foreign key and representing
the relationship as 1:N. Since each instance of $et$ may be
associated through $fk$ with at most one instance of $es$, an
attribute-less opposite \texttt{Reference} $r_i$ with upper
bound one is synthesized in $et$. Its lower bound is one when
$fk$ participates in the primary key of $t$, and zero
otherwise.

Let $kt=\texttt{map}(\texttt{pk}(t))$ be the \texttt{Key}
associated with the primary key of $t$. Rule~RU7.3 may process
either a foreign key that does not participate in the primary
key or the identifying foreign key of a non-aggregable weak
table.

When $\mathrm{FkInPk}(t,fk)$ holds, the attributes in
$A_{fk}$ must contribute to both $et.features$ and
$kt.attributes$, since Rule~RU3 does not generate attributes
for foreign key columns. 

\vspace{-0.5cm}
{\fontsize{9}{10}\selectfont
\[
\begin{aligned}
\mathrm{FkInPk}(t,fk)
       \Rightarrow &\{\,
             et.features = et.features \cup A_{fk} \, ,\\
&\, kt.attributes = kt.attributes \cup A_{fk}\,\}\\ 
\end{aligned}
\]
}

\vspace{-0.5cm}
Furthermore, since $r_t$ targets $et$, its associated attributes must identify instances of $et$. Therefore, $A_{r_t}$ is derived from the complete set of primary key attributes in
$kt.attributes$, including those contributed by foreign keys participating
in that key. 
For each attribute $at_k \in k_t.attributes$, a fresh attribute
$at_r$ and an associated \texttt{PrimitiveType} $pt_r$ are
created and associated with $r_t$.

Let $A_{r_t}$ denote the set of these newly created attributes:

\vspace{-0.5cm}
{\fontsize{9}{10}\selectfont
\[
\begin{aligned}
A_{r_t} &= \{\, at_r \,\mid \\ 
&\{\, at_r.name = 
    \texttt{concat}(at_k.name, \text{`\_'}, et.name),\\
   &\;\; at_r.optional = at_k.optional,\\
   &\;\; pt_r.name = at_k.type.name,\\
   &\;\; at_r.type = pt_r,
\,\}: \forall at_k \in k_t.attributes
\}
\end{aligned}
\]
}

Rule~RU7.3 is formalized as follows: 

\vspace{-0.5cm}
{\fontsize{9}{10}\selectfont
\[
\begin{aligned}
&fk \rightarrow (r_t,r_i) \;\|\;\{\\
&\quad r_t.name = \texttt{plural}(t.name),\\
&\quad r_t.refsTo = et,\\
&\quad r_t.lowerBound = 0,\;
       r_t.upperBound = n,\\
&\quad r_t.attributes = A_{r_t},\\
&\quad r_i.name =
       \texttt{concat}(r_t.name,\text{`\_opposite'}),\\
&\quad r_i.refsTo = es,\\
&\quad \mathrm{FkInPk}(t,fk)
       \Rightarrow r_i.lowerBound = 1,\\
&\quad \neg\mathrm{FkInPk}(t,fk)
       \Rightarrow r_i.lowerBound = 0,\\
&\quad r_i.upperBound = 1,\\
&\quad r_i.attributes = \emptyset,\\
&\quad r_i.opposite = r_t,\;
       r_t.opposite = r_i
\;\}
\end{aligned}
\]
}

The resulting \texttt{Reference} $r_t$ and the attributes in
$A_{r_t}$ are added to $es.features$, whereas the synthesized
opposite reference $r_i$ is added to $et.features$.

In the running example, the non-unique foreign key
\texttt{album\_fk} in the non-aggregable table
\texttt{song} is mapped to the multi-valued reference
\texttt{songs} in the entity type \texttt{album}, whose
target is \texttt{song}. The attributes associated with this
reference are fresh local copies of the attributes that
identify \texttt{song}. An attribute-less opposite reference
in \texttt{song} records that each song may belong to at most
one album.

%NUEVA ALTERNATIVA RU7.4

\ruletitle{RU7.4}{Non-Unique Foreign Key in a Table Without a Primary Key}

This variant applies when $fk$ is a non-unique foreign key
owned by a table $t$ satisfying $\neg\mathrm{HasPK}(t)$.
Since $t$ has no primary key, the reverse representation used
in Rule~RU7.3 cannot be constructed. Therefore, the foreign key
direction is preserved and the same mapping constraints as in
Rule RU7.2 apply.

The running example does not contain tables without a primary key.
However, this situation occurs in the TPC-C evaluation case
(Section~\ref{sec:experimental-setup}), 
where table \texttt{history} has no primary key
and its foreign keys are therefore handled by this variant.

\subsection{U-Schema to Document Mapping}
\label{subsec:uschema-to-document}

This section defines the mapping from \uschema{} to the Document metamodel.
The auxiliary functions introduced in
Section~\ref{sec:mapping-rule-notation} are used with the same meaning
as in the relational to \uschema{} mapping.
Figure~\ref{fig:running-example-mapping} is also used to illustrate the mappings.

\ruletitle{UD1} {Schema Mapping}
A \texttt{USchema} $uS$ is mapped to a \texttt{DocumentSchema} $dS$ preserving its name.

{\fontsize{9}{10}\selectfont
\[
uS \rightarrow dS \;\|\; \{\, dS.name = uS.name \,\}
\]
}

The elements of $dS$ (i.e., its \textit{documents} and \textit{types}) are
defined by the following rules. 
%\jesus{Initially, $dS.documents$ is empty, i.e., $dS.documents = \emptyset$.}
%\mjose{¿hace falta decirlo? en el mapping anterior no se hace. Creo que se puede suponer sin problema. no?}

\ruletitle{UD2}{Root Entity Type Mapping}
%Each root \texttt{EntityType} $e$ 
Each \texttt{EntityType} $e$ in $uS$ that satisfies \textit{e.root}==true
is mapped to a \texttt{DocumentType} $d$ with the same name.

{\fontsize{9}{10}\selectfont
\[
e \rightarrow d \;\|\; 
\{\, %e.\text{root} = \mathrm{true},\; 
d.name = e.name\;
%d \in dS.types,\;
%d \in dS.\text{documents} 
\}
\]
}

$d$ is added to the $dS.documents$ collection, where $dS$
is the document schema obtained by Rule~UD1, i.e.,
$dS = \texttt{map}(uS)$.
The features of $e$ (i.e., \textit{e.features}) will be mapped to 
\texttt{Property} elements as established in rules UD3 to UD6, 
and will be added to \textit{d.properties}.

In the running example, the entity types \texttt{song}, \texttt{album}, 
and \texttt{app\_user} are mapped to \texttt{DocumentType} instances,
whereas those with \textit{root}=false are handled by rule UD6.

\ruletitle{UD3}{Attribute Mapping}

Let $at$ be an \texttt{Attribute} that belongs to a schema type $st$
(either an entity type or a relationship type) and is \emph{not} part
of a reference (i.e., $at.references$ is \texttt{NULL}).
Then, $at$ is mapped to a \texttt{Field} $f$ preserving its name and
with the corresponding type.

{\fontsize{9}{10}\selectfont
\[
\begin{aligned}
at \rightarrow f \;\|\;\{\, 
%\text{empty}(at.\text{references}),\; \\ %\land \\
%&\hspace{0.8cm}(\text{empty}(at.\text{key}) \lor \neg at.\text{key.isID}), \\
&f.name = at.name,\;\\
&f.type = \texttt{map}(at.type)
\,\}
\end{aligned}
\]
}

As shown in Figure~\ref{fig:running-example-mapping}, all attributes are
mapped to fields, with the exception of \texttt{song\_id} in entity type 
\texttt{playlist\_song}, and \texttt{song\_id\_song} in entity type \texttt{album},
which participate in references \texttt{song} and \texttt{songs} respectively.

\ruletitle{UD4}{Key Mapping}

Both simple and composite keys are mapped to a single identifier field.
Let $k$ be a \texttt{Key} in an entity type $et$ such that $k.isID$ is true.
If $k$ consists of a single attribute $at$, then the
corresponding field $f$=\texttt{map}($at$) is marked as a key field.
Otherwise, $k$ is mapped to a \emph{new} \texttt{Field} $f$ also marked as a key.
{This field acts as a single identifier derived from the values of
the attributes composing the key.
The attribute components of $k$ are mapped to their corresponding
fields according to Rule~UD3 and are treated as regular fields
in the target schema.

\vspace{-0.5cm}
{\fontsize{9}{10}\selectfont
\[
\begin{aligned}
k \rightarrow f\;\|\, \{& \mid k.attributes \mid \,== 1 \Rightarrow \{\,f=\texttt{map}(at), \\
&\hspace{3.5cm} at \in k.attributes\,\}, \\
&\mid k.attributes \mid \,> 1 \Rightarrow f.name = \text{`\_id'},\\
%& f.name = \texttt{concat}(\text{map}(e).name,\text{`\_id'}), \\
&\, f.isKey = \text{true}
\,\}
\end{aligned}
\]
}

When a composite key is mapped to a derived identifier field,
the generated field has a \texttt{PrimitiveType} with datatype
\texttt{STRING}.

In our example, fields corresponding to attributes belonging to keys with \textit{isID}==true 
are marked with \textit{isKey}=true in the document model. The remaining key, 
\texttt{user\_name\_ak}, which enforces the uniqueness of the \texttt{name} 
attribute, is not preserved, as each document type can only contain a single identifier.

\ruletitle{UD5}{Reference Mapping}

This rule maps a U-Schema \texttt{Reference} $r_U$ representing a
direct link between two \texttt{EntityType} instances to a document
\texttt{Reference} $r_D$. The mapping preserves whether the
reference is single- or multi-valued and whether the same target
instance may be referenced by multiple source instances.

References associated with a \texttt{RelationshipType}, i.e.,
those for which $r_U.isFeaturedBy$ is not null, are mapped
together with their corresponding relationship type by Rule~UD7.
As explained in Section~\ref{sec:metamodels}, synthesized opposite
references in \uschema{} are auxiliary elements used to represent
inverse multiplicity. They are not mapped independently, but are
inspected to determine the target shareability of their corresponding
direct references. In these mappings, a synthesized reference is
identified by the following predicate:

\vspace{-0.5cm}
{\fontsize{9}{10}\selectfont
\[
\begin{aligned}
\mathrm{SynthesizedReference}(r) \equiv {}
&r.attributes=\emptyset \;\land\\
&r.opposite\neq\texttt{NULL}\;\land\\
&r.opposite.attributes\neq\emptyset
\end{aligned}
\]
}

The mapping applies to $r_U$ provided that

{\fontsize{9}{10}\selectfont
\[
\begin{aligned}
(r_U.isFeaturedBy ==  \texttt{NULL}) \, \land  \\
\neg\mathrm{SynthesizedReference}(r_U)
\end{aligned}
\]
}

\vspace{0.05cm}
The attributes associated with $r_U$ are not mapped to separate
\texttt{Field} elements. Their identifying role is instead
represented by the target identifier stored in the document
reference, following the single-field identifier convention
introduced in Section~\ref{sec:metamodels}.
The direct upper multiplicity is represented through the type of the
document reference: a single-valued reference is represented by a
\texttt{PrimitiveType}, whereas a multi-valued reference is
represented by an \texttt{Array}. The inverse upper multiplicity is
used to determine the value of \texttt{targetShareability}.
%\jesus{Let $k_t$ be the identifier \texttt{Key} of $r_U.refsTo$,
%and let $f_{id}=\texttt{map}(k_t)$ be the corresponding
%identifier \texttt{Field} in the target document type.}

\vspace{-0.5cm}
{\fontsize{9}{10}\selectfont
\[
\begin{aligned}
 r_U & \rightarrow r_D\;\|\;\{\\
& r_D.name = r_U.name,\\
& r_D.target = \texttt{map}(r_U.refsTo),\\
& r_U.upperBound=1 \\
& \quad \Rightarrow \texttt{class}(r_D.type)=\texttt{PrimitiveType},\\
& r_U.upperBound>1 %\\
%& \quad 
  \Rightarrow \texttt{class}(r_D.type)=\texttt{Array},\\
& r_U.opposite==\texttt{NULL}\\
& \quad \Rightarrow r_D.targetShareability=\text{true},\\
& r_U.opposite\neq\texttt{NULL} \\
& \quad \Rightarrow r_D.targetShareability=\\
& \qquad \qquad \qquad (r_U.opposite.upperBound>1)
\;\}
\end{aligned}
\]
}

The primitive datatype of the reference $r_D$ is the datatype 
of the identifier field of its document type target. 
For a multi-valued reference, this datatype is used as the element
type of the generated \texttt{Array}.

The generated reference $r_D$ is added to the property collection
of the document or embedded structure corresponding to
$r_U.owner$.

In the running example, the references \texttt{pls\_song} in entity type
\texttt{playlist\_song} and \texttt{songs} in entity type
\texttt{album} satisfy the conditions of this rule.
The single-valued reference \texttt{pls\_song} has no synthesized
opposite reference. It is therefore mapped to a document reference
of type \texttt{PrimitiveType} with \textit{targetShareability}==true, 
indicating that the same song may be referenced by multiple playlist entries.

The multi-valued reference \texttt{songs} has a synthesized opposite
reference with upper bound one. It is consequently mapped to a
document reference of type \texttt{Array} with
\textit{targetShareability}=false, indicating that each song may
belong to at most one album.

References \texttt{listening\_user} and
\texttt{listening\_song} are associated with a
\texttt{RelationshipType} and are therefore handled by Rule~UD7.

\ruletitle{UD6} {Aggregate (and non-root Entity Type) Mapping}

Aggregations are mapped as embedded documents. %, as follows.
Let $ag$ be an \texttt{Aggregate} that belongs to an entity type $e$
and is specified by a non\mbox{-}root \texttt{EntityType} $e_{nr}$ 
(i.e. $ag.specifiedBy = e_{nr}$).
Then $ag$ and $e_{nr}$ are mapped to an \texttt{Embedded} element $em$. 
The features in $e_{nr}$ are then recursively mapped to properties added to $em.aggregates$ 
%and to $d.properties$ 
by applying the corresponding rules UD3 to UD6.

\vspace{-0.5cm}
{\fontsize{9}{10}\selectfont
\[
\begin{aligned}
%(ag, e_{nr}) \rightarrow \,& em \;\|\; 
ag &\rightarrow \, em \;\|\; 
\{\,\\
& em.name = ag.name,\\
% &em.aggregates \leftarrow e_{nr}.features,\\
% &em.aggregates = \text{map}(e_{nr}.features),\\
% &em.aggregates = \{e\;|\;e = \text{map}(x),\forall x \in e_{nr}.features\},\\
&em.aggregates = \{p\;|\;x \rightarrow p,\forall x \in e_{nr}.features\},\\
&em.isMany = \neg\; (ag.upperBound ==1)
\,\}
\end{aligned}
\]
}

In the \uschema{} model, the entity type \texttt{app\_user} aggregates 
\texttt{playlist} which, in turn, aggregates \texttt{playlist\_songs}. 
In the document model, these aggregates are mapped to embedded objects to preserve this nesting relationship.   

\ruletitle{UD7}{RelationshipType Mapping}

Each \texttt{RelationshipType} $rt$ is mapped to an independent
\texttt{DocumentType} $d$. The attributes owned by $rt$ are mapped
to \texttt{Field} elements by Rule~UD3 and added to $d.properties$.
In addition, a \texttt{Field} is generated to serve as the
identifier of $d$.

The references in $rt.references$ are not mapped by Rule~UD5.
Although each such reference is owned by one of the participating
entity types, its association with $rt$ indicates that it represents
a participant in the relationship. Therefore, each reference is
mapped by this rule to a document \texttt{Reference} contained in
$d$.

Under the relationship representation adopted in this mapping, the
same participating entity instance may occur in multiple relationship
instances. Therefore, the generated document references have
\textit{targetShareability}=true.

Let

\vspace{-0.5cm}
{\fontsize{9}{10}\selectfont
\[
P_{rt} =
\{\,p \mid x \rightarrow p:\forall x\in rt.features\,\}
\]
}

denote the properties generated from the features of $rt$.

\vspace{-0.5cm}
{\fontsize{9}{10}\selectfont
%\hilight{FORMULA CAMBIADA PARA EXPRESAR EL CONJUNTO DE REFERENCIAS CREADAS}
\[
\begin{aligned}
rt \rightarrow 
\bigl(d,f,R_D & \bigr)\;\|\;\{\\
&\quad d.name = rt.name,\\
&\quad f.name =
       \texttt{concat}(rt.name,\texttt{`\_id'}),\\
&\quad \texttt{class}(f.type)=
       \texttt{PrimitiveType},\\
&\quad f.type.datatype=\texttt{STRING},\\
&\quad f.isKey=\texttt{true},\\
&\quad \{\,r_i \rightarrow r_D^i,\\
    &\qquad r_D^i.name = r_i.name,\\
    &\qquad r_D^i.target =
           \texttt{map}(r_i.refsTo),\\
    &\qquad \texttt{class}(r_D^i.type)=
           \texttt{PrimitiveType},\\
    &\qquad r_D^i.targetShareability =
           \text{true} \\
&\quad \}:\forall r_i\in rt.references,\\
&\quad R_D  = \{\,r_D^i\,\},\\
%&\quad R_D =
       %\{\,r_D^i\mid r_i\in rt.references\,\},\\
%&\quad \mjose{\text{sustituir la anterior por:} R_D = \{\,r_D^i\,\},}\\
&\quad d.properties =
       P_{rt}\cup\{\,f\,\}\cup R_D
\;\}
\end{aligned}
\]
}

% \[
% \begin{aligned}
% &rt \rightarrow
% \bigl(d,f,\{r_D^i\}\bigr)\;\|\;\{\\
% &\quad d.name = rt.name,\\
% &\quad f.name =
%        \texttt{concat}(rt.name,\texttt{`\_id'}),\\
% &\quad \mathrm{class}(f.type)=
%        \texttt{PrimitiveType},\\
% &\quad f.type.datatype=\texttt{STRING},\\
% &\quad f.isKey=\texttt{true},\\
% &\quad \{\, r_i \rightarrow r_D^i,\\
% &\qquad r_D^i.name = r_i.name,\\
% &\qquad r_D^i.target =
%        \texttt{map}(r_i.refsTo),\\
% &\qquad \mathrm{class}(r_D^i.type)=
%        \texttt{PrimitiveType},\\
% &\qquad r_D^i.targetShareability=
%        \texttt{true}
%        \,\}:\forall r_i\in rt.references,\\
% &\quad d.properties =
%        P_{rt}\cup\{\,f\,\}\cup%\\&\qquad
%        \{\,r_D^i\mid r_i\in rt.references\,\}
%        \\&\quad \mjose{ \{\,r_D^i ++++++++++++++\}}
% \;\}
% \end{aligned}
% \]
The primitive datatype of each generated reference is the datatype
of the identifier field of its target document type.

The generated document type $d$ is added to $dS.documents$, where
$dS=\texttt{map}(rt.owner)$.

In the running example, the relationship type
\texttt{listening} is mapped to a document type with the same name.
The relationship attributes \texttt{plays\_count} and
\texttt{status} are mapped to regular fields, and the fresh field
\texttt{listening\_id} acts as its identifier. Its references to
\texttt{app\_user} and \texttt{song} are mapped to single-valued
document references with \texttt{targetShareability}=true, since the
same user or song may participate in multiple listening instances.

\ruletitle{UD8}{Type Mapping}

The current transformation supports the \uschema{} data types that have
a direct representation in the document model.
A \texttt{PrimitiveType} in \uschema{} is mapped to a \texttt{PrimitiveType}, 
whose \texttt{datatype} is obtained by applying \texttt{primitiveTypeMap}
to the \uschema{} type name.
A \texttt{PList} or \texttt{PSet} is mapped to an \texttt{Array}, whose
element \texttt{type} is the \texttt{type} corresponding to the element type
of the \uschema{} collection.
The distinction between list and set semantics is not preserved
in the document model.
% A \texttt{PMap} is also represented as a document
% \texttt{Array} whose element type corresponds to the map
% value type; the key type is not represented by the current
% document metamodel.

\section{Evaluation}
\label{sec:validation}

The empirical evaluation focuses on the relational-to-document
migration scenario addressed in this work. Schema and data are
migrated for three cases: the Music Streaming case
study, the TPC-C benchmark, and the Northwind sample database.
Together, their source schemas exhibit a broad range of the
relational patterns addressed by the mappings.

% The evaluation examines three complementary dimensions:
% \textit{(i)} validation of the generated schema models against expected
% models, \textit{(ii)} preservation of schema information and data
% semantics, and \textit{(iii)} operational feasibility and scaling behavior
% of the complete migration process.
% For the third dimension, three database instances of increasing size,
% denoted by \setS{}, \setM{}, and \setL{},
% were generated for both Music Streaming and TPC-C.

The evaluation is guided by the following research questions:

\begin{description}[
  leftmargin=1.2em,
  style=nextline,
  font=\normalfont\itshape
]

\item[\textit{RQ1. Schema-transformation chain validation.}]
Do the implemented schema transformations generate the expected
schema models for the evaluated cases, including when the intermediate
\uschema{} model is adapted?

\item[\textit{RQ2. Schema-information preservation.}]
To what extent does the migration preserve the schema information
required to represent the source database and to reconstruct its
relational structure through the Document-to-\uschema{} and
\uschema{}-to-Relational mappings?

\item[\textit{RQ3. Data-semantic preservation.}]
Do queries executed over the migrated document database produce
results equivalent to those obtained from the source relational
database?

\item[\textit{RQ4. Operational feasibility and scaling behavior.}]
How does the complete migration process behave as data volume
increases, and what additional execution cost does schema-driven
semantic transformation introduce with respect to a flat data copy?

\end{description}

\paragraph{Evaluation methodology}
To address the research questions, the evaluation combines four
complementary groups of activities. 
RQ1 is assessed through incremental
testing of the M2M transformations, complementary tests using schemas
that combine multiple mapping patterns, comparison of generated target
schemas with expected models, and evaluation of trace-aware schema adaptation.
RQ2 is addressed through round-trip schema reconstruction.
RQ3 is assessed by comparing the results of corresponding SQL and
MongoDB queries. Finally, RQ4 is addressed through end-to-end execution
measurements over increasing data volumes and a flat-copy baseline.
For both Music Streaming and TPC-C, three database instances of
increasing size, denoted by \setS{}, \setM{}, and \setL{}, were used
for the scaling experiments.

% \mjose{Discutir si en esta sección se podría quitar algo de redundancia. 
% Por ejemplo, aquí (en el párrafo siguiente) se presenta *de forma extensa* todo lo que se verá en esta sección.
% En la siguiente (Methodology), se vuelve a explicar. 
% Y en las siguientes (6.3 a 6.6.) se detalla, a veces repitiendo lo ya contado (dos veces) previamente.
% Quizá el párrafo que sigue se pueda eliminar, puesto que prácticamente se repite en el 6.1 Methodology. O fusionar con él. Y ese 6.1 debería reducirse al mínimo, pues todo está muy detallado en 6.3 y siguientes.}

% The M2M transformations are evaluated through incremental rule-by-rule 
% testing and complementary tests using schemas that combine multiple 
% mapping patterns.
% Schema comparison and round-trip schema reconstruction, in which a
% relational schema is transformed to the document model and subsequently
% reconstructed as a relational schema, provide evidence concerning 
% schema-information preservation, whereas query-result equivalence is 
% used to assess data-semantic preservation. Finally, end-to-end execution
% measurements and a flat-copy baseline are used to characterize
% operational feasibility and scaling behavior.

% Finally, end-to-end execution measurements, a flat-copy baseline, and a batch-size sensitivity
% analysis are used to evaluate operational feasibility and scaling
% behavior.

The remainder of this section presents 
%the evaluation methodology (Section~\ref{sec:validation-objectives}), 
the experimental setup
(Section~\ref{sec:experimental-setup}), and the results concerning
incremental development and testing of the M2M transformations
(Section~\ref{sec:transformation-testing}), schema-information preservation
(Section~\ref{sec:schema-validation}), operational feasibility and scaling behavior
(Section~\ref{sec:data-migration-performance}), and data-semantic preservation
(Section~\ref{sec:semantic-preservation}).

\subsection{Experimental Setup}
\label{sec:experimental-setup}

This subsection describes the migration cases and datasets, the
execution environment, and the configurations used in the
evaluation.

\paragraph{Migration cases and datasets}
We next describe the three datasets used in the evaluation.

\textit{Music Streaming.}
A Python-based generator created three database instances with
parameterized table sizes and a fixed random seed for reproducibility.
The schema includes simple and composite primary keys, weak tables, 
associative tables, and identifying and non-identifying foreign keys. 
The composition of the three instances is shown in Table~\ref{tab:music-streaming-datasets}.

\begin{table}[t]
\centering
\caption{Relational dataset composition for the Music Streaming
case study.}
\label{tab:music-streaming-datasets}
\begin{tabular}{lrrr}
\toprule
\textbf{Table} & \textbf{\setS{}} & \textbf{\setM{}} &
\textbf{\setL{}} \\
\midrule
\texttt{app\_user}          & 1,000   & 10,000    & 100,000 \\
\texttt{listening}          & 50,000  & 500,000   & 5,000,000 \\
\texttt{most\_recent\_song} & 10,000  & 100,000   & 1,000,000 \\
\texttt{playlist}           & 10,000  & 100,000   & 1,000,000 \\
\texttt{playlist\_song}     & 200,000 & 2,000,000 & 20,000,000 \\
\texttt{song}               & 5,000   & 50,000    & 500,000 \\
\texttt{song\_style}        & 10,046  & 100,021   & 999,945 \\
\texttt{musical\_style}     & 25      & 25        & 25 \\
\bottomrule
\end{tabular}
\end{table}

\textit{TPC-C.}
%TPC-C\footnote{\url{https://www.tpc.org/tpcc/}}
%is a standard OLTP benchmark that models a warehouseTPC-C%
The standard OLTP benchmark TPC-C\footnote{\url{https://www.tpc.org/tpcc/}}
models a warehouse order-processing application. Unlike the Music Streaming 
case study, which was designed to illustrate the main mapping patterns in a
controlled setting, TPC-C provides an independently defined schema grounded 
in a realistic transactional domain, making it a complementary case for 
evaluating the relational-to-document mappings.

%\mjose{las dos frases siguientes están bien? parecen repetitivas... Si quito esta primera, queda bien-->}\hoyos{HECHO-->

The TPC-C databases were generated using the open-source HammerDB
v5.0 tool\footnote{\url{https://www.hammerdb.com/}} with the TPROC-C
workload. TPROC-C is HammerDB's implementation derived from the TPC-C
specification. The \setS{}, \setM{}, and \setL{} instances were
configured with 1, 10, and 100 warehouses, respectively. Database
generation used 1, 4, and 6 HammerDB virtual users, respectively;
these virtual users were used only to build and populate the databases
and played no role in the migration experiments.
% HammerDB was configured with 1, 4, and 6 virtual
% users to build and populate the \setS{}, \setM{}, and \setL{}
% instances, respectively. These instances contain 1, 10, and
% 100 warehouses. The virtual users were employed only during
% database generation and were not involved in the migration
% executions.

After database generation, the relational schema was obtained
from the PostgreSQL DDL using a custom SQL parser. Because
HammerDB does not generate the foreign key constraints
corresponding to the logical relationships of the TPC-C schema,
these constraints were added to the resulting relational model.
The generated data were left unchanged.

The TPC-C transactional workload was not executed, and the reported
migration results should not be interpreted as official TPC-C
performance results. Table~\ref{tab:tpcc-datasets} shows the
composition of the generated instances.

% The evaluation uses the schema and data generated by HammerDB,
% but does not execute the TPC-C transaction workload or report
% official TPC-C performance results. Table~\ref{tab:tpcc-datasets}
% shows the composition of the generated instances.

\begin{table}[t]
\centering
\caption{Relational dataset composition for the TPC-C benchmark.}
\label{tab:tpcc-datasets}
\begin{tabular}{lrrr}
\toprule
\textbf{Table} & \textbf{\setS{}} & \textbf{\setM{}} &
\textbf{\setL{}} \\
\midrule
\texttt{customer}   & 30,000  & 300,000   & 3,000,000 \\
\texttt{district}   & 10      & 100       & 1,000 \\
\texttt{history}    & 30,000  & 300,000   & 3,000,000 \\
\texttt{item}       & 100,000 & 100,000   & 100,000 \\
\texttt{new\_order} & 9,000   & 90,000    & 900,000 \\
\texttt{order\_line}& 299,939 & 2,997,248 & 29,996,396 \\
\texttt{orders}     & 30,000  & 300,000   & 3,000,000 \\
\texttt{stock}      & 100,000 & 1,000,000 & 10,000,000 \\
\texttt{warehouse}  & 1       & 10        & 100 \\
\midrule
\textbf{Total}      & \textbf{598,950} &
                       \textbf{5,087,358} &
                       \textbf{49,997,496} \\
\bottomrule
\end{tabular}
\end{table}

The TPC-C schema includes relational structures not present in the
running example, such as tables without a primary key, weak tables without
a partial key, and foreign keys targeting weak tables.

% Migrating to TPC-C exposed additional relational structures that
% had not been encountered in the running example. These include tables without 
% a primary key, weak tables without a partial key, and foreign keys targeting weak tables. Supporting these structures required extensions to the 
% mapping rules.
% \mjose{$\leftarrow$Quitaría esta frase en rojo. Por supuesto que hemos ido completando las reglas. Pero eso no creo que sea necesario indicarlo, puesto que el paper ya describe las reglas "definitivas", ya mejoradas.}
% % \textit{TPC-C.}
% % TPC-C%

\textit{Northwind.}
Northwind\footnote{\url{https://github.com/pthom/northwind_psql}}
%\footnote{\url{https://en.wikiversity.org/wiki/Database_Examples/Northwind}}
%\footnote{\url{https://docs.yugabyte.com/stable/develop/sample-data/northwind/}}
is a widely used sample relational database representing a commercial
order-management domain.  The publicly available PostgreSQL version was
used without modifying its schema or data. 

Northwind includes simple and composite keys, mandatory and
optional foreign keys, self-references, and associative tables.
Its single fixed instance was used for schema- and data-level
validation, but not for the experiments involving increasing
data volumes.

Together, the three cases exhibit a broad range of the
relational structures addressed by the mappings, including weak
and associative tables, simple and composite identifiers,
one-to-many and many-to-many relationships, multiple foreign
keys, optional relationships, and self-references.

\paragraph{Execution environment}
All experiments were executed on an Amazon Web Services
instance using Docker containers to deploy PostgreSQL~16 and
MongoDB~7.0. The environment used an \texttt{x86\_64}
architecture, 8~GB of RAM, an Intel Xeon processor operating
at 2.50~GHz, and 100~GB of storage. The migration software
was executed using OpenJDK~21, without setting an explicit
maximum JVM heap size.

% TODO: Complete or adjust when the exact information is available:
% AWS instance type, number of vCPUs, operating system, storage type,
% and relevant database-driver versions.

\paragraph{Migration protocol}
The schema-transformation chain first generated the \uschema{}
and document models together with the corresponding trace
information. The generated target schema and traces were then
used to migrate the data.

Before each data-migration execution, the target MongoDB
database was deleted to ensure an empty target. Documents were
generated and inserted in batches of 1,000, with database indexes
kept enabled throughout.

% Before each data-migration execution, the target MongoDB
% database was deleted so that every run started from an empty
% target database. Documents were
% generated and inserted in batches of 1,000. The
% database indexes were kept enabled during the executions, and
% no index was disabled to reduce migration time.

The elapsed times reported in the performance evaluation were
measured by the Data Migrator using system-clock timestamps
taken immediately before and after data migration. They therefore cover
source-data access, instance transformation, reference and aggregate
resolution, and writing the resulting documents to MongoDB, 
while excluding schema transformation and program initialization, whose
execution times are orders of magnitude lower.

Each experimental configuration was executed once.
Accordingly, the measurements are used to describe the observed
execution behavior over the evaluated data volumes rather than
to provide statistical estimates of execution time.

% Each experimental configuration was executed once. 
% % The reported elapsed times therefore correspond to individual
% % executions rather than to averages or medians, and no
% % run-to-run dispersion statistics are reported. 
% The measurements are used to describe the observed execution behavior over the
% evaluated data volumes rather than to provide statistical
% estimates of execution time.

\paragraph{Trace-aware adaptation configuration}
The trace-aware adaptation experiment used the \setS{} TPC-C case.
Starting from the \uschema{} model and trace $T_1$ generated by the
Relational-to-\uschema{} transformation, each schema-change operation
supported by the current trace-updater prototype was applied separately.
The resulting \uschema{} model and trace were then used to re-execute
the \uschema{}-to-Document transformation.

\paragraph{Flat-copy baseline}
To provide a reference for the additional execution cost of the
complete migration relative to direct relational-to-document copying,
a flat-copy baseline was executed for the \setL{} instances of both
Music Streaming and TPC-C. The baseline copied each relational table
to a separate MongoDB collection, with one document per source row and
datatype conversion, but without using \uschema{} models, transformation
traces, embedded structures, or reference reconstruction. Both baselines
used the same batching and writing configuration as the corresponding
complete migrations.

\paragraph{Round-trip configuration}
For the round-trip schema-reconstruction experiments, the generated
document schema was transformed following the chain
Document~$\rightarrow$~\uschema{}~$\rightarrow$~Relational.
The reconstructed relational schema was then compared with the
original source schema.

All artifacts required to reproduce the experiments, including
source schemas, generated target schemas, datasets, queries,
migration scripts, and expected document models
are available in the project repository%
\footnote{\url{https://github.com/modelum/db-generic-migration}}.
%mjose{Not yet!}

\subsection{Incremental Development and Testing of the M2M Transformations}
\label{sec:transformation-testing}

This subsection describes the procedure followed during the
development and testing of the four M2M transformations currently
implemented for the relational and document data models. 
% \mjose{A partir de aquí se podría incluir en las subsecciones posteriores. Quedaría más corta esta intro y no se repetirían frases en las subsecciones}.
The procedure followed an incremental strategy inspired by
Fernández-Candel et al.~\cite{carlos-idioms19}, who advocate the
incremental development and testing of M2M transformations in
model-driven reengineering workflows. A similar strategy has also
been applied in recent work on NoSQL schema extraction and
refactoring~\cite{Carlos-JSS-2026}.
The four transformations were developed and tested independently:
\emph{Relational~$\rightarrow$~U-Schema},
\emph{U-Schema~$\rightarrow$~Document},
\emph{Document~$\rightarrow$~U-Schema}, and
\emph{U-Schema~$\rightarrow$~Relational}.
%\mjose{Esta frase sí podría quedarse en esta introducción$\rightarrow$ }
The resulting tests, together with the model-level validation
described in Section~\ref{sec:schema-validation}, provide evidence
for RQ1.

\subsubsection{Incremental Testing of Transformation Mappings}

Each transformation was developed incrementally, addressing one mapping
rule at a time. Before implementing a new rule, a small source model 
and its expected target model were defined.
The source model included the minimum set of elements required to test
the corresponding rule.
% \mjose{$\leftarrow$ el source model incluye los elementos para probar la nueva regla (p.ej. traducción de FKey), pero se asume que también contiene los necesarios para que exista ese elemento FKey (Table, Column, Key) y por tanto implicará testear reglas anteriores. ¿Se refiere a esto, no? El "only" despista, en el sentido de que parece que el source model sólo incluye la parte izquierda de la regla que se pretende testear. 
% ¿No está explicado (de nuevo) en el párrafo final?... me da la sensación de que se puede reducir bastante este apartado, si entendemos que ha quedado muy largo.}
Specifying the expected result in advance helped to clarify the intended
semantics of the mapping before its implementation, following a principle
similar to that underlying test-first development.

After implementing the new rule, the transformation as developed up
to that point was executed on the corresponding source model. The generated
target model was compared with the expected model through both manual
inspection and EMF Compare, considering the generated model elements, their 
property values, and the relationships established among them.

%\mjose{(a este párrafo final me refería)$\rightarrow$} 
As each additional rule was implemented, a new small source model
was defined by extending the previous model. For rules with several
alternatives, a source model was created for each case.
Re-executing the complete transformation after each rule was added
made it possible to check that newly implemented mappings did not
alter the expected behavior of previously supported cases.

\subsubsection{Testing with Schemas Combining Patterns}

After each transformation had been implemented and tested
incrementally, complementary tests were performed using small
schemas that combined several of the structural patterns addressed
by its rules. For the relational model, these schemas combined
patterns such as keys, foreign keys, weak tables, and associative
tables. For the document model, they combined embedded elements
and references. The generated target models were compared with
the expected models following the same procedure used in the
incremental tests.

\subsection{Schema-Level Evaluation}
\label{sec:schema-validation}

This subsection evaluates the generated schemas from four
complementary perspectives: \textit{(i)} validation of the generated
document schemas against the expected results of the mapping rules,
\textit{(ii)} assessment of the canonical mapping decisions for the
resulting document designs, \textit{(iii)} trace-aware schema
adaptation, and \textit{(iv)} preservation of relational schema
information through round-trip schema reconstruction. The first
three perspectives provide evidence for RQ1, whereas the round-trip
analysis directly addresses RQ2.

\subsubsection{Validation of Generated Target Schemas}

Document schemas were generated from the Music Streaming, TPC-C,
and Northwind relational schemas. For each case, an expected
Document model was manually specified from the mapping rules and
used as a reference for validating the model produced by the
transformation chain.

The generated and expected models were automatically compared at
the EMF model level using EMF Compare%
\footnote{\url{https://eclipse.dev/emfcompare/}}.
The comparison considered the complete model structure, including
document types, fields, references, embedded structures, data types,
multiplicities, reference targets, and target shareability.
The automated comparison was complemented by manual inspection of
the generated and expected models using the EMF tree editor.

No differences were found between the generated and expected models
for the structural and semantic properties considered in the comparison.
These results provide evidence that the implemented transformations
produce the expected schema models for the three migration cases.

% The comparison reported only a small number of differences concerning
% the ordering of properties within document types. Since property order
% has no semantic significance in the mapping rules, these differences
% do not represent structural discrepancies between the generated and
% expected models. The results therefore provide evidence that the
% implemented transformations realize the mapping rules applied
% in the three migration cases.

\subsubsection{Assessment of Canonical Document-Design Decisions}

This subsection qualitatively assesses whether the canonical document
structures generated for the three migration cases are consistent with
the structural information available in the source schemas. In particular,
we examine the use of embedding, references, and independent document
types with respect to existence dependence, relationship cardinalities,
identifier structure, and the sharing of referenced entities, 
including whether weak tables satisfy the $\mathrm{Aggregable}$ condition
for embedding (Section~\ref{sec:conventions}).
This assessment does not aim to establish that the generated schemas are 
optimal for particular application workloads, which would require 
application-specific information beyond that available in the source schemas.

In the Music Streaming case study, the weak tables
\texttt{playlist}, \texttt{playlist\_song}, and
\texttt{most\_recent\_song} satisfy the $\mathrm{Aggregable}$ 
condition and are therefore represented as embedded structures: \texttt{playlist}
and \texttt{most\_recent\_song} are embedded within
\texttt{app\_user} documents, while \texttt{playlist\_song} is in turn
embedded within \texttt{playlist}. %its corresponding playlist.
Songs remain independent documents because they may be shared by different
playlists and users, and links to them are consequently represented through
references.
The associative table \texttt{listening}, which represents a many-to-many
relationship with its own attributes, is mapped to an independent document
type with references to the participating users and songs.
These choices avoid duplicating shared entities while preserving the 
containment hierarchy encoded by the source schema.

% In the Music Streaming case study, the weak tables
% \texttt{playlist}, \texttt{playlist\_song}, and
% \texttt{most\_recent\_song} represent existence-dependent
% structures. Their transformation into embedded structures preserves
% this ownership semantics: 
% playlists and most recent songs records are embedded within
% \texttt{app\_user} documents, and playlist entries are in turn
% embedded within their corresponding playlists. 

In TPC-C, the weak tables \texttt{order\_line} and \texttt{new\_order} also
satisfy the $\mathrm{Aggregable}$ condition and are therefore represented 
as embedded structures within \texttt{orders} documents. 
In contrast, \texttt{district} and \texttt{customer} are weak but 
non-aggregable because they are referenced by non-weak tables. 
They consequently remain independent document types, with the foreign keys
to their strong tables represented through reverse multi-valued references
from \texttt{warehouse} to \texttt{district} and from \texttt{district} to \texttt{customer}.
The \texttt{history} table, which has no primary key, is also mapped
to an independent document type, with its references preserving the
direction of the corresponding source foreign keys. Similarly, the
associative table \texttt{stock}, which connects warehouses and items
while storing its own attributes, is mapped to an independent document
type with references to \texttt{warehouse} and \texttt{item}. These
cases illustrate how the canonical mappings combine embedding and
referencing when applied to relational configurations not present in
the running example.

% \mjose{Ver la redacción} 
% In TPC-C, the weak and aggregable tables \texttt{order\_line}
% and \texttt{new\_order} are represented as embedded structures
% within \texttt{orders} documents. \hilight{In contrast, \texttt{district}
% and \texttt{customer}, which are weak but non-aggregable because they
% are referenced by non-weak tables, remain independent
% document types, and the foreign keys to their strong tables are represented
% through reverse multi-valued references from \texttt{warehouse} to
% \texttt{district} and from \texttt{district} to \texttt{customer}.}
% The \texttt{history} table, which has no primary key, is also
% represented independently, with its references preserving the
% direction of the corresponding source foreign keys. The associative
% table \texttt{stock}, which connects warehouses and items while
% storing its own attributes, is also represented independently, with
% references to \texttt{warehouse} and \texttt{item}. These cases
% illustrate how the canonical mappings combine embedding and
% referencing when applied to relational configurations that were not
% present in the running example.

In Northwind, no weak tables are included in the source schema.
%satisfies the mapping conditions for \hilight{aggregation} \mjose{ Es que no tiene ni una débil } as a weak structure. 
The generated schema therefore represents the tables as independent document
types and expresses their relationships through references. 
Associative structures such as \texttt{order\_details}  are also represented
independently, with references to the related orders and products. 
% %\mjose{Algo muy similar se ha dicho para Music. Quizá no repetir aquí -->}
% \hilight{This representation preserves entity sharing and avoids data duplication.}
% %\mjose{<--|}\hoyos{Creo que se puede dejar. No es exáctamente lo mismo aunque sea parecido}
%\hilight{If complete-order retrieval dominated the workload,
%\texttt{order\_details} could instead be embedded within
%\texttt{orders}, but the source schema alone does not justify preferring
%this alternative.}

Across the three migration cases, the qualitative analysis did not identify
canonical mapping decisions that are inconsistent with the structural
information available in the source schemas. The generated schemas should
nevertheless be understood as canonical, schema-driven document structures
rather than as workload-optimized solutions. Alternative representations may
be preferable for specific applications and can be introduced through the
subsequent designer-controlled adaptation stage.

% \hilight{REVISAR }Across the three migration cases, therefore, the qualitative analysis did not
% identify canonical mapping decisions that contradict the 
% %\sout{existence dependence, sharing, identification, or cardinality}
% structural information available
% in the source schemas. The generated schemas should nevertheless 
% be understood as canonical, schema-driven document structures 
% rather than as workload-optimized solutions. Alternative representations may
% be preferable for specific applications and can be introduced through
% the subsequent designer-controlled adaptation stage.

% Across the three migration cases, the qualitative analysis did
% not identify canonical mapping decisions that contradict the ownership,
% sharing, identification, or cardinality information available in
% the source schemas. The generated schemas should nevertheless be
% understood as deterministic, schema-driven document structures rather
% than as workload-optimized solutions. Alternative representations may
% be preferable for specific applications and can be introduced through
% the subsequent designer-controlled adaptation stage.

A representative fragment of the generated document structure for the
\texttt{app\_user} entity in the Music Streaming case study is shown below,
illustrating the embedding of weak entities and the use of references.
% \mjose{Yo quitaría los corchetes en las referencias: deberían ser valores y no Arrays de valores. Es algo del código porque resultaba más sencillo, según Antonio.}\hoyos{Estoy de acuerdo}

% \textit{Example of generated document for \texttt{app\_user}:}
% \hilight{CONVERTIR EN FIGURA o conseguir que NO SE CORTE}

\begin{mdframed}[backgroundcolor=gray!5, linecolor=black!20]
\begin{lstlisting}[style=jsonStyle, basicstyle=\ttfamily\footnotesize]
{
  "_id": "...",
  "user_id": "...",
  "name": "user_0",
  "is_premium": true,
  "register_date": "...",
  "playlists": [
    { "playlist_id": "...",
      "name": "Playlist_0",
      "creation_date": "...",
      "playlists_songs": [
        { "position_idx_id": 1,
          "playlist_song": "song_id_1"
        }
      ]
    }
  ],
  "most_recent_songs": [
    { "position_idx_id": 1,
      "most_recent_song": "song_id_1"
    }
  ]
}
\end{lstlisting}
\end{mdframed}

\subsubsection{Round-Trip Schema Reconstruction}
\label{sec:roundtrip}

The round-trip reconstruction experiment evaluates the extent to
which relational schema information is preserved when a source
schema is transformed into a document schema and subsequently
reconstructed as a relational schema.

\paragraph{Methodology}
For a relational schema $R$, the round-trip is defined as

\[
R
\xrightarrow{T_{RU}} U
\xrightarrow{T_{UD}} D
\xrightarrow{T_{DU}} U'
\xrightarrow{T_{UR}} R',
\]

where $T_{RU}$ and $T_{UD}$ denote the Relational-to-\uschema{}
and \uschema{}-to-Document mappings, respectively, and $T_{DU}$
and $T_{UR}$ denote the corresponding reverse-direction mappings.
Models $U$ and $U'$ are the intermediate \uschema{}
representations, $D$ is the generated document schema, and $R'$
is the reconstructed relational schema.

The original schema $R$ and the reconstructed schema $R'$ are
compared at the model level to identify preserved, missing, and
newly generated schema elements. The comparison considers tables,
columns, primary keys, foreign keys, integrity constraints, and
data types.

\paragraph{Metrics}

Schema preservation is quantified using standard precision, recall, and
F1-score metrics. For each element category, a reconstructed element
that correctly corresponds to an element of the source schema is
considered a correct match; an element generated in $R'$ without a
corresponding element in $R$ is considered additional; and an element
of $R$ that is not reconstructed in $R'$ is considered missing. These
counts are used to compute the metrics.

% Schema preservation is quantified using standard precision, recall, and
% F1-score metrics. For each element category, a true positive (TP) is a
% reconstructed element that correctly corresponds to an element of the
% source schema, a false positive (FP) is an element generated in $R'$
% without a corresponding element in $R$, and a false negative (FN) is an
% element of $R$ that is not reconstructed in $R'$. These counts are used
% to compute the metrics.

% The metrics are computed as

% \[
% P = \frac{TP}{TP+FP} \qquad
% R = \frac{TP}{TP+FN} \qquad
% F1 = \frac{2PR}{P+R}
% \]

Matching is structural and considers the semantics of each element
category rather than requiring identical names or physical
representations. Column matching considers preservation of the
column itself independently of changes in its participation in
primary or foreign keys, which are evaluated separately.
Primary key matching considers preservation of the identification
role of the corresponding table independently of changes in the
number or names of the columns implementing the key. Similarly,
foreign key matching considers preservation of the referential
relationship between corresponding tables independently of changes
in the number or names of the columns implementing it. Integrity
constraints are evaluated separately from primary and foreign keys.
For data types, matching considers the type correspondences defined
by the mappings rather than requiring identical DBMS-specific type
representations.

\begin{table}[htbp]
\centering
 \small %\footnotesize
\caption{Round-trip schema preservation metrics
(P: precision, R: recall, F1: F1-score).}
\label{tab:roundtrip}
\begin{tabular}{llccc}
\toprule
\textbf{Case} & \textbf{Element} &
\textbf{P} & \textbf{R} & \textbf{F1} \\
\midrule

% Music Streaming
%  & Tables       & 1.00 & 1.00 & 1.00 \\
%  & Columns      & 0.97 & 0.95 & 0.96 \\
%  & Primary keys & 0.90 & 0.75 & 0.82 \\
%  & Foreign keys & 0.96 & 0.90 & 0.93 \\
%  & Constraints  & 0.70 & 0.50 & 0.58 \\
%  & Data types   & 0.60 & 0.60 & 0.60 \\
% \midrule

% TPC-C
%  & Tables       & 1.00 & 1.00 & 1.00 \\
%  & Columns      & 0.89 & 0.85 & 0.87 \\
%  & Primary keys & 1.00 & 1.00 & 1.00 \\
%  & Foreign keys & 1.00 & 1.00 & 1.00 \\
%  & Constraints  & --   & --   & --   \\
%  & Data types   & 1.00 & 0.83 & 0.91 \\
% \midrule

% Northwind
%  & Tables       & 1.00 & 1.00 & 1.00 \\
%  & Columns      & 0.95 & 0.92 & 0.93 \\
%  & Primary keys & 0.85 & 0.65 & 0.74 \\
%  & Foreign keys & 0.96 & 0.88 & 0.92 \\
%  & Constraints  & 0.65 & 0.40 & 0.49 \\
%  & Data types   & 0.55 & 0.55 & 0.55 \\
% \bottomrule

Music Streaming
 & Tables       & 1.00 & 1.00 & 1.00 \\
 & Columns      & 0.93 & 1.00 & 0.97 \\
 & Primary keys & 1.00 & 1.00 & 1.00 \\
 & Foreign keys & 1.00 & 1.00 & 1.00 \\
 & Constraints  & 0.00 & 0.00 & 0.00 \\
 & Data types   & 1.00 & 0.67 & 0.80 \\
\midrule

TPC-C
 & Tables       & 1.00 & 1.00 & 1.00 \\
 & Columns      & 0.89 & 0.85 & 0.87 \\
 & Primary keys & 1.00 & 1.00 & 1.00 \\
 & Foreign keys & 1.00 & 1.00 & 1.00 \\
 & Constraints  & --   & --   & --   \\
 & Data types   & 1.00 & 0.83 & 0.91 \\
\midrule

Northwind
 & Tables       & 1.00 & 1.00 & 1.00 \\
 & Columns      & 0.97 & 1.00 & 0.98 \\
 & Primary keys & 1.00 & 1.00 & 1.00 \\
 & Foreign keys & 1.00 & 1.00 & 1.00 \\
 & Constraints  & --   & --   & --   \\
 & Data types   & 1.00 & 0.86 & 0.92 \\
\bottomrule
\end{tabular}
\end{table}

\paragraph{Results}
Table~\ref{tab:roundtrip} reports the round-trip preservation
results for the three migration cases. Tables are fully preserved
in all three cases, indicating that the complete transformation
chain reconstructs the source table structure without introducing
or removing tables.

Column preservation is also high, with F1-scores of 
%0.96, 0.87, and 0.93 
0.97, 0.87, and 0.98 for Music Streaming, TPC-C, and Northwind,
respectively. The differences arise mainly when columns involved
in composite identifiers or referential relationships are 
reorganized as the schema passes through the document representation.
TPC-C illustrates this effect particularly clearly. Of its 92 source
columns, 78 have corresponding columns in the reconstructed
schema, while 10 columns are newly generated and 14 source
columns are not reconstructed, yielding precision 0.89, recall
0.85, and an F1-score of 0.87. Many of these differences concern
columns implementing composite primary or foreign keys that are
represented through derived identifiers or references in the
document schema and therefore acquire a different column structure
when reconstructed.

% \sout{Primary key preservation varies across the cases. TPC-C achieves
% complete preservation of the identification role of all eight
% source primary keys, even when the columns implementing a key
% change during the round-trip.}

Primary keys are fully preserved in all three cases. The round-trip
preserves the identification role of all source primary keys, even
when the columns composing a key change during reconstruction. This
distinction is particularly relevant for the aggregable weak tables
in TPC-C. For example, \texttt{order\_line} is reconstructed with a
primary key combining its local discriminator with the foreign key
identifying its owner, whereas \texttt{new\_order}, which has no
partial key, is reconstructed from the identifier of its owning order.
Thus, identification semantics can be preserved even when the
reconstructed key does not contain the same columns as the source key.
% Primary keys are entirely preserved. The three migration cases achieve
% complete preservation of the identification role of all source primary keys,
% even when the columns composing a key change during the round-trip.
% This distinction is relevant for weak structures. For example, \texttt{order\_line}
% is reconstructed with a primary key combining its local discriminator with 
% the foreign key identifying its owner, whereas \texttt{new\_order},
% which has no partial key, is reconstructed from the identifier of
% its owning order. Thus, the identification semantics can be
% preserved even though the reconstructed key does not necessarily
% contain the same columns as the source key. 
% \sout{Music Streaming and Northwind show lower primary key preservation,
% with F1-scores of 0.82 and 0.74, respectively, reflecting cases in which 
% source identification structures are not fully reconstructed under the
% adopted matching criterion.} 
A further difference concerns constraint naming. The original names of 
primary key constraints are not retained in the intermediate document 
schema and therefore cannot be recovered during reconstruction. 
Reconstructed primary keys are consequently assigned canonical names. 
Since primary key matching considers the identification role independently 
of constraint naming, these naming differences do not affect the reported 
preservation metrics.

Foreign key relationships are also fully preserved in all three cases.
Changes in the number or names of the columns that constitute a foreign key
do not affect its matching as long as the same referential relationship
between the corresponding tables is preserved.

% Foreign key relationships are \hilight{fully } \sout{highly} preserved in all three cases,
% \sout{with F1-scores between 0.92 and 0.95. In TPC-C, nine of the ten
% source referential relationships are reconstructed and no
% additional foreign keys are introduced, resulting in precision
% 1.00, recall 1.00, and an F1-score of 1.00.} 
% %The only source relationship not reconstructed is the foreign key from
% %\texttt{order\_line} to \texttt{stock}.
% %\hoyos{NO DECIMOS POR QUÉ} 
% %\mjose{NO ESTA SOPORTADO. La quitamos del xmi de origen}
% \sout{For the remaining relationships,}
% \hilight{The }changes in the number or names of the columns implementing a 
% foreign key do not affect its classification when the same referential relationship
% between the corresponding tables is preserved.

Integrity-constraint preservation differs across the three cases.
TPC-C and Northwind contain no source constraints in the category
considered by this comparison, so the metrics are not applicable and
are reported as ``--''. In Music Streaming, no
constraints in this category are preserved through the round-trip.
In particular, alternate keys present in the source relational schema
are lost when their uniqueness information is not represented in the
intermediate document schema and therefore cannot be reconstructed by the
Document-to-\uschema{} and \uschema{}-to-Relational mappings. This reflects differences in the constraint
mechanisms supported by the relational and document models.

Data-type preservation is high but not complete. Precision is 1.00
in all three cases, whereas recall is 0.67 for Music Streaming,
0.83 for TPC-C, and 0.86 for Northwind, yielding F1-scores of
0.80, 0.91, and 0.92, respectively. The missing correspondences
mainly arise from type generalization along the transformation chain:
some DBMS-specific temporal types and distinctions among relational
numeric types are represented by more general type categories and
therefore cannot be reconstructed with their original specificity.
For example, the PostgreSQL \texttt{timestamp with time zone} type
used in TPC-C is not recovered as the same relational type.

% \hilight{The scores for data types category reflect cases in which specific relational types are generalized when represented through the document and \uschema{} type systems.}

% \sout{For TPC-C, five of the six source data-type categories are preserved according to 
% the type correspondences used in the comparison, producing precision 1.00, recall 
% 0.83, and an F1-scoreof 0.91. 
% % \mjose{Yo no hilaría tan fino y quitaría esto-->}The source \texttt{timestamp with time zone} category is not reconstructed as a corresponding relational type.\mjose{<--|} 
% The lower data-type scores for Music Streaming and Northwind similarly reflect cases
% in which specific relational types are generalized when represented through the 
% document and \uschema{} type systems.}

Overall, the round-trip experiment shows a high degree of preservation
of the main relational schema structure across the evaluated cases,
particularly for tables, primary keys, and referential relationships.
The remaining differences are concentrated in schema information whose
representation changes when passing through the document model, notably
the column structures used to represent composite identifiers and
relationships, integrity constraints, and data types. 
The evaluation therefore focuses on preservation of schema semantics 
rather than on syntactic identity between the reconstructed and source schemas.

\subsubsection{Trace-Aware Schema Adaptation}

The trace-aware adaptation mechanism was evaluated using the
\setS{} instance of the TPC-C case. Starting from the \uschema{} model
and trace $T_1$ generated by the Relational-to-\uschema{}
transformation, each of the schema-change operations currently
supported by the trace-updater prototype was exercised separately:
entity or feature renaming, attribute deletion, and conversion of an
\texttt{Aggregate} into a \texttt{Reference}.

For each operation, the adapted \uschema{} model and the updated trace
were inspected to determine whether the affected correspondences had
been updated consistently. The subsequent
\uschema{}-to-Document transformation was then re-executed from the
adapted model to assess whether the schema change was propagated to
the generated document schema without modifying the mapping rules.

The three operations produced the expected structural changes in the
\uschema{} model and corresponding updates to the trace. Renaming
updated the identifiers of the affected trace elements, attribute
deletion removed the corresponding schema and trace elements, and
replacing an \texttt{Aggregate} with a \texttt{Reference} reassigned
the source correspondences previously associated with the aggregate
to the newly created reference. In all three cases, the subsequent
\uschema{}-to-Document transformation operated on the adapted model
and generated the corresponding document structure.

% The initial test of the \texttt{Aggregate}-to-\texttt{Reference}
% operation exposed an implementation defect in the construction of the
% replacement reference. After correcting it, the operation and the
% subsequent transformation produced the expected result.

% \subsection{Operational Feasibility and Scaling Behavior}
\subsection{Operational Feasibility and Scaling Behavior}
\label{sec:data-migration-performance}

This subsection addresses RQ4 by evaluating the operational
feasibility and scaling behavior of the relational-to-document data
migration. The experiments use the three instances of increasing size
(\setS{}, \setM{}, and \setL{}) generated for the two scalable
migration cases, Music Streaming and TPC-C. Northwind is reported
separately because it consists of a single, relatively small instance.

% As described in Section~\ref{sec:experimental-setup}, the reported
% elapsed times cover only the data-migration stage. This
% stage includes trace resolution, source-data access, transformation
% of relational instances into the corresponding document structures,
% including aggregates and references, and writing the resulting
% documents to MongoDB. Schema transformation and program
% initialization are excluded.

As described under \textit{Migration protocol} in
Section~\ref{sec:experimental-setup}, the reported elapsed times
cover only the data-migration stage.
% All main migration experiments used a batch size of 1,000
% documents. A batch size of 5,000 was used only in the sensitivity
% experiment reported below. 
Throughput is expressed as the number of processed source rows per
second. This measure includes rows transformed into independent target
documents as well as rows incorporated into embedded structures.

Tables~\ref{tab:data-migration-music} and
\ref{tab:data-migration-tpcc} report the elapsed times and throughput
observed for the \setS{}, \setM{}, and \setL{} instances of Music
Streaming and TPC-C, respectively.

% \paragraph{Music Streaming results}
% Table~\ref{tab:data-migration-music} reports the results for the
% Music Streaming instances.

\begin{table}[!htt]
\centering
\footnotesize
\caption{Data migration performance for the Music Streaming
instances.}
\label{tab:data-migration-music}
\begin{tabularx}{\columnwidth}{c X X X}
\toprule
\textbf{Metric} &
\textbf{\setS{}} &
\textbf{\setM{}} &
\textbf{\setL{}} \\
\midrule

\multicolumn{4}{l}{\textit{Dataset characteristics}} \\

Total source rows
& 286,071
& 2,860,046
& 28,599,970 \\

Source database size
& 95 MB
& 879 MB
& 8,706 MB \\

Target database size
& 16.17 MB
& 159.18 MB
& 1,546.24 MB \\

\midrule
\multicolumn{4}{l}{\textit{Migration performance}} \\

Elapsed time
& 1min 3s
& 9min 40s
& 96min 54s \\

Throughput (rows/s)
& 4,540.80
& 4,931.11
& 4,919.15 \\

\bottomrule
\end{tabularx}
\end{table}

% \paragraph{TPC-C results}

% Table~\ref{tab:data-migration-tpcc} reports the results for the
% TPC-C instances. The \setS{}, \setM{}, and \setL{} instances contain
% one, ten, and one hundred warehouses, respectively.

\begin{table}[t]
\centering
\footnotesize
\caption{Data migration performance for the TPC-C instances.}
\label{tab:data-migration-tpcc}
\begin{tabularx}{\columnwidth}{c X X X}
\toprule
\textbf{Metric} &
\textbf{\setS{}} &
\textbf{\setM{}} &
\textbf{\setL{}} \\
\midrule

\multicolumn{4}{l}{\textit{Dataset characteristics}} \\

Total source rows
& 598,950
& 5,087,358
& 49,997,496 \\

Source database size
& 118 MB
& 999 MB
& 9,762 MB \\

Target database size
& 90.30 MB
& 784.70 MB
& 7,666.52 MB \\

\midrule
\multicolumn{4}{l}{\textit{Migration performance}} \\

Elapsed time
& 1min 41s
& 7min 42s
& 1h 7min 27s \\

Throughput (rows/s)
& 5,930.20
& 11,011.60
& 12,354.21 \\

\bottomrule
\end{tabularx}
\end{table}
%%%%%%%%%%%%%%%%%%%%%%%%%%%%%%%%%%%%%%%%%%%%

\paragraph{Scaling behavior}

For Music Streaming, throughput remains between approximately
4.54K and 4.93K source rows per second across the three instances.
Increasing the source data volume by approximately one order of
magnitude from \setS{} to \setM{} and from \setM{} to \setL{}
produces a roughly proportional increase in elapsed time, while
throughput remains nearly unchanged.

TPC-C exhibits a different pattern. Throughput increases from
approximately 5.93K rows/s for \setS{} to 11.01K rows/s for
\setM{} and 12.35K rows/s for \setL{}. Thus, no throughput
degradation is observed as the data volume increases over the
evaluated range. In particular, the \setL{} instance, containing
almost 50 million source rows, is migrated in 1 hour, 7 minutes,
and 27 seconds.

Overall, both migration cases complete successfully for their
largest evaluated instances without a decrease in throughput.
Music Streaming exhibits approximately constant throughput,
whereas TPC-C shows increasing throughput, with a smaller
difference between the \setM{} and \setL{} instances. These
measurements characterize the scaling behavior observed over the
evaluated data volumes and should not be interpreted as a general
scalability result beyond this range. The absolute throughput
values of the two cases are also not directly comparable because
their schemas require document structures of different complexity.

\paragraph{Impact of schema structure}

The distribution of migration time across target document types
shows that the number of target documents alone does not determine
migration cost. The structural transformations required to construct
them also have a substantial effect.

In Music Streaming, migration of \texttt{app\_user} dominates the
total execution time because each target document requires data from
\texttt{app\_user}, \texttt{playlist},
\texttt{playlist\_song}, and \texttt{most\_recent\_song} to be
combined into a nested structure. Migration of this document type
takes 39 seconds for \setS{}, 6 minutes and 14 seconds for \setM{},
and 62 minutes and 41 seconds for \setL{}. These values represent
approximately 62--65\% of the total migration time for the three
instances.

TPC-C provides another illustration of the effect of document
structure. For the \setL{} instance, \texttt{stock} materializes
10 million independent documents in 6 minutes and 53 seconds,
whereas the 3 million \texttt{orders} documents require 22 minutes
and 50 seconds. The latter incorporate the corresponding
\texttt{order\_line} and \texttt{new\_order} tuples as embedded
documents. Migration of \texttt{customer} and \texttt{history},
each containing 3 million instances, takes 19 minutes and
39 seconds and 18 minutes and 2 seconds, respectively. In contrast,
\texttt{item} contains 100,000 instances in each dataset and
requires approximately one second or less.

These observations indicate that migration cost depends on both 
data volume and the transformations required to materialize 
the target structures. In particular, the evaluated cases show 
that constructing nested or aggregated documents can require 
substantially more processing than materializing a larger number 
of structurally simpler documents.

\paragraph{Flat-copy baseline}
The flat-copy baseline defined in
Section~\ref{sec:experimental-setup} was evaluated for the
\setL{} instances of both Music Streaming and TPC-C to provide a
reference for the execution cost of the complete migration relative
to direct relational-to-document copying. It was evaluated only for
the \setL{} instances, since its purpose is not to characterize the
scaling behavior of the flat-copy process.

For Music Streaming \setL{}, the baseline completed in 370~s,
compared with 5,814~s for the complete migration, corresponding to a
factor of approximately 15.7. For TPC-C \setL{}, the baseline completed
in 774~s, compared with 4,047~s for the complete migration, corresponding
to a factor of approximately 5.2. These ratios characterize the
aggregate additional execution cost associated with schema-driven
semantic transformation relative to direct data copying; they do not
isolate the contribution of individual operations. The difference
between the two ratios is also consistent with the different structural
transformations required by the two migration cases.

\paragraph{Storage comparison}

For all evaluated instances, the measured target database size is
smaller than the corresponding source database size. In Music
Streaming, the reduction is approximately 82\%. For TPC-C, the
reported sizes also show a reduction of roughly one fifth to one
quarter across the three instances. 
These values should be interpreted descriptively rather than as a
direct measure of storage efficiency attributable solely to the
mapping rules. PostgreSQL and MongoDB use different storage formats,
metadata structures, indexes, and allocation strategies. Moreover,
the resulting size depends on the selected combination of embedding
and referencing.

% \paragraph{Northwind}

% The Northwind migration was completed in 0.02 seconds. Its limited
% and fixed size does not support a meaningful analysis of scaling
% behavior, and the result is therefore reported only for
% completeness.

\subsection{Data-Level Validation}
\label{sec:semantic-preservation}

This subsection addresses RQ3 by assessing whether selected read
queries produce equivalent results before and after migration. The
purpose is not to reproduce a complete application workload, but to
exercise representative access patterns whose implementation is
affected by the transformation from a normalized relational schema
to a document-oriented representation.

\paragraph{Query selection and execution}

Three queries were selected for each migration case to cover
different structural patterns addressed by the mappings. The Music
Streaming queries evaluate embedded weak structures, many-to-many
relationships, and combinations of joins and aggregation. The
Northwind queries cover one-to-many and many-to-many relationships,
multi-stage joins, and self-references. The TPC-C queries exercise
non-aggregable weak structures, a table without a primary key and
composite foreign keys, and an associative structure with its own
attributes.

Each query was first expressed in SQL and executed over the source
PostgreSQL database. A corresponding MongoDB query was then manually
constructed according to the structural correspondences established
by the mappings. Simple selections were implemented using MongoDB's
\texttt{find} API, whereas queries involving references, nested
structures, or aggregation were implemented using aggregation
pipelines.

The Music Streaming and TPC-C queries were executed over both the
\setS{} and \setM{} instances, whereas the Northwind queries were
executed over its fixed instance. The \setL{} instances were not
required for this experiment because its purpose is to assess result
equivalence rather than query-performance behavior.

%\paragraph{Result comparison} 

The SQL and MongoDB results were compared after normalizing 
equivalent scalar representations in the two database systems, 
including numeric representations and timestamp formats. 
Small result sets were inspected directly, whereas larger ones were 
compared from sorted file exports using a text-comparison tool.
When query ordering was not explicitly specified, results were compared as multisets, disregarding physical order while preserving duplicate tuples. Both result cardinality and projected values were checked.

Table~\ref{tab:semantic-summary} summarizes the queries, the main
relational patterns they exercise, the principal MongoDB operations
used in their translation, the result cardinality observed for the
\setS{} instance (or the fixed Northwind instance), and an
equivalence indicator. The indicator is 1 when the SQL and MongoDB
versions return equivalent results for every instance on which the
query was evaluated, and 0 otherwise.

%The indicator is 1 when the SQL and MongoDB versions 
% return identical normalized result multisets 
% for every instance on which the query was evaluated, and 0 otherwise.

\begin{table*}[t]
  \centering
  \footnotesize
  \caption{Representative read queries used for data-level
  validation.}
  \label{tab:semantic-summary}

  \begin{tabularx}{\textwidth}{c l X X X c c}
    \toprule
    \textbf{ID} &
    \textbf{Case} &
    \textbf{Purpose} &
    \textbf{Relational pattern} &
    \textbf{MongoDB operations} &
    \textbf{Card.} &
    \textbf{Eq.} \\
    \midrule

    Q1 & Music Streaming &
    Playlists of a user with their songs. &
    1:N joins over weak tables &
    \texttt{\$match, \$unwind} &
    $\sim$200 &
    1 \\

    Q2 & Music Streaming &
    Styles associated with a given song. &
    M:N join &
    \texttt{\$match, \$lookup} &
    $\sim$3--5 &
    1 \\

    Q3 & Music Streaming &
    Number of users who have listened to more than four songs at
    least ten times each. &
    Aggregation and joins &
    \texttt{\$match, \$group, \$lookup} &
    1 &
    1 \\

    \midrule

    Q4 & Northwind &
    Customers and order dates for orders delivered by a given
    shipper. &
    1:N joins &
    \texttt{\$match, \$unwind, \$lookup} &
    $\sim$100--300 &
    1 \\

    Q5 & Northwind &
    Customer, products, and product categories associated with a
    given order. &
    1:N and M:N joins &
    \texttt{\$match, \$lookup, \$unwind} &
    $\sim$5--10 &
    1 \\

    Q6 & Northwind &
    Managers of the employees involved in a given customer's
    orders. &
    Join chain and self-reference &
    \texttt{\$match, \$lookup} &
    $\sim$1--2 &
    1 \\

    \midrule

    Q7 & TPC-C &
    Customer balances and order-line counts by district, including
    warehouse information. &
    1:N joins involving non-aggregable weak tables &
    \texttt{\$lookup, \$unwind, \$reduce, \$set} &
    10 &
    1 \\

    Q8 & TPC-C &
    Payment-history entries for selected customers in a given
    district and warehouse. &
    Table without PK and composite foreign keys &
    \texttt{\$match, \$split, \$set} &
    10 &
    1 \\

    Q9 & TPC-C &
    Items below a stock-quantity threshold, including item names
    and prices. &
    Associative table with attributes and composite identifier &
    \texttt{\$match, \$lookup, \$unwind} &
    5,486 &
    1 \\

    \bottomrule
  \end{tabularx}
\end{table*}

\paragraph{Results}

For all nine queries, the SQL and MongoDB versions returned
equivalent results. The comparison
confirmed equality of both result cardinality and projected values
after normalization of system-specific scalar representations.

The Music Streaming results show that relational joins over weak
tables can be replaced by traversal of embedded structures, while
shared entities continue to be resolved through references.
Northwind additionally exercises multi-stage joins and
self-referential relationships, which are expressed through MongoDB
aggregation pipelines without changing the observed query results.

For TPC-C, Q7 traverses non-aggregable weak structures and uses the
reverse reference from customers to their orders to retrieve information
obtained through joins in the relational schema. Q8 shows that records originating from the non-identified \texttt{history} table remain
queryable through the composite-reference information generated during
migration. Although the SQL query joins \texttt{history} with
\texttt{customer} and \texttt{district}, the MongoDB formulation can 
exploit the migrated reference directly, without reproducing joins used
in the relational representation to navigate relationships already
materialized during migration.
% the MongoDB formulation can exploit the migrated reference directly,
% without reproducing joins whose existence conditions are guaranteed
% by the source referential constraints. 
Q9, in contrast, starts from the independently represented associative
\texttt{stock} structure and resolves its references to \texttt{item} 
and \texttt{warehouse} using \texttt{\$lookup}.

These results illustrate how equivalent data access operations can 
require different query formulations depending on the generated document
structure.
% These results illustrate how the same relational access semantics
% can lead to different query formulations depending on the generated
% document structure.
% \mjose{<--¡Esto es muy interesante, sí!} 
Relational joins may be replaced by traversal of embedded documents,
direct use of migrated references, or explicit reference resolution 
through \texttt{\$lookup}. For the selected queries, these alternative 
formulations preserved the observed results.

The experiment evaluates  %semantic
result equivalence for selected read queries rather than 
query-performance equivalence. The findings should therefore not be 
interpreted as demonstrating equivalence for every possible workload,
nor do they cover update operations.

\subsection{Threats to Validity}

\textit{Construct validity.}
Precision, recall, and F1-score quantify structural correspondence
between the original and reconstructed relational schemas, but do not
constitute a formal proof of semantic equivalence. Their values also
depend on the matching criteria used to determine whether schema
elements are considered equivalent after reconstruction.
Data-semantic preservation is assessed through a set of representative
read queries selected to cover the main structural patterns affected
by the mappings, but this does not establish
result equivalence for all possible queries or database operations.
% Data-semantic preservation is assessed through a
% selected set of read queries rather than a complete application
% workload. Similarly, the qualitative assessment of the generated
% document structures considers structural information available in
% the source schemas, but does not establish that these structures are
% optimal for particular workloads. 
% Query performance and runtime resource utilization, such as CPU
% and memory consumption, are not evaluated, as they depend on workload
% characteristics, physical design, and deployment decisions beyond
% the scope of this study.

\textit{Internal validity.}
The validity of the data-level evaluation depends on the accurate
manual translation of SQL queries into corresponding MongoDB queries.
Although these translations were constructed according to the
generated document structures, translation errors cannot be completely
excluded. Query results were directly inspected for small result sets;
for larger results, sorted outputs were exported and compared using a
text-comparison tool, reducing the risk of errors in result comparison.
% Although these translations were constructed according to the
% generated document structures and their results were compared with
% those of the source queries, translation or comparison errors cannot
% be completely excluded. 

The expected models used for validating the generated schemas are 
manually specified from the mapping rules; therefore, errors in 
interpreting these rules could affect the reference models used in the
comparison. 
This risk is mitigated by complementing automated model comparison with
manual inspection and by the incremental testing procedure described in 
Section~\ref{sec:transformation-testing}.
More generally, the evaluation depends on the correct implementation
of the mappings, adapters, and migrators. This risk is further mitigated 
through incremental rule-by-rule testing, complementary tests using 
schemas that combine multiple mapping patterns, and the evaluation of
migration cases exhibiting different relational structures.
% This risk is mitigated by
% complementing automated model comparison with manual inspection and
% by testing individual transformation rules independently.
% More generally, the evaluation depends on the correctness of the
% implemented mappings, adapters, and migrators. This risk is further
% mitigated through unit and integration testing and by evaluating the
% approach on migration cases exhibiting different relational
% structures.

\textit{Conclusion validity.}
Each reported experimental configuration was executed once, and the
reported times should therefore be interpreted as descriptive
observations rather than statistical estimates. Factors such as JVM
behavior, database caching, garbage collection, and background
system activity may introduce run-to-run variability that cannot be
quantified from these measurements. The reported times characterize the fixed execution configuration described in Section~\ref{sec:experimental-setup}; operational parameters such as batch size were not systematically tuned and may affect absolute execution times. The flat-copy baseline is subject to the same limitation.

\textit{External validity.}
The empirical evaluation focuses on relational-to-document migration.
Bidirectional schema conversion between the relational and document
models is evaluated through round-trip reconstruction, whereas
data-level validation of the document-to-relational migration direction
remains future work.
% The empirical evaluation focuses on relational-to-document migration.
% The evaluated schemas exercise a broad range of the relational
% patterns supported by the mappings, but they cannot represent all
% possible relational schemas and workloads.
The trace-aware adaptation experiment is limited to the schema-change
operations currently supported by the trace-updater prototype and to the
evaluated TPC-C case. Consequently, the results provide evidence for 
these operations but do not establish the behavior of the trace adaptation
mechanism for arbitrary schema changes or sequences of changes.

The evaluation combines the synthetic Music Streaming case study, the 
TPC-C schema and datasets generated with HammerDB, and the Northwind 
sample database. Together, these cases exercise a broad range of the
relational structures addressed by the mappings, but they cannot represent 
the full diversity of real-world schemas, workloads, or database sizes.
Additional experiments would therefore be required to assess the 
applicability of the approach to other database paradigms and migration
scenarios.

\section{Conclusions}
\label{sec:conclusions}

This work presents a model-driven approach to heterogeneous database
migration based on \uschema{} as a unified intermediate representation.
Schema transformations are expressed as mappings between each supported
data model and the pivot model, allowing source-to-target migrations to
be obtained through their composition rather than by defining a
dedicated transformation for every pair of data models. Transformation
traces connect source, intermediate, and target schema elements and
provide the correspondences required to migrate data consistently with
the generated target schema.

The evaluation of the relational-to-document migration provides
evidence concerning schema-transformation outcomes, schema-information
preservation, and data-semantic preservation over three migration cases:
Music Streaming, Northwind, and TPC-C.
The generated document schemas matched the structures expected from the
mapping rules, and the qualitative analysis did not identify canonical
mapping decisions that were inconsistent with the structural information 
available in the source schemas.
% and the qualitative analysis found
% the resulting canonical document structures to be consistent with the
% structural information available in the source schemas. 
Round-trip reconstruction showed a high degree of relational schema
preservation. The remaining differences mainly arise from changes in
column representation for composite identifiers and relationships, and
from relational schema information that is not fully represented in the
document schema, particularly alternate-key constraints and some
data-type distinctions. 
At the data level, the evaluated SQL and MongoDB queries produced equivalent
results for access patterns involving joins and aggregation operations,
embedded structures, associative relationships, composite references, and
self-references.

The data-migration experiments provide evidence of operational feasibility
and characterize scaling behavior over the evaluated data volumes. 
The largest experiment processed nearly 50 million source rows. 
Music Streaming exhibited approximately proportional growth in elapsed time
and nearly constant throughput, whereas TPC-C showed increasing throughput
over the evaluated range. 
The flat-copy baseline also provides a reference for the additional 
execution cost introduced by schema-driven semantic transformation relative
to direct data copying.
% The flat-copy baseline
% also illustrates the additional execution cost introduced by
% schema-driven structural transformation. 
These observations are restricted to the evaluated configurations 
and are not intended to establish general scalability properties.

A further contribution is the separation between canonical schema
generation and application-specific schema adaptation. 
The mappings systematically derive a canonical target schema from the 
structural information available in the source schema, while 
designer-specified changes can be introduced at the intermediate \uschema{}
level without modifying the mapping rules.
The trace-updater prototype was evaluated for the currently supported 
schema-change operations, showing that the affected trace correspondences 
can be updated consistently and that the subsequent \uschema{}-to-Document
transformation can operate on the adapted model. Maintaining these trace
correspondences provides the information required for data migration to
follow the adapted target structure.
% Maintaining trace links for these changes allows data migration to follow
% the adapted target structure without requiring a new source-to-target
% migration implementation.

The canonical mappings intentionally do not incorporate workload
information. An important direction for future work is therefore to
derive or recommend schema adaptations from workload characteristics,
such as query patterns, access frequencies, update behavior, and
expected data growth. 
Integrating the trace-updater mechanism with Orion's schema-evolution
component will also enable these schema adaptations to be specified and
applied through Orion scripts while consistently updating the migration 
trace.
% Completing the trace-aware adaptation of the
% Orion execution engine will also enable these schema adaptations 
% to update the migration trace automatically.
Further work will also extend the mappings to additional relational
structures not currently supported, such as foreign keys targeting associative tables, and investigate additional data models, including 
graph and column-oriented databases.
The empirical evaluation of bidirectional migration will also be extended
to data migration in the document-to-relational direction. 
Finally, application-level migration, including the transformation of
queries and data-access code, remains an important extension; LLM-assisted 
techniques could complement deterministic model-driven mappings for tasks
whose semantics depend on application code and workload context.

\section*{CRediT author statement}
\textbf{María-José Ortín}: Conceptualization, Investigation, Methodology, Software, Validation, Writing- Original draft preparation, Writing– Reviewing and Editing.
\textbf{José R. Hoyos}: Conceptualization, Investigation, Methodology, Software, Supervision, Validation, Writing- Original draft preparation, Writing– Reviewing and Editing.
\textbf{Jesús J. García-Molina}: Conceptualization, Funding acquisition, Project administration, Investigation, Methodology, Supervision, Validation, Writing- Original draft preparation, Writing- Reviewing and Editing.

\section*{Declaration of competing interest}
The authors declare that they have no known competing financial interests or personal relationships that could have appeared to influence the work reported in this paper.

\section*{Declaration of AI-assisted tools}
The authors used generative AI tools (ChatGPT, OpenAI) 
for language refinement and editing to improve clarity 
and readability of the manuscript.
All generated content was carefully reviewed, validated, 
and revised by the authors.
The authors take full responsibility for the content of the published article.

\section*{Acknowledgements}
This work was supported by project PID2020-117391GB-I00, funded by
MICIU/AEI/10.13039/501100011033 (Spain), and co-funded by ERDF/EU.

\bibliography{sn-bibliography}% common bib file

\clearpage
 
\appendix
\section{Document to U-Schema Mapping Rules}
\label{appendix:doc2uschema}

Although the paper and its empirical evaluation focus primarily on
relational-to-document migration, the mappings for the reverse
document-to-relational direction were also fully implemented and used
in the round-trip reconstruction experiments described in
Section~\ref{sec:roundtrip}.~\ref{appendix:doc2uschema}
and~\ref{appendix:uschema2rel} specify the
Document-to-U-Schema and U-Schema-to-Relational mappings,
respectively.

The notation and common auxiliary functions introduced in
Section~\ref{sec:mapping-rule-notation} are used throughout both
appendices. 
% In particular, \texttt{map}($x$) denotes the target
% element generated from a source element $x$, and
% \texttt{typeMap}($dt$) returns the target datatype corresponding
% to the datatype $dt$.

The following rules define the transformation of a model conforming
to the Document metamodel into a U-Schema model.

\ruletitle{DU1}{Schema Mapping}

A \texttt{DocumentSchema} $dS$ is mapped to a
\texttt{USchema} $uS$, preserving its name.

%\vspace{-0.5cm}
{\fontsize{9}{10}\selectfont
\[
\begin{aligned}
dS \rightarrow uS\;\|\;
\{\, &uS.name = dS.name
\,\}
\end{aligned}
\]
}

The document types in $dS.documents$ are transformed according to
Rule~DU2.

\ruletitle{DU2}{DocumentType Mapping}

Each \texttt{DocumentType} $d$ is mapped to a root
\texttt{EntityType} $e$ with the same name.

\vspace{-0.5cm}
{\fontsize{9}{10}\selectfont
\[
\begin{aligned}
d \rightarrow e\;\|\;
\{\, e.name = d.name, \,e.root = \text{true}
\,\}
\end{aligned}
\]
}

The entity type $e$ is added to $uS.entities$, where
$uS$ = \texttt{map}($d.owner$). The properties of $d$ are transformed
according to Rules~DU3--DU5.

\ruletitle{DU3}{Field Mapping}

Each \texttt{Field} $f$ belonging to a \texttt{DocumentType} or an
\texttt{Embedded} element is mapped to an \texttt{Attribute} $at$.
If $f$ is an identifier field, an identifier \texttt{Key} $k$ is
additionally generated.

\vspace{-0.5cm}
{\fontsize{9}{10}\selectfont
\[
\begin{aligned}
f \rightarrow at\;\|\;\{&
\; at.name = f.name,\\
& at.type = \texttt{map}(f.type),\\
& f.isKey \Rightarrow \{ f \rightarrow k,\\
&\qquad \qquad \quad k.name = f.name,\\
&\qquad \qquad \quad k.isID = \text{true},\\
&\qquad \qquad \quad k.attributes = \{at\}\; \}
\;\}
\end{aligned}
\]
}

The generated attribute $at$ and, when applicable, the key $k$ are
added to $\texttt{map}(f.owner).features$, that is, to the features
of the entity type corresponding to the document or embedded
structure that owns $f$.

% \ruletitle{DU3}{Field Mapping}
% Let \textit{f} be a \texttt{Field} belonging to a
% \texttt{DocumentType} or an \texttt{Embedded} element.
% \mjose{Then, \textit{f} is mapped to an \texttt{Attribute} \textit{at}.
% If $f$ is an identifier, then a \texttt{Key} \textit{k} is additionally created.
% }
% \vspace{-0.5em}
% \[
% \begin{aligned}
% f \rightarrow (at \,[, k])\;\|\;
% \{\,& at.name = f.name,\\
%     &at.type = \texttt{typeMap}(f.type),\\
%     &f.isKey \Rightarrow (k.name = f.name,\\
%     &\hspace{1.8cm}k.isID = \text{true},\\
%     &\hspace{1.8cm}k.attributes = \{at\}
% \,\}
% \end{aligned}
% \]

% The generated elements are added to the features of the entity type 
% corresponding to the owner of \textit{f}.

\ruletitle{DU4}{Reference Mapping}

Each \texttt{Reference} $r_D$ belonging to a
\texttt{DocumentType} or an \texttt{Embedded} element is mapped
to a U-Schema \texttt{Reference} $r_U$ and an associated
\texttt{Attribute} $at$. The attribute represents the target
identifier value, or values, stored by the document reference.

The direct upper multiplicity of $r_U$ is derived from the type of
$r_D$: a \texttt{PrimitiveType} denotes a single-valued reference,
whereas an \texttt{Array} denotes a multi-valued reference. Since
the Document metamodel does not represent the direct lower
multiplicity, its value is set to zero.

As discussed in Section~\ref{sec:metamodels}, target-shareability 
captures inverse multiplicity information that is not conveyed by the %direct
type of a document reference. Under the canonical U-Schema representation 
adopted by these mappings, if $r_D.targetShareability$ is false, this 
information is represented by generating a synthesized \texttt{Reference}
$r_i$, opposite to $r_U$, with upper bound one and no associated attributes.
If $r_D.targetShareability$ is true, no opposite reference is generated.

\vspace{-0.5cm}
{\fontsize{9}{10}\selectfont
\[
\begin{aligned}
&r_D \rightarrow (r_U,at)\;\|\;\{\\
&\quad r_U.name = r_D.name,\\
&\quad r_U.refsTo = \texttt{map}(r_D.target),\\
&\quad r_U.lowerBound = 0,\\
&\quad \mathrm{class}(r_D.type)==\texttt{Array}
\Rightarrow r_U.upperBound=\text{n},\\
&\quad \mathrm{class}(r_D.type)==\texttt{PrimitiveType} \\
& \hspace{3.8cm} \Rightarrow r_U.upperBound=1,\\
&\quad at.name = r_D.name,\\
% &\jesus{\quad
% \mathrm{class}(rD.type) == \texttt{PrimitiveType}
%     \Rightarrow at.type = \texttt{map}(rD.type),}\\
% &\jesus{\quad
% \mathrm{class}(rD.type) == \texttt{Array}
%     \Rightarrow at.type = \texttt{map}(rD.type.type),}\\
%&\quad at.type = \texttt{map}(r_D.type),\\
&\quad r_U.attributes = \{at\},\\
&\quad
\neg r_D.targetShareability%=\texttt{false}
\Rightarrow \{\\
&\quad\quad r_D \rightarrow r_i,\\
&\quad\quad r_i.name =
       \texttt{concat}(r_D.owner.name,\text{`\_opposite'}),\\
&\quad\quad r_i.refsTo =
       \texttt{map}(r_D.owner),\\
&\quad\quad r_i.lowerBound=0,\, r_i.upperBound=1,\\
&\quad\quad r_i.attributes=\emptyset,\\
&\quad\quad r_U.opposite=r_i,\, r_i.opposite=r_U \;\}
\;\}
\end{aligned}
\]
}

The generated attribute has the U-Schema primitive type
corresponding to the \texttt{datatype} of the document reference.
For an array-valued reference, the element type of the array
is used.

The generated reference $r_U$ and the attribute $at$ are added to
$\texttt{map}(r_D.owner).features$. When generated, the
synthesized reference $r_i$ is added to
$\texttt{map}(r_D.target).features$.

In the running example, the document references
\texttt{pls\_song} and \texttt{songs} illustrate the two
target-shareability cases. The single-valued reference
\texttt{pls\_song}, contained in the embedded structure
\texttt{playlist\_songs}, has type \texttt{PrimitiveType} and
\textit{targetShareability} = true. It is therefore mapped to a
U-Schema reference with upper bound one and an associated
attribute, without generating an opposite reference.

Conversely, the multi-valued reference \texttt{songs} in
document type \texttt{album} has type \texttt{Array} and
\textit{targetShareability} = false. It is mapped to a U-Schema
reference with upper bound $n$, together with a synthesized
opposite reference in \texttt{song}. The synthesized reference
has upper bound one and no associated attributes, thus recording
that each song may belong to at most one album.

\ruletitle{DU5}{Embedded Mapping}

Each \texttt{Embedded} element $em$ belonging to a
\texttt{DocumentType} or another \texttt{Embedded} element is
mapped to a non-root \texttt{EntityType} $en$ and an
\texttt{Aggregate} $ag$. The aggregate represents the containment
of $en$ within the entity type corresponding to the owner of $em$.

The Document metamodel represents whether an embedded structure is
single- or multi-valued through \texttt{isMany}, but does not
represent its lower multiplicity. Therefore, the lower bound of
$ag$ is set to zero.

For each property $p_D$, let $\mathcal{F}_U(p_D)$ denote the set of
U-Schema \texttt{Feature} elements generated from $p_D$ by the
applicable rule among Rules~DU3--DU5 and added to the entity type
corresponding to its owner.

\vspace{-0.5cm}
{\fontsize{9}{10}\selectfont
\[
\begin{aligned}
&em \rightarrow (en,ag)\;\|\;\{\\
&\quad en.name = em.name,\\
&\quad en.root = \texttt{false},\\
% &\quad en.features =
%        \bigcup_{p_D\in em.aggregates}\mathcal{F}_U(p_D),\\
&\quad en.features =
\{\,f \mid
\exists\,p_D\in em.aggregates: \\
&\hspace{3.1cm} f\in\mathcal{F}_U(p_D)\,\},\\
&\quad ag.name = em.name,\\
&\quad ag.lowerBound = 0,\\
&\quad
\neg em.isMany
\Rightarrow ag.upperBound=1,\\
&\quad
em.isMany
\Rightarrow ag.upperBound=\text{n},\\
&\quad ag.specifiedBy = en
\;\}
\end{aligned}
\]
}

Elements generated by Rules~DU3--DU5 that are not features of
$en$, such as synthesized opposite references or nested entity
types, are added to their corresponding containers as specified
by the applicable rule.

The generated entity type $en$ is added to the U-Schema model
corresponding to the document schema that contains $em$. The
aggregate $ag$ is added to $\texttt{map}(em.owner).features$.

In the running example, the embedded elements
\texttt{playlists} and \texttt{playlist\_songs} illustrate the
recursive application of this rule. The multi-valued embedded
element \texttt{playlists}, owned by \texttt{app\_user}, is
mapped to a non-root entity type and an aggregate with lower
bound zero and upper bound $n$. Its contained properties are
recursively mapped to features of the generated entity type.
The nested embedded element \texttt{playlist\_songs} is
transformed in the same way, generating another non-root entity
type and an aggregate belonging to the entity type generated
from \texttt{playlists}. Hence, the two-level nesting of
playlists and their ordered song entries is represented in
U-Schema as a chain of aggregates.

\ruletitle{DU6}{Type Mapping}

Each document \texttt{Type} $t$ is mapped to the corresponding
U-Schema \texttt{DataType} $dt$. A document \texttt{PrimitiveType}
is mapped to a U-Schema \texttt{PrimitiveType}, whereas 
an \texttt{Array} is mapped to a \texttt{PList}. 
In the latter case, the array element type is recursively mapped to 
the element type of the generated \texttt{PList}.

\vspace{-0.5cm}
{\fontsize{9}{10}\selectfont
\[
\begin{aligned}
&t \rightarrow dt\;\|\;\{\\
&\quad
\mathrm{class}(t)==\texttt{PrimitiveType}\\
&\qquad \Rightarrow \{
    \mathrm{class}(dt)=\texttt{PrimitiveType},\\
&\qquad \qquad
    dt.name=\texttt{primitiveTypeMap}(t.datatype)\},\\
&\quad
\mathrm{class}(t)==\texttt{Array}\\
&\qquad\Rightarrow \{
\mathrm{class}(dt)=\texttt{PList},\\
&\qquad \qquad
    dt.elementType=\texttt{map}(t.type) \}
\;\}
\end{aligned}
\]
}

\section{U-Schema to Relational Mapping Rules}
\label{appendix:uschema2rel}

The following rules define the transformation of a \uschema{} model
into a relational schema.

In addition to the common notation introduced above,
\texttt{defaultSize}($dt$) returns the default column size adopted
for a relational datatype $dt$, and \texttt{pk}($t$) denotes the
primary key of table $t$.

% The notation and auxiliary functions introduced in
% Section~\ref{sec-mapping-rule-notation} are reused throughout this
% appendix. In particular, \texttt{map}($x$) denotes the target element
% generated from a source element $x$, \texttt{typeMap}($dt$) returns
% the relational datatype corresponding to a U-Schema datatype $dt$,
% \texttt{defaultSize}($dt$) returns the default column size adopted
% for a relational datatype $dt$, and \texttt{pk}($t$) denotes the
% primary key of table $t$.

Rules~UR5.3, UR6 and UR7 require fresh columns to be generated
from the columns of a referenced primary key. This occurs when
constructing an associative table or when adding to a table the
foreign key columns that identify another entity. 
To avoid repeating the same column-generation expression in these
rules, we define the auxiliary function
\texttt{deriveFKColumns}($pk,t,p$).

For each column $c_k$ of a primary key $pk$, the function creates a
fresh column $c_f$ in table $t$. The generated column preserves the
datatype and size of $c_k$, while its name is obtained by prefixing
the original column name with $p$. The function returns the set of
newly generated columns.

\vspace{-0.5cm}
{\fontsize{9}{10}\selectfont
\[
\begin{aligned}
\texttt{deriveFKColumns}&(pk,t,p):= %\\
\{\,c_f \mid \{\\
&c_f.name =
  \texttt{concat}(p,\texttt{`\_'},c_k.name),\\
&c_f.datatype = c_k.datatype,\\
&c_f.size = c_k.size,\\
&c_f.defaultValue = \texttt{NULL}\\
%&\qquad \hilight{QUITAR} t.columns = t.columns \cup \{c_f\}:\forall c_k\in pk.columns   \\
%&\qquad \hilight{c_f \in t.columns} \\
&\}:\forall c_k\in pk.columns
%&\quad
\;\}
\end{aligned}
\]
}
The generated columns are distinct model elements rather than copies
of the original key columns. Only the structural properties required
to represent the foreign key are derived from them. Their
nullability is established by the rule in which the function is
invoked.

\ruletitle{UR1}{USchema Mapping}

A \texttt{USchema} $uS$ is mapped to a
\texttt{RelationalSchema} $rS$, preserving its name.

{\fontsize{9}{10}\selectfont
\[
\begin{aligned}
uS \rightarrow rS\;\|\;
\{\, &rS.name = uS.name
\,\}
\end{aligned}
\]
}

The entity and relationship types contained in $uS.entities$ and
$uS.relationships$, respectively, are transformed according to
Rules~UR2--UR8.

%\paragraph{Rule R2 -- EntityType Mapping}
\ruletitle{UR2}{EntityType Mapping}

Each \texttt{EntityType} $e$, whether root or non-root, 
is mapped to a \texttt{Table} $t$ with the same name.

{\fontsize{9}{10}\selectfont
\[
\begin{aligned}
e \rightarrow t\;\|\;
\{\, &t.name = e.name
\,\}
\end{aligned}
\]
}

The table $t$ is added to $rS.tables$, where
$rS$=\texttt{map}($e.owner$). The features of $e$ are transformed
according to Rules~UR3--UR6.

%\paragraph{Rule R3 -- Attribute Mapping}
\ruletitle{UR3}{Attribute Mapping}

Each \texttt{Attribute} $at$ whose type is a
\texttt{PrimitiveType} and that is not associated with a
\texttt{Reference} (i.e., $at.references$ is \texttt{NULL}) is mapped to a
\texttt{Column} $c$. Attributes associated with a
\texttt{Reference} are instead handled when that reference is mapped,
since its mapping determines the table in which their corresponding
columns must be created.

\vspace{-0.5cm}
{\fontsize{9}{10}\selectfont
\[
\begin{aligned}
at \rightarrow \,& c\,\|\,\{\\
    &c.name = at.name,\\
    &c.datatype = \texttt{primitiveTypeMap}(at.type.name),\\
    &c.size = \texttt{defaultSize}(c.datatype),\\
    &c.nullable = \text{true},\\
    &c.defaultValue = \texttt{NULL}
\,\}
\end{aligned}
\]
}

The generated column $c$ is added to
$\texttt{map}(at.owner).columns$, that is, to the table corresponding
to the schema type that owns $at$. Its \textit{nullable} value is
subsequently set to false by Rule~UR4 if $at$ participates
in an identifier key. 
% \mjose{[MJ] Quitaría la frase que sigue, pq ya está explicado en la intro de este mapping}
% Here, \texttt{defaultSize} returns 
% the default size adopted for the specified relational datatype.

%\paragraph{Rule R4 -- Key Mapping}
\ruletitle{UR4}{Key Mapping}

% Each U-Schema \texttt{Key} $k$ is mapped to a relational
% \texttt{Key} $rk$. The value of $k.isID$ determines whether $rk$
% is a primary or an alternate key. If $rk$ is a primary key, all
% its columns are marked as non-nullable.

Each U-Schema \texttt{Key} $k$ is mapped to a relational
\texttt{Key} $rk$. The value of $k.isID$ determines whether $rk$
is a primary or an alternate key. A canonical constraint name
is assigned to primary keys, whereas the names of alternate
keys are preserved. If $rk$ is a primary key, all its columns
are marked as non-nullable.

Let $t=\texttt{map}(k.owner)$ be the table corresponding to the
entity type that owns $k$, and let

\vspace{-0.5cm}
{\fontsize{9}{10}\selectfont
\[
C_k =
\{\,\texttt{map}(at)\mid at\in k.attributes\,\}
\]
}
denote the set of relational columns corresponding to the
attributes of $k$.
% \[
% \begin{aligned}
% &k \rightarrow rk\;\|\;\{\\
% &\quad rk.isPK = k.isID,\\
% &\quad
% \bigl(\neg k.isID \lor k.owner.root\bigr)
% \Rightarrow rk.columns = C_k,\\
% &\quad
% \bigl(k.isID \land \neg k.owner.root\bigr)
% \Rightarrow C_k \subseteq rk.columns,\\
% &\quad k.isID \Rightarrow
% rk.constraintName =
% \texttt{concat}(t.name,\texttt{`\_pk'}),\\
% &\quad \neg k.isID \Rightarrow
% rk.constraintName = k.name,\\
% &\quad rk.isPK \Rightarrow
% \bigl(
% \{\,c_i.nullable=\texttt{false}\,\}:
% \forall c_i\in rk.columns
% \bigr)
% \;\}
% \end{aligned}
% \]
% \mjose{Mejorable? Algo más concisa:}

\vspace{-0.5cm}
{\fontsize{9}{10}\selectfont
\[
\begin{aligned}
&k \rightarrow rk\;\|\;\{\\
&\quad rk.isPK = k.isID,\\
&\quad
(\neg k.isID \lor k.owner.root)
\Rightarrow rk.columns = C_k,\\
% &\hilight{\quad
% (k.isID \land \neg k.owner.root)
% \Rightarrow C_k \subseteq rk.columns,}\\
%&\qquad \mjose{\text{(homogeneidad con otras reglas)en vez de lo rojo, poner:}}\\
&\quad (k.isID \land \neg k.owner.root) \\
    &\hspace{2cm} \Rightarrow rk.columns=rk.columns \cup C_k,\\
&\quad \neg k.isID \Rightarrow rk.constraintName = k.name,\\
&\quad \, k.isID \\
    &\qquad \Rightarrow \{\,rk.constraintName = \\
  &\hspace{4cm}\texttt{concat}(t.name,\texttt{`\_pk'}),\\
  &\qquad \quad \;\; \{\,c_i.nullable=\text{false}\,\}: \forall c_i\in rk.columns
  \,\}
\;\}
\end{aligned}
\]
}
The generated key $rk$ is added to $t.keys$. For the identifier
key of a non-root \texttt{EntityType}, $C_k$ represents only its
local identifier; Rule~UR6 completes $rk.columns$ with the
foreign key columns that identify the owner entity.

The columns in $C_k$ are generated by Rule~UR3 or, when their
corresponding attributes are associated with a \texttt{Reference},
by Rule~UR5.

% \mjose{¿Por qué hablar de Document si este mapping es de USchemaToRelational?VER SI SIRVE PARA ROUNDTRIP}
% Since document schemas retain identifier fields but not the original
% names of relational key constraints, UR4 assigns a canonical name
% to primary keys. Alternate keys are not reconstructed from document
% schemas; when they are present in the input U-Schema model, their
% names are preserved.

%\paragraph{Rule R5 -- Reference Mapping}
\vspace{1ex}
\ruletitle{UR5}{Reference Mapping}

This rule maps a U-Schema \texttt{Reference} $ref$ representing a
direct link between two \texttt{EntityType} instances to either a
foreign key or an associative table in the relational schema.

References associated with a \texttt{RelationshipType}, i.e.,
those for which $ref.isFeaturedBy$ is not null, are mapped
together with their corresponding relationship type by Rule~UR7.
Synthesized opposite references are auxiliary elements used to record
inverse multiplicity and are not mapped independently, as explained in
Section~\ref{sec:metamodels}. The following predicate identifies
synthesized references.

\vspace{-0.5cm}
{\fontsize{9}{10}\selectfont
\[
\begin{aligned}
\mathrm{SynthesizedReference}(r) \equiv {}&
r.attributes = \emptyset \;\land\\
&r.opposite \neq \texttt{NULL} \;\land\\
&r.opposite.attributes \neq \emptyset
\end{aligned}
\]
}

Let $t$ = \texttt{map}($ref.owner$) and
$tr$ = \texttt{map}($ref.refsTo$) be the tables corresponding to the
source and target entity types of $ref$, respectively. 
Let also $pkt$ = \texttt{pk}($t$) and $pkr$ = \texttt{pk}($tr$) denote
their primary keys.

Three variants are distinguished according to the direct and inverse
multiplicities represented by $ref$ and its optional synthesized
opposite reference: \textit{(i)} a single-valued reference,
representing an N:1 relationship when no opposite exists, and a 1:1
relationship when the opposite has upper bound one;
\textit{(ii)} a multi-valued reference whose opposite has upper bound one,
representing a 1:N relationship; and \textit{(iii)} a multi-valued reference with
no opposite, representing an M:N relationship.

\vspace{1ex}
\textbf{UR5.1}: \emph{Single-Valued Reference (N:1 or 1:1).}
%\jesus{@mariajo debes comprobar si el cambio atendiendo a tu petición es correcto,
%he dejado la definición de fkColumns porque la tuya es menos legible} 
%\mjose{OK}

If $ref.upperBound$ == 1, the reference is mapped to a foreign key
$fk$ in table $t$ that references $tr$, thus preserving the
direction of the reference. This variant represents an N:1
relationship when no synthesized opposite reference exists, and a
1:1 relationship when $ref$ has a synthesized opposite reference
with upper bound one.

% The columns of $fk$ are obtained from the attributes associated
% with $ref$. If $ref$ has no associated attributes, they are instead
% derived from the columns of the referenced primary key $pkr$.
% The auxiliary function \texttt{fkColumns}, defined below, captures
% these two cases. When $ref$ has no associated attributes, the nullability of the
% columns derived from the referenced primary key is determined by
% the lower bound of $ref$.

The columns of $fk$ are determined by the auxiliary function
\texttt{fkColumns}.
%\texttt{fkColumns($ref,pk,t$)} \mjose{Yo quitaría los argumentos y dejaría solo el nombre de la función}
When $ref$ has associated attributes, each attribute is mapped to a fresh 
column preserving its name and datatype and translating its optionality into 
column nullability.
% \jesus{lo ultimo se dice porque deriveFKColumns que se usa en UD5.3, UD6 y UD7,
% no incluye la asignación de nullable pero Cfk sí, y he procurado hacer los cambios mínimos}
% \mjose{OK!}

% \mjose{¿traer aqui la definición de fkColumns y $C_{fk}$?}

Let
\vspace{-0.5cm}
{\fontsize{9}{10}\selectfont
\[
\begin{aligned}
&C_{ref}=\{\,c_i \mid
\{\, at_i \rightarrow c_i,\\
    &\quad c_i.name = at_i.name, \\
    &\quad c_i.datatype = \texttt{primitiveTypeMap}(at_i.type.name),\\
    &\quad c_i.size = \texttt{defaultSize}(c_i.datatype),\\
    &\quad c_i.nullable = at_i.optional,\\
    &\quad c_i.defaultValue = \texttt{NULL}
\,\}:\forall at_i \in ref.attributes
\,\}
\end{aligned}
\]
}

The correspondence $at_i \rightarrow c_i$ ensures that
$\texttt{map}(at_i)$ is defined for Rule~UR4 when an
attribute associated with $ref$ participates in a U-Schema
\texttt{Key}.

The auxiliary function \texttt{fkColumns} is defined as follows:

\vspace{-0.5cm}
{\fontsize{9}{10}\selectfont
\[
\begin{aligned}
ref.attributes \neq \emptyset
&\Rightarrow
\texttt{fkColumns}(ref,pk,t):=C_{ref},\\
ref.attributes = \emptyset
&\Rightarrow
\texttt{fkColumns}(ref,pk,t):={}\\[-1mm]
&\texttt{deriveFKColumns}(pk,t,ref.name)
\end{aligned}
\]
}

When $ref$ has no associated attributes,
\texttt{deriveFKColumns} does not determine column nullability;
in Rule~UR5.1 it is therefore derived from the lower bound of $ref$.

In the 1:1 case, the columns of $fk$ %the foreign key 
must additionally be subject to a uniqueness constraint. Such a 
constraint may already be represented by a U-Schema \texttt{Key} whose
attributes coincide with those associated with $ref$. 
The following predicate identifies this situation:

\vspace{-0.5cm}
{\fontsize{9}{10}\selectfont
\[
\begin{aligned}
&\mathrm{KeyForReference}(ref) \equiv\\
&\quad \exists \,k \in ref.owner.features:\\
&\qquad \qquad \{\,\texttt{class}(k)=\texttt{Key} \, \land \,\\
&\qquad \qquad \; \; k.attributes==ref.attributes\,\}
\end{aligned}
\]
}

The mapping is defined as follows:

\vspace{-0.5cm}
{\fontsize{9}{10}\selectfont
\[
\begin{aligned}
ref &\rightarrow (fk\,[,uk])\;\|\;\{\\
&fk.constraintName = ref.name,\\
& fk.refsTo = pkr,\\
& fk.columns =
       \texttt{fkColumns}(ref,pkr,t),\\
&ref.attributes == \emptyset \Rightarrow \{ \\
&\qquad \, c_i.nullable = (ref.lowerBound == 0)\\
&\qquad \}: \forall c_i \in fk.columns,\\
& fk.onDelete = \texttt{NO\_ACTION},\\
& fk.onUpdate = \texttt{CASCADE},\\
&\bigl(\;
       ref.opposite \neq \text{NULL}\\
&\,\land ref.opposite.upperBound == 1\\
&\,\land\neg\mathrm{KeyForReference}(ref)\bigr) 
    \Rightarrow \{\,ref \rightarrow uk,\\
&\quad uk.constraintName =%\\&\qquad
       \texttt{concat}(ref.name,\text{`\_ak'}),\\
&\quad uk.isPK = \text{false},\\
&\quad uk.columns = fk.columns%\\&
\,\}
\;\}
\end{aligned}
\]
}

The generated foreign key $fk$ is added to $t.fks$, and the
columns in $fk.columns$ are added to $t.columns$. If the optional
key $uk$ is generated, it is added to $t.keys$. Otherwise, in the
1:1 case, the corresponding uniqueness constraint is produced by
Rule~UR4 from the existing U-Schema \texttt{Key}, thereby avoiding
the creation of a duplicate relational key.

In the U-Schema model generated for the running example, 
the single-valued reference 
\texttt{song} in the entity type \texttt{playlist\_song} 
targets the entity type \texttt{song} and
has no synthesized opposite reference. When this U-Schema model is
mapped back to the relational model, the reference represents an
N:1 relationship and is mapped to a foreign key in table
\texttt{playlist\_song} that references the primary key of table
\texttt{song}, thereby preserving the direction of the U-Schema
reference. The associated attribute \texttt{song\_id} is mapped to
the corresponding foreign key column. The U-Schema model of the
running example does not contain a single-valued reference with a
synthesized opposite and therefore does not illustrate the 1:1
case of this variant.

% In the running example, the single-valued reference
% \texttt{song} in the entity type \texttt{playlist\_song} targets
% the entity type \texttt{song} and has no synthesized opposite
% reference. It therefore represents an N:1 relationship and is
% mapped to a foreign key in table \texttt{playlist\_song} that
% references the primary key of table \texttt{song}, preserving the
% direction of the U-Schema reference. The associated attribute
% \texttt{song\_id} is mapped to the corresponding foreign key
% column. The running example does not include a single-valued reference with
% a synthesized opposite and therefore does not illustrate the 1:1
% case of this variant.

\vspace{1ex}
\textbf{UR5.2}: \emph{Multi-Valued Reference (1:N).}

If $ref.upperBound$>1 and $ref$ has a synthesized opposite reference 
whose upper bound is one, then $ref$ represents a 1:N
relationship. Each instance of $t$ may be linked to multiple
instances of $tr$, whereas each instance of $tr$ may be linked to
at most one instance of $t$.

Since a relational 1:N relationship is represented by a foreign key
on the N-side, a foreign key $fk$ is created in $tr$ that references
the primary key $pkt$ of $t$, thus reversing the direction of
$ref$.

% The attributes associated with $ref$ identify the target instances
% in $tr$ and therefore cannot be used as the columns of the reversed
% foreign key. Instead, fresh columns are derived from the components
% of $pkt$ and added to $tr$. Let $C_{fk}$ denote this set of columns:
% \mjose{¿Por qué no usa deriveFKColumns?}

The attributes associated with $ref$ identify the target instances
in $tr$ and therefore cannot be used as the columns of the reversed
foreign key. Instead, fresh columns are derived from the components
of $pkt$ using \texttt{deriveFKColumns}. Their nullability is
determined by the lower bound of the synthesized opposite reference.
% \[
% \begin{aligned}
% C_{fk}=\{\,c_i \mid
% \{\, &c_i.name =
%        \texttt{concat}(t.name,\texttt{`\_'},ct_i.name),\\
%     &c_i.datatype = ct_i.datatype,\\
%     &c_i.size = ct_i.size,\\
%     &c_i.nullable =
%        (ref.opposite.lowerBound = 0),\\
%     &c_i.defaultValue = \texttt{NULL}
% \,\}:\forall ct_i \in pkt.columns
% \,\}
% \end{aligned}
% \]

The mapping is then defined as follows:

\vspace{-0.5cm}
{\fontsize{9}{10}\selectfont
\[
\begin{aligned}
&ref \rightarrow fk\;\|\;\{\\
& \; fk.constraintName =
       \texttt{concat}(t.name,\text{`\_fk'}),\\
&\; fk.refsTo = pkt,\\
&\; fk.columns =
       \texttt{deriveFKColumns}(pkt,tr,t.name),\\
&\;
\{\,c_i.nullable =
       (ref.opposite.lowerBound == 0)\,\}\\[-1mm]
&\qquad :\forall c_i \in fk.columns,\\
&\; fk.onDelete = \texttt{NO\_ACTION}, fk.onUpdate = \texttt{CASCADE}
\;\}
\end{aligned}
\]
}
% \[
% \begin{aligned}
% &ref \rightarrow fk\;\|\;\{\\
% &\quad fk.constraintName =
%        \texttt{concat}(t.name,\texttt{`\_fk'}),\\
% &\quad fk.refsTo = pkt,\\
% &\quad fk.columns = C_{fk},\\
% &\quad fk.onDelete = \texttt{NO\_ACTION},\\
% &\quad fk.onUpdate = \texttt{CASCADE}
% \;\}
% \end{aligned}
% \]

The generated foreign key $fk$ is added to $tr.fks$, and the columns
in $fk.columns$ are added to $tr.columns$. The
synthesized opposite reference is not mapped independently; it is
only used to determine that the inverse upper bound is one and,
through its lower bound, the nullability of the generated
foreign key columns.

In the running example, the multi-valued reference
\texttt{songs} in the entity type \texttt{album} targets
\texttt{song} and has a synthesized opposite reference with upper
bound one. It is therefore mapped to a foreign key in table
\texttt{song} that references the primary key of
\texttt{album}. This reverses the direction of the U-Schema
reference and reconstructs the relational 1:N relationship.

\vspace{1ex}
\textbf{UR5.3}: \emph{Multi-Valued Reference (M:N).}

If $ref.upperBound>1$ and $ref$ has no synthesized opposite
reference, then $ref$ represents an M:N relationship. Each instance
of $t$ may be linked to multiple instances of $tr$, and the same
instance of $tr$ may be linked to multiple instances of $t$.

Since an M:N relationship cannot be represented by a foreign key in
either of the two participating tables, a new associative table $m$
is created. This table contains a foreign key $fk_t$ referencing the
primary key $pkt$ of $t$, a foreign key $fk_{tr}$ referencing the
primary key $pkr$ of $tr$, and a composite primary key $pk_m$ formed
by the columns of both foreign keys.

The columns of $fk_t$ are newly generated from the columns in $pkt$,
whereas those of $fk_{tr}$ are obtained through
\texttt{fkColumns}. The latter function uses the attributes
associated with $ref$ when they exist and otherwise derives the
columns from $pkr$.

\vspace{-0.5cm}
{\fontsize{9}{10}\selectfont
\[
\begin{aligned}
&ref \rightarrow (m,fk_t,fk_{tr},pk_m)\;\|\;\{\\
&\; m.name =
       \texttt{concat}(t.name,\texttt{`\_'},tr.name),\\
&\; fk_t.constraintName =
       \texttt{concat}(t.name,\text{`\_fk'}),\\
&\; fk_t.refsTo = pkt,\\
&\; fk_t.columns = \texttt{deriveFKColumns}(pkt,m,t.name),\\
&\; fk_t.onDelete = \texttt{NO\_ACTION}, fk_t.onUpdate= \texttt{CASCADE},\\
&\; fk_{tr}.constraintName =
       \texttt{concat}(tr.name,\text{`\_fk'}),\\
&\; fk_{tr}.refsTo = pkr,\\
&\; fk_{tr}.columns =
       \texttt{fkColumns}(ref,pkr,m),\\
&\; fk_{tr}.onDelete = \texttt{NO\_ACTION},\\
&\; fk_{tr}.onUpdate= \texttt{CASCADE},\\
&\; pk_m.constraintName =
       \texttt{concat}(m.name,\text{`\_pk'}),\\
&\; pk_m.isPK = \text{true},\\
&\; pk_m.columns =
       fk_t.columns \cup fk_{tr}.columns,\\
&\;
\{\,c_i.nullable=\text{false}\,\}:
\forall c_i \in pk_m.columns
\;\}
\end{aligned}
\]
}

The generated table $m$ is added to $rS.tables$, where
$rS$=\texttt{map}($ref.owner.owner$). The foreign keys $fk_t$ and
$fk_{tr}$ are added to $m.fks$, and the primary key $pk_m$ is added
to $m.keys$. The columns generated for both foreign keys are added
to $m.columns$.

Because all foreign key columns participate in the composite primary
key, they are marked as non-nullable. The composite key also ensures
that each pair of referenced instances occurs at most once in the
associative table.

% \mjose{OJO: esto no está en el running example (figura 12, de colores), quizá hacer referencia a la figura 11}
In the conceptual model of the running example shown in
Figure~\ref{fig:music-uml}, \texttt{song} and
\texttt{musical\_style} participate in an M:N relationship: a song
may have multiple styles, and the same style may be associated with
multiple songs. In the corresponding U-Schema model, this
relationship is represented by the multi-valued reference
\texttt{styles} from \texttt{song} to \texttt{musical\_style},
with no synthesized opposite reference. Although this reference is
omitted in Figure~\ref{fig:running-example-mapping} for space reasons,
Rule~UR5.3 maps it to an associative table containing foreign keys
to \texttt{song} and \texttt{musical\_style}, together with a
composite primary key formed by the columns of both foreign keys.

\ruletitle{UR6}{Aggregate Mapping}

Each \texttt{Aggregate} $ag$ is mapped to a foreign key $fk$ in
the table $tg$ corresponding to the non-root \texttt{EntityType}
referenced by $ag.specifiedBy$. This foreign key references the
table $t$ corresponding to the owner entity type of $ag$.

The columns of $fk$ are included in the primary key of $tg$,
thereby defining $tg$ as a weak table. The identifier key
mapped by Rule~UR4 provides the columns corresponding to the local
identifier of the non-root entity type, whereas this rule completes
the primary key with the columns identifying its owner.

Let $t$ = \texttt{map}($ag.owner$),
$pkt$ = \texttt{pk}($t$), 
$tg$ = \texttt{map}($ag.specifiedBy$), and
$pkg$ = \texttt{pk}($tg$).

\vspace{-0.5cm}
{\fontsize{9}{10}\selectfont
\[
\begin{aligned}
&ag \rightarrow fk\;\|\;\{\\
&\quad fk.constraintName =
       \texttt{concat}(t.name,\text{`\_fk'}),\\
&\quad fk.refsTo = pkt,\\
&\quad fk.columns =
       \texttt{deriveFKColumns}(pkt,tg,t.name),\\
&\quad fk.onDelete = \texttt{NO\_ACTION},%\\&\quad 
fk.onUpdate = \texttt{CASCADE},\\
%&\quad \hilight{fk.columns \subseteq pkg.columns,}\\
&\quad pkg.columns = pkg.columns \cup fk.columns,\\
&\quad
\{\,c_i.nullable=\text{false}\,\}:
\forall c_i \in fk.columns
\;\}
\end{aligned}
\]
}
The generated foreign key $fk$ is added to $tg.fks$, and the
columns in $fk.columns$ are added to $tg.columns$. Since these
columns are also included in $pkg.columns$, they form, together
with the columns corresponding to the local identifier of
$ag.specifiedBy$, the composite primary key of the generated
weak table.

In the running example, the aggregate \texttt{playlists} owned by
\texttt{app\_user} and specified by the non-root entity type
\texttt{playlist} is mapped to a foreign key in table
\texttt{playlist} that references the primary key of
\texttt{app\_user}. The generated foreign key columns are combined
with the local identifier \texttt{playlist\_id} to form the
composite primary key of \texttt{playlist}.
Similarly, the aggregate \texttt{playlist\_songs} is mapped to a
foreign key in table \texttt{playlist\_song} that references
\texttt{playlist}. Its columns and the local identifier 
\texttt{position\_idx} together form the primary key of the nested weak table.

\ruletitle{UR7}{RelationshipType Mapping}

Each \texttt{RelationshipType} $rt$ is mapped to a new associative
\texttt{Table} $m$. Each reference in $rt.references$ represents
one of the entity types participating in the relationship and is
mapped to a foreign key in $m$.

For each $r_i \in rt.references$, let
$t_i$=\texttt{map}($r_i.refsTo$) be the table corresponding to its
target entity type, and let $pk_i$=\texttt{pk}($t_i$) denote the
primary key of that table.

\vspace{-0.5cm}
{\fontsize{9}{10}\selectfont
\[
\begin{aligned}
rt& \rightarrow (m,rk)\;\|\;\{\\
&m.name = rt.name,\\
&\{\;r_i \rightarrow fk_i,\\
    &\; fk_i.constraintName = r_i.name,\\
    &\; fk_i.refsTo = pk_i,\\
    &\; fk_i.columns =
       \texttt{deriveFKColumns}(pk_i,m,r_i.name),\\
    &\; fk_i.onDelete = \texttt{NO\_ACTION},\\
    &\; fk_i.onUpdate = \texttt{CASCADE} \\
&\}:\forall r_i\in rt.references,\\
&rk.constraintName =
       \texttt{concat}(m.name,\text{`\_pk'}),\\
&rk.isPK = \texttt{true},\\
&rk.columns =
       \bigcup_{fk_i \in m.fks} fk_i.columns,\\
&\{\,c.nullable=\text{false}\,\}:
\forall c\in rk.columns
\;\}
\end{aligned}
\]
}

The table $m$ is added to $rS.tables$, where
$rS=\texttt{map}(rt.owner)$. Each generated foreign key $fk_i$ is
added to $m.fks$, and its columns are added to $m.columns$. The
primary key $rk$ is added to $m.keys$.

The attributes in $rt.features$ are transformed into columns by
Rule~UR3 and added to $m.columns$. The primary key of $m$ consists
of the columns of all the generated foreign keys, so each
combination of participating entity instances identifies at most one
relationship instance.

In the running example, the relationship type
\texttt{listening} is mapped to an associative table with the same
name. Its references to \texttt{app\_user} and \texttt{song} are
mapped to foreign keys referencing the primary keys of the
corresponding tables. The columns of both foreign keys form the
composite primary key of \texttt{listening}, while the relationship
attributes \texttt{plays\_count} and \texttt{status} are mapped to
regular columns.

\ruletitle{UR8}{Primitive Type Mapping}

For this transformation, \texttt{primitiveTypeMap} maps the supported
\uschema{} primitive type names to relational datatype values as follows:
\texttt{String}$\rightarrow$\texttt{VARCHAR},
\texttt{Integer}$\rightarrow$\texttt{INTEGER},
\texttt{Double}$\rightarrow$\texttt{DOUBLE\_PRECISION},
\texttt{Decimal}$\rightarrow$\texttt{NUMERIC},
\texttt{Boolean}$\rightarrow$\texttt{BOOLEAN},
\texttt{Date}$\rightarrow$\texttt{DATE},
\texttt{Time}$\rightarrow$\texttt{TIME}, and
\texttt{Datetime}$\rightarrow$\texttt{TIMESTAMP}.

% For the \uschema{}-to-Relational transformation,
% \texttt{primitiveTypeMap} maps the supported \uschema{}
% \texttt{PrimitiveType} names to relational datatype values
% as follows:

% % Each U-Schema \texttt{PrimitiveType} is mapped to the
% % corresponding SQL data type as follows:
% \vspace{-0.4em}
% \[
% {\small
% \begin{aligned}
% \texttt{String}   &\rightarrow \texttt{VARCHAR}\\
% \texttt{Integer}  &\rightarrow \texttt{INTEGER}\\
% \texttt{Double}   &\rightarrow \texttt{DOUBLE PRECISION}\\
% \texttt{Decimal}  &\rightarrow \texttt{NUMERIC}\\
% \texttt{Boolean}  &\rightarrow \texttt{BOOLEAN}\\
% %\end{aligned}
% %\]
% %\[
% %\begin{aligned}
% \texttt{Date}     &\rightarrow \texttt{DATE}\\
% \texttt{Time}     &\rightarrow \texttt{TIME}\\
% \texttt{Datetime} &\rightarrow \texttt{TIMESTAMP}
% \end{aligned}
% } %end small
% \]
The size, precision, or scale of the generated relational columns,
when applicable, is determined by the default values adopted by the
rules that create those columns. Structured U-Schema types, such as
\texttt{PList}, \texttt{PSet}, \texttt{PTuple}, and \texttt{PMap},
are not currently supported by this transformation.

\end{document}